\documentclass{jpp}
\usepackage{graphicx}
\usepackage[dvipsnames]{xcolor}
\usepackage{subcaption}

\usepackage[utf8]{inputenc}
\usepackage[T1]{fontenc}
\usepackage{amsmath}
\usepackage{flexisym}
\usepackage{wasysym}
\usepackage{natbib}
\usepackage{enumerate}   
\usepackage{braket}
\usepackage{tabularx}
\usepackage{booktabs}
\usepackage{adjustbox}
\usepackage{hhline}

\newcommand{\be}{\begin{equation}}
\newcommand{\ee}{\end{equation}}
\newcommand{\beq}{\begin{equation}}
\newcommand{\eeq}{\end{equation}}
\newcommand\bea{\begin{eqnarray}}
\newcommand\eea{\end{eqnarray}}

\usepackage{flexisym}
\usepackage{wasysym}
\usepackage{natbib}
\usepackage{enumerate}   
\usepackage{braket}

\newcommand{\figref}[1]{\figurename~\ref{#1}}

\usepackage[final]{hyperref} 
\hypersetup{
	linkcolor=black,        
	citecolor=black,        
	filecolor=magenta,     
	urlcolor=black         
}

\shorttitle{Island kinetic equation}

\title{Statistical description of coalescing magnetic islands via magnetic reconnection}

\author{Muni Zhou\aff{1},
     David H. Wu\aff{2}, 
      Nuno F. Loureiro\aff{1} 
\and Dmitri A. Uzdensky\aff{3}}

\affiliation{\aff{1}Plasma Science and Fusion Center$,$ Massachusetts Institute of Technology$,$ Cambridge$,$ MA 02139$,$ USA
\aff{2} Department of Physics, California Institute of Technology$,$ Pasadena$,$ CA 91125$,$ USA
\aff{3}Center for Integrated Plasma Studies$,$ Physics Department$,$ UCB-390$,$ University of Colorado$,$ Boulder$,$ CO 80309$,$ USA}

\begin{document}

\maketitle

\begin{abstract}
The physical picture of interacting magnetic islands provides a useful paradigm for certain plasma dynamics in a variety of physical environments, such as the solar corona, the heliosheath, and the Earth's magnetosphere.
In this work, we derive an island kinetic equation to describe the evolution of the island distribution function (in area and in flux of islands) subject to a collisional integral designed to account for the role of magnetic reconnection during island mergers. 
This equation is used to study the inverse transfer of magnetic energy through the coalescence of magnetic islands in 2D. 
We solve our island kinetic equation numerically for three different types of initial distribution: delta-distribution, Gaussian and power-law distribution.
The time evolution of several key quantities is found to agree well with our analytical predictions: magnetic energy decays as $\tilde t^{-1}$, the number of islands decreases as $\tilde t^{-1}$, and the averaged area of islands grows as~$\tilde t$, where $\tilde t$ is the time normalized to the characteristic reconnection time scale of islands. 
General properties of the distribution function and the magnetic energy spectrum are also studied.
Finally, we discuss the underlying connection of our island-merger models to the (self-similar) decay of magnetohydrodynamic turbulence.
\end{abstract}

\section{Introduction}
\label{sec:introduction}
Magnetic flux tubes are often observed in the heliosphere and inferred to exist in some space and astrophysical systems. 
These structures have been observed in, for example, the solar wind \citep{moldwin1995ulysses,moldwin2000small,cartwright2010heliospheric,hu2018automated}, the Earth's magnetosphere \citep{borg2012,oieroset2016}, the solar corona \citep{zhang2012,wang2006}, and are likely to be present in the heliosheath ~\citep{stone2005voyager,stone2008asymmetric,opher2011magnetic}. 
The dynamics of magnetic flux tubes are thought to be important for many physical phenomena such as plasma heating in the solar corona \citep{parker1972,parker1983part2,parker1988nanoflares,galsgaard1996a,holman2003electron,velli1999,dmitruk1999,Klimchuck2006solving,Klimchuk2008} and the acceleration of anomalous cosmic rays (ACRs)~\citep{lazarian2009model,drake2010magnetic}.

In the heliosphere, various observations show evidence of merging of magnetic flux tubes. 
{\it In-situ} measurements by spacecraft such as the Wind, ACE and Ulysses find small-scale flux tubes to be ubiquitous in the solar wind~\citep{moldwin1995ulysses,moldwin2000small,cartwright2010heliospheric,hu2018automated}.
As the flux tubes are convected radially outward by the solar wind, quantities such as their number, magnetic field strength and inverse scale length decrease beyond what is naturally expected from simple expansion due to diverging outflows.
It is possible that flux tube merging plays a role in explaining these observations, as shown by the correlation of flux tube structures as they travel past spacecraft in different locations \citep{hu2019}. 

In the heliosheath (the interface between the heliosphere and the interstellar medium), 
recent observations from the two Voyager spacecraft~\citep{stone2005voyager,stone2008asymmetric} strongly suggest that this region is composed of a turbulent sea of magnetic flux tubes~\citep{opher2011magnetic,scheoffler2011}.
Understanding the dynamics of these flux tubes is thus critical to developing a model of the heliosheath, as well as inferring its efficiency at accelerating particles. Indeed, magnetic reconnection in this ``sea'' of flux tubes has been proposed as an important mechanism for the acceleration of ACRs~\citep{lazarian2009model,drake2010magnetic}, providing motivation to understand the merging of flux tubes using a statistical approach. 

The solar corona is populated by magnetic flux tubes of widely different sizes whose ends (footpoints on the solar surface) are constantly being shuffled by the turbulent motion of the photosphere. 
While a detailed understanding of solar coronal dynamics still eludes us, flux tube dynamics (including processes like Taylor relaxation, reconnection and wave-driven activity) is believed to be a key element in determining the nature of its turbulent state~\citep{parker1983}--- and, in particular, in addressing longstanding problems such as the origin of significant populations of supra-thermal electrons ~\citep{galsgaard1996a,holman2003electron}, and the anomalously high temperatures in the corona~\citep{parker1972,parker1983part2,parker1988nanoflares,velli1999,dmitruk1999,Klimchuck2006solving,Klimchuk2008}. 
Similar problems exist in accreting systems, where the hot and tenuous corona above an accretion disk can be heated through the interaction of a large number of magnetic flux loops tied to the disk~\citep{Galeev1979structured,uzdensky2008statistical}.

{\it In situ} evidence of magnetic reconnection at the bow shock and in the turbulent magnetosheath of the Earth has been provided~\citep{retino2007situ,phan2018electron,gingell2019observations}. 
Large-scale reconnection between the interplanetary magnetic field and that of the Earth generates magnetic flux tubes, which are often discussed in the form of their two-dimensional (2D) counterpart: magnetic islands (plasmoids). How these magnetic islands evolve after reconnection ceases, and how they change the structure of the magnetic field in the magnetosphere, is still not clear. 

The configuration of interacting flux tubes is also relevant to the magnetic fields generated by the Weibel instability, which typically possess a filamentary structure~\citep{weibel1959,burton1959}.
Such Weibel fields can be produced in, for example, a collisionless shock~\citep{medvedev1999generation,kato2008,spitkovsky2008}, relevant to a wide range of systems such as gamma-ray burst (GRB) jets and supernova explosions. 
A recent study has found that the Weibel instability can also be triggered by generic plasma motions as simple as a shear flow to produce the ``seed'' magnetic fields~\citep{zhou2021spontaneous}, supporting the conjecture that the Weibel instability is a plausible key ingredient of magnetogenesis~\citep{schlicheiser2003,Lazar2009}.
However, it is not {\it a priori} clear whether such kinetic-scale magnetic fields can grow to larger scales via flux-tube coalescence and survive on long time scales before they are dissipated.
Similar questions exist in the context of GRB prompt emission and afterglow, such as whether the Weibel fields generated in a relativistic shock can last long enough to explain the observed powerful synchrotron emission~\citep{medvedev1999generation,silva2003interpenetrating,medvedev2004long,ruyer2018}.

Because of the importance of flux tubes in these physical environments, their dynamics has been intensively studied theoretically~\citep{gruzinov2001gamma,linton2001reconnection,medvedev2004long,kato2005saturation,linton2006reconnection,lyutikov2017a,lyutikov2017b,zrake2017turbulent,zhou2019magnetic,zhou2020multi} and experimentally~\citep{yamada1997study,jara2016laboratory,furno2005coalescence,intrator2013flux,gekelman2016pulsating}.
Recently, the role of magnetic flux tubes in plasmoid-mediated turbulence has also been investigated both theoretically and numerically~\citep{loureiro2017collisionless,loureiro2017role,boldyrev2017magnetohydrodynamic,mallet2017disruption,comisso2018magnetohydrodynamic,dong2018role}. 
Despite being widely explored, there remain important questions regarding the dynamics of a system composed of a large number of flux tubes. 
These include the identification of the main underlying physical processes governing their interaction, the statistics of interacting flux tubes in different astrophysical contexts, and the time evolution of macroscopic quantities (such as the strength and length scale of magnetic fields) as a result of flux tube interaction.  

Many of the theoretical and numerical studies addressing the aforementioned questions involving large numbers of flux tubes (or magnetic islands in 2D) have focused on their generation in reconnection by the plasmoid instability~\citep{loureiro2007instability}. The plasmoid instability in elongated current sheets with Lundquist number $S \gtrsim 10^4$ has been shown to be crucial for understanding the fast reconnection rate in resistive magnetohydrodynamic (MHD)~\citep{lapenta2008self,bhattacharjee2009fast,samtaney2009formation,huang2010scaling,uzdensky2010fast,loureiro2012magnetic}. In the kinetic regime, plasmoid generation is also observed in simulations [e.g., \cite{drake2006island,daughton2011}], and is thought to play an essential role in particle energization [e.g., \cite{drake2012power,cerutti2013,dahlin2014,sironi2014,Guo2015,werner2016,petropoulou2016blazar,Sironi2016plasmoids,li2019formation,Guo2019,uzdensky2020relativistic,Hakobyan2020a}].

While direct numerical simulations are an important tool for understanding the dynamics of plasmoids, studies at realistic scales are infeasible due to limited computational resources. 
For example, the length of the current sheet in solar corona is about $10^6 d_i$ (where $d_i$ is the ion skin depth), while PIC simulations of sub-relativistic reconnection with realistic mass ratios only have lengths of tens of $d_i$ [e.g.,~\cite{le2013,cazzola2016,liu2014}] in 2D, while reduced mass ratios are required to perform simulations with similar length scales in 3D~\citep{daughton2011,liu2013bifurcated,dahlin2017,li2019formation}.
As for the relativistic regime, simulations reach lengths of hundreds of $d_i$ with electron-proton plasmas~\citep{Werner2018,ball2018electron} and thousands of $d_e$ with electron-positron plasmas [e.g.,~\cite{sironi2014,Sironi2016plasmoids,werner2016,Guo2019,guo2020magnetic}].
This motivates the development of statistical models to study the phase-space distributions of large numbers of plasmoids.
A heuristic model based on the one-dimensional multilevel plasmoid hierarchy induced by plasmoid instability in a reconnection layer was proposed by~\citet{uzdensky2010fast} and numerically tested by~\citet{loureiro2012magnetic}.
Boltzmann-type equations to describe the evolution of plasmoid distributions in reconnection have been developed by~\citet{fermo2010statistical, huang2012distribution}, and a conceptually similar approach has also been used to study the distribution of magnetic loops in an accretion disk corona~\citep{uzdensky2008statistical}. Other statistical approaches have been used by more recent work [e.g.,~\citet{lingam2018maximum}].

In this work, we adopt a similar statistical approach and derive a Boltzmann-type island kinetic equation (IKE).
Our IKE is extended from previous statistical models~\citep{fermo2010statistical,huang2012distribution} to a more general case involving a large number of randomly distributed and freely-moving islands, rather than islands constrained to one-dimensional (1D) motions within a reconnecting current sheet. 
One major difference in our IKE concerns the dynamics of merging magnetic islands. While previous models assume island merging to be instantaneous, we, instead, calculate the time taken for islands to merge based on reconnection models, and evaluate its effects on the distribution function.
Additionally, whereas previous models focus on the steady state, we are more interested in the time evolution of the system; for example, how macroscopic quantities such as island number, energy density, and magnetic field length scale change with time.
In particular, we investigate two fundamental questions regarding the self-dynamics of magnetic islands: (1) how do magnetic islands born at small spatial scales evolve to macro-scale objects? and (2) how efficient is the associated inverse transfer of magnetic energy in this process?

In~\citet{zhou2019magnetic,zhou2020multi}, we derived a solvable analytical model for 2D and 3D systems (named the 
``hierarchical model'' in what follows) to describe the evolution of initially small-scale magnetic fields via their successive hierarchical coalescence enabled by magnetic reconnection.
Our hierarchical model is based on the conservation of mass and of magnetic flux during the merging process and, therefore, can be applied to MHD-scale fields, regardless of plasma collisionality (it can also be generalized to kinetic-scale fields with an adequate replacement of the ideal invariants of the system).
Our theory identifies magnetic reconnection as the key mechanism enabling the field evolution and determining the properties of such evolution: magnetic energy is found to decay as $\tilde t^{-1}$, where $\tilde t$ is the time normalized to the (appropriately defined) reconnection time scale; the correlation length of the field grows as $\tilde t^{1/2}$, and the number density of magnetic islands evolves as $\tilde t^{-1}$.

The hierarchical model was directly verified by 2D and 3D reduced-MHD (RMHD) simulations that we carried out~\citep{zhou2019magnetic,zhou2020multi} and its applicability to the inverse transfer phenomenon in isotropic turbulent MHD systems was suggested by~\citet{bhat2021inverse}. 
This reconnection-controlled evolution of MHD turbulence is also observed and more systematically studied by~\citet{hosking2021}, with a focus on the role of invariants.
Despite the good agreement between our hierarchical model and RMHD simulations, some questions remain which can be addressed by the IKE studied in this paper. 
The hierarchical model has a strict assumption that only mergers between similar islands (or flux tubes in 3D) are taken into consideration, and our RMHD simulations start with identical islands.  
It is not clear whether the scaling laws of energy decay and growth of length scale will change significantly with a non-trivial distribution of islands that can freely interact.
A study of the interaction of an ensemble of non-identical islands with fewer constraints is thus still needed. This can be modeled by the IKE. 
Also, in our 3D RMHD simulations, although the dynamics of the system are shown to be dominated by the merging of magnetic islands, other nonlinear processes, such as the direct cascade through self-developed turbulence, also play a role. 
Therefore, it is challenging to identify the nature and underlying mechanisms of the inverse energy transfer observed in such complex systems. 
Using the IKE, which only contains the dynamics of merger and the convective motion of magnetic islands, enables us to isolate the contribution of the island dynamics in the overall evolution of the system.

In this paper, we apply our IKE to study the evolution of a distribution of magnetic islands whose dynamics is dominated by their merger enabled by magnetic reconnection. In Section~\ref{sec:merger_IKE}, we describe the statistical approach and the inclusion of a finite merging time; this is followed by a discussion of the numerical implementation of the IKE in Section~\ref{sec:numerical}. 
We then study the evolution of several different initial island distributions and analyse the results in Section~\ref{sec:NR}. 
An interpretation of these numerical results using a scaling theory based on invariants and a discussion on the relevance of our IKE to the decaying turbulence problem are presented in Section~\ref{sec:scaling_law}.
Finally, we conclude in Section~\ref{sec:conclusion}. 
A comparison of our IKE with previous models is described in Appendix~\ref{sec:comparison}, a numerical convergence study is described in Appendix~\ref{sec:convergence}, and a derivation of multi-scale rules of successive merger is included in Appendix~\ref{sec:multi_scale_rules}.

\begin{figure}
\centering
\includegraphics[width=1.0\linewidth]{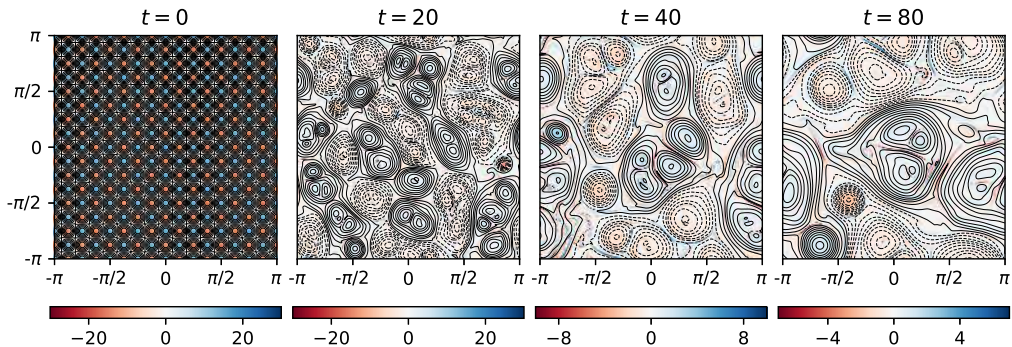}
\caption{Evolution of current density (colors) and magnetic flux (contours) from a 2D (reduced) MHD simulation. As time evolves, islands merge and form progressively larger structures. This figure is reproduced from ~\citet{zhou2019magnetic} for illustration purposes.}
\label{fig:mhd_contour}
\end{figure}

\section{Merger-dominant Island Kinetic Equation}
\label{sec:merger_IKE}
In this section we first introduce our Boltzmann-type island kinetic equation (IKE), focusing on how the coalescence of islands is included in this model. While earlier studies of island distributions [e.g.~\cite{fermo2010statistical,huang2012distribution}] focus on islands in 1D current sheets, we study a sea of interacting islands in two spatial dimensions (2D). This leads to differences between our model and those studies (discussed in Appendix~\ref{sec:comparison}).

In this work, we are primarily interested in a 2D system with volume-filling interacting magnetic islands. To provide an intuitive physical picture of the kinds of systems that our work aims to address, we show in Figure~\ref{fig:mhd_contour} a visualisation of current and magnetic flux from one of the 2D RMHD simulations reported in~\citet{zhou2019magnetic}. The system is tightly packed with islands growing in size through successive mergers and exhibits chaotic dynamics due to the random motion of a large number of islands. 
Such (idealised) system provides a useful paradigm to investigate magnetically-dominated decaying turbulence, relevant to a wide range of astrophysical systems, as discussed in Section~\ref{sec:introduction}. 
In this work, we develop an IKE to describe the dynamics of interacting magnetic islands, thereby illuminating the underlying physical mechanisms governing the dynamics of decaying turbulence.
We note that although the system of volume-filling islands is the main focus of our study, our IKE can also describe systems only partially filled with islands, as we will discuss in Sections~\ref{sec:macro} and~\ref{sec:delta}. 

Consider, therefore, a 2D system in which each island can be characterized by its area $A$ and the magnetic flux $\psi$ it encloses.
A time-dependent distribution function of islands $f(A,\psi,t)$ is introduced to describe an ensemble of magnetic islands in the $(A, \psi)$ phase space. 
The distribution function is normalized as $ \int f(A,\psi,t) dA d\psi = N(t)$ where $N(t)$ is the total number of islands at any given time~$t$.
No spatial location information of islands is contained in this distribution function.
We assume the islands to be randomly and isotropically distributed in space, leading to a uniform number density of islands in real space $n(t) = N(t)/L^2$, where $L$ is the length scale of the system.
The distribution function is considered as the most fundamental quantity of the system, from which macroscopic quantities (total magnetic energy, number of islands, etc), as well as magnetic energy spectrum, can be derived.
The evolution of $f(A,\psi,t)$ is governed by the IKE:
\begin{equation}
    \frac{\partial f}{\partial t}+\frac{\partial \left(\dot{\psi} f\right)}{\partial \psi}+\frac{\partial \left(\dot{A}f\right)}{\partial A} = \mathcal{C}_{\text{new}}+\mathcal{C}_{\text{loss}}+\mathcal{C}_{\text{merge}}.
    \label{eq:boltzmann_full}
\end{equation}

This equation was originally postulated for islands in a 1D reconnecting current sheet \citep{fermo2010statistical}, in which the ongoing reconnection causes the flux and area of the islands to grow. The second and third terms on the left-hand side describe this effect. In this work, we study an ensemble of merging islands in free space instead of those within a large-scale reconnecting curent sheet. For each isolated island, flux $\psi$ and area $A$ are both conserved quantities, i.e., $\dot{\psi}$ and $\dot{A}$ are zero. We therefore neglect the two convective terms on the left-hand side in the remainder of the paper.

On the right-hand side, three operators are introduced to take into account different processes. The operator $\mathcal{C}_{\text{new}}$ represents the generation of new islands in the system, $\mathcal{C}_{\text{loss}}$ represents the loss of islands either by advection out of the system or by resistive dissipation, and $\mathcal{C}_{\text{merge}}$ represents the change of the distribution function resulting from the coalescence of islands. 

In this work, we study an effectively infinite (or periodic) system with small enough resistivity that the resistive decay of islands can be ignored; thus, we assume $\mathcal{C}_{\text{loss}}=0$.
As for $\mathcal{C}_{\text{new}}$: we investigate reconnection both in the MHD regime and in the collisionless regime;
and, in both regimes, we for simplicity assume constraints that prevent the formation of secondary plasmoids during island-merger events, thus ensuring that $\mathcal{C}_{\text{new}}=0$. In the MHD reconnection regime, this means that the Lundquist number associated with each island is relatively low, $S \lesssim 10^4$~\footnote{The Lundquist number of islands, $S$, is preserved in mergers of identical islands and therefore, if islands start with $S \lesssim 10^4$, the system will remain in the single-X-point regime even as islands grow~\citep{zhou2019magnetic,zhou2020multi}.
In the case of mergers between non-identical islands, according to the merger rules that we will explain later in the paper, the Lundquist number of the resulting island will be the same as that of the previous island with larger magnetic flux. 
Therefore, we expect reconnection to stay in the same regime even if we relax the constraint of identical-islands merger. } (we note that it is nonetheless possible for the global Lundquist number, defined with the system size $L$, to be large).
In the collisionless reconnection regime, the radius of the islands is required to be $R\lesssim 40 d_i$~\citep{ji2011phase}. 
\footnote{
Our model in its current form does not apply to the plasmoid regime of reconnection in which the reconnecting current sheets are unstable to the formation of plasmoids. }
Our restriction to non-plasmoid-generating reconnection regimes stems from the fact that possible plasmoid generation and early-time dynamics are effectively constrained to a 1D current sheet between merging islands, while the IKE that we study evolves $f$ in 2D (as described below in Section~\ref{sec:sinkterm}). Therefore, the dynamics of the merging islands and the generated secondary plasmoids cannot be considered self-consistently using this IKE. 
Nevertheless, we do not think this is a major limitation, for two reasons. First, regarding the reconnection rate, in MHD the average dimensionless reconnection rate in the plasmoid-mediated regime is $\sim 0.01$~\citep{huang2010scaling,uzdensky2010fast,loureiro2012magnetic}, similar to the Sweet-Parker reconnection rate at $S = 10^4$; while in collisionless reconnection the presence of plasmoids does not appear to affect the average value of the reconnection rate \citep{daughton2007collisionless,daughton2011}, which is $\sim 0.1$.
Second, the new plasmoids would be trapped between the merging islands and would eventually be absorbed by the resulting islands [see, e.g.,~\citet{lyutikov2017b}], so their effect on the system would be small, only affecting the small-$A$ and small-$\psi$ parts of the distribution function.
The lifetime of these new plasmoids would also be at most the same as the lifetime of their parent islands, so they would not cause large changes in $f$, the evolution of macroscopic quantities, or the energy spectrum.
In the rest of the paper, we will thus focus on the merger in the Sweet-Parker regime or the single X-line collisionless regime, but we do not expect significant differences even if reconnection between islands is in the plasmoid regime.\label{footnote_plasmoids}
We will make it more explicit in Section.~\ref{sec:tau_rec} what we mean by the collisionless regime and the reason this regime is of interest for this IKE model.

In summary, we assume the change of the distribution function $f(\psi,A)$ is only caused by the coalescence of islands, represented by the collision operator $\mathcal{C}_{\text{merge}}$. 
Indeed, the coalescence of islands effectively acts like a collision in phase space. Due to the change of areas and fluxes, the islands vanish at their original positions in the phase space defined by $\psi$ and $A$ before the coalescence, and emerge at new positions.
Therefore, the collision operator can be divided into a sink term $\mathcal{C}_{\text{Snk}}(\psi,A,t)$, and a source term $\mathcal{C}_{\text{Src}}(\psi,A,t)$, and the simplified IKE can be written as:
\begin{equation}
    \frac{\partial f(\psi,A,t)}{\partial t} = \mathcal{C}_{\text{Snk}}(\psi,A,t)+\mathcal{C}_{\text{Src}}(\psi,A,t).
    \label{eq:boltzmann}
\end{equation}

The sink term $\mathcal{C}_{\text{Snk}}(\psi,A,t)$ represents the disappearance of islands with area $A$ and flux $\psi$ at time $t$ due to their coalescence with other islands with arbitrary area $A'$ and flux~$\psi'$.
The source term $\mathcal{C}_{\text{Src}}(\psi,A,t)$ represents the emergence of islands with area $A$ and flux $\psi$, due to coalescence between islands with area $A'$ ($A'<A$), flux $\psi$ and islands with area $A-A'$, flux $\psi'$ ($\psi'<\psi$).
These selection rules are implemented through the limits of the collision integrals, as we will explain in the following subsections. 
An important feature of our collision operator $\mathcal{C}_{\text{merge}}$ is that it is non-instantaneous, as explained in Section~\ref{sec:Csnk}.

\subsection{Expression for \texorpdfstring{$\mathcal{C}_{\rm {Snk}}(\psi,A,t)$}{2}}
\label{sec:sinkterm}
We assume that any single coalescence event takes place between two islands only, i.e., we effectively treat island-island interaction as binary.
This assumption imposes an exclusion principle on islands undergoing merging. Once an island starts to merge with another island, it is unavailable for other merging events until the current coalescence process finishes.~\footnote{The assumption of binary mergers excludes the possibility of multi-island clustering in our IKE. This could possibly lead to a slower evolution in our model than in the realistic case. However, the reasonably good agreement between the numerical results obtained by solving our IKE (Section~\ref{sec:delta}) and those from direct numerical simulations~\citep{zhou2019magnetic} suggests that the effects from the simultaneous coalescence of multiple islands are not significant.}
When one island with area $A$ and flux $\psi$ merges with another island with area $A'$ and flux $\psi'$, the resulting island will possess an area $A+A'$ and a flux with the larger value between $\psi$ and $\psi'$~\citep{fermo2010statistical}.

The rate of change of $f$ can be calculated using the 2D hard-sphere scattering model.
According to our normalization, the number density of islands (in $A-\psi$ space) with flux $\psi$ and area $A$ is $f(\psi,A,t)$. The number of islands with arbitrary $A'$ and $\psi'$ is given by $ \int_0^\infty dA' \int_0^\infty d\psi'  f(\psi',A',t)$.
The probability rate for two islands to meet and merge in real 2D space is given by $\sigma \delta v /L^2$, where the cross section for an island with area $A$ interacting with an island with area $A'$ is $\sigma(A,A^\prime) = \sqrt{A}+\sqrt{A'}$, multiplied by a geometric factor of order unity, which we take to be one for simplicity.\footnote{Islands may, of course, have different polarities (i.e.~the currents generating the magnetic field in the islands are in opposite directions), in which case they repel instead of merging. So, in fact, the probability of two random islands merging is $1/2$, rather than $1$. This factor of $1/2$ is neglected in the IKE, as are other factors expected to be of order unity, such as may arise from the fact that islands are not necessarily circular.}
The relative velocity between two islands, $\delta v$, is approximated with the larger Alfv\'en velocity of each of them, $\delta v = \text{max}\{\psi/\sqrt{A},\psi'/\sqrt{A'}\}$.~\footnote{A scale-by-scale equipartition between magnetic energy and kinetic energy is assumed, and the random motion velocities of islands are thus related to their magnetic fields, i.e., taken to be Alfv\'enic. 
The reasoning underlying this assumption is as follows. 
In~\citet{zhou2019magnetic} we found that the ratio of the (box-averaged) magnetic-to-kinetic energy approximately remains a constant of order of unity throughout the evolution. 
Since that system is well described by our hierarchical coalescence model, in which energy is mostly concentrated at a single (time-evolving) scale, the constancy of the box-averaged magnetic-to-kinetic energy ratio implies the same for the energy-containing scale. 
It is true that different magnetic and kinetic power-law spectra are found in that work, seemingly indicating that the energy-density ratio is scale-dependent. 
However, as argued there, the $k^{-2}$ magnetic energy spectrum is a feature of magnetic reversals at current sheets (i.e., a Burgers' spectrum), and the shallower kinetic spectrum reflects the outflow of reconnection sites. In other words, in the 2D case, the spectra do not meaningfully represent the energy distribution over length scales.
In 3D (with a strong guide field), where the contribution of current sheets to the energy spectra is less dominant, we do find the same power-law slope for both magnetic and kinetic spectra despite the system still being magnetically dominated~\citep{zhou2020multi}. 
These considerations lead us to approximate the island advection velocity with the Alfv\'en velocity as there is only a scale-independent constant factor of order unity between the two velocities.}
With such an approximation, the maximum possible error in the relative velocity is $\delta v$. 

Using the model described above, the sink term $\mathcal{C}_{\text{Snk}}(\psi,A,t)$, which equals the decreasing rate of $f(\psi,A)$, can be expressed as:
\begin{equation}
\begin{split}
\mathcal{C}_{\text{Snk}}(\psi,A,t) = 
-\frac{1}{L^2} f(\psi,A,t)\int_0^\infty dA' \int_0^\infty d\psi'\ f(\psi',A',t) \sigma(A,A^\prime) \delta v.
\label{eq:sink}
\end{split}
\end{equation}

We perform a simple dimensional analysis to show that our IKE (dimensionally $\partial_t f \sim \mathcal{C}_{\rm snk}$) has the standard form $\partial_t f \sim \nu f$ where $\nu$ is a characteristic changing rate of $f$. The mean-free path of the system is $\lambda_{\rm mfp}\sim 1/(n \sigma)$, where $n=N/L^2=(\int f dA d\psi)/L^2$ is the island density in configuration space. Plugging the scaling $\int f dA d\psi \sim L^2/(\lambda_{\rm mfp} \sigma)$ into Eq.~\eqref{eq:sink} we obtain $\mathcal{C}_{\rm snk}(A,\psi,t) \sim (\delta v/\lambda_{\rm mfp}) f(\psi,A,t)$, and hence the characteristic decreasing rate of $f$ is as expected: $\nu \sim \delta v/\lambda_{\rm mfp}$. The mean-free path is determined both by the density and typical cross-section of islands. In the case of volume-filling islands, the density of islands is $n=N/L^2 \sim 1/\braket{A}$ where $\braket{A}$ is the typical area of the islands. The typical cross-section of the islands is thus $\sqrt{\braket{A}}$, giving rise to $\lambda_{\rm mfp} \sim 1/(n \sigma) \sim \sqrt{\braket{A}}$. This is consistent with the intuition that when the system is packed with islands, each island will interact with an adjacent one, and $\lambda_{\rm mfp}$ should be comparable to the size of islands. 
In this case, the convection time of islands ($\sim \lambda_{\rm mfp}/v_A$) is the same as their local Alfv\'en time ($\sim \sqrt{\braket{A}}$), where $v_A$ is the local Alfv\'en velocity determined by the magnetic field of islands.
In our IKE, $\lambda_{\rm mfp}$ can be adjusted by tuning the size and density of islands, i.e., changing the ``volume-filling fraction'' of islands in the system. We will define this quantity in Section~\ref{sec:macro}. 

\subsection{Time delay between \texorpdfstring{$\mathcal{C}_{\rm {Snk}}(\psi,A,t)$}{2} and \texorpdfstring{$\mathcal{C}_{\rm {Src}}(\psi,A,t)$}{2}}
\label{sec:Csnk}
The source term $\mathcal{C}_{\text{Src}}(\psi,A,t)$ can be obtained in a similar way, except that there is an additional feature that we now explain. 
Any given merging event is described by both a sink term and a source term in the collision operator.
In traditional collision theory,  collisions (between particles) are regarded as instantaneous, since the particle interaction time is short compared to the particle flight time between collisions. 
In our island coalescence problem, in contrast, the coalescing process is slow, taking a long time compared to the advection of islands.
The coalescence time of islands thus becomes an important dynamical time scale in our system. 
Therefore, it is necessary to consider the time difference between the sink term and source term for any given merger, where the former corresponds to the start of a merger and the latter corresponds to the end.
The time delay for a given merging event is denoted by $\tau_{\rm rec}$, which depends on the areas and fluxes of both islands. The expression for $\tau_{\rm rec}$ will be given in Section~\ref{sec:tau_rec} and the way we treat the islands during the merger will be described in Section~\ref{sec:missing_islands}.

We assume that the decrease in the number of islands happens as soon as the two islands meet in real space. However, the increase in the number of islands happens when the ``parent'' islands finish the merging process. 
That is, for a merging event between two islands characterised by ($\psi$,$A$) and ($\psi'$,$A'$) (with $\psi'<\psi$), starting at time $t$, the sink term acts to decrease $f(\psi,A,t)$ and $f(\psi',A',t)$ immediately, while the source term acts to increase $f(\psi, A+A', t+\tau_{\rm rec})$ when islands finish merging at $t+\tau_{\rm rec}(\psi,A,\psi',A')$. Equivalently, for a source term increasing $f(\psi,A,t)$ at time $t$, the corresponding sink term resulting from the same merging event decreases $f(\psi,A',t-\tau_{\rm rec})$ and $f(\psi',A-A',t-\tau_{\rm rec})$, when the islands with $(\psi,A')$ and $(\psi',A-A')$ start to merge at $t-\tau_{\rm rec}(\psi,A',\psi',A-A')$. 

The way we implement this time delay $\tau_{\rm rec}$ in the source term $\mathcal{C}_{\text{Src}}(\psi,A,t)$ is as follows. At any given time $t$ and given point $(\psi,A)$ in the phase space, we trace back the history of island distributions at all previous times $t-\tau$ ($\tau \in [0,t]$) and consider all possible pairs of islands ($\psi,A'$) and ($\psi',A-A'$), from which the new islands can be formed. 
The merging times $\tau_{\rm rec}(\psi, A', \psi', A-A')$ of these islands are calculated. Those islands whose merging time $\tau_{\rm rec}$ matches $\tau$ contribute to the source term.

Following the above description, the source term can thus be formally expressed as
\begin{equation}
\begin{split}
    \mathcal{C}_{\text{Src}}(\psi,A,t)& = \frac{1}{L^2} \int_0^t d\tau \int_0^A dA'\  f(\psi,A',t-\tau)\int_0^\psi d\psi'  \\ &f(\psi',A-A',t-\tau)\, \sigma(A-A^\prime, A^\prime)\, \delta v\, \delta[\tau-\tau_{\rm rec}(\psi, A', \psi', A-A')]. 
    \label{eq:Csrc}
\end{split}
\end{equation}

We note that introducing the delay of islands' merger time is important to obtain the correct dynamical time scale of the evolution of the system. With an instantaneous collision operator, the system would evolve on the Alfv\'enic time scale, whereas with the consideration of merger time delay, the system evolves on the reconnection time scale [which has indeed been shown to be the correct evolution time via direct numerical simulations of this problem~\citep{zhou2019magnetic,zhou2020multi,bhat2021inverse,hosking2021}].

\subsection{Reconnection time \texorpdfstring{$\tau_{\rm rec}$}{2}}
\label{sec:tau_rec}
The time for two islands to merge, $\tau_{\rm rec}$, is determined by the physics of the reconnection process. 
In this subsection, we first consider the collisional, low Lundquist number regime $S\lesssim 10 ^4$, and calculate the reconnection time $\tau_{\rm rec}$ based on the Sweet-Parker (SP) reconnection model~\citep{sweet_neutral_1958,parker_sweet_1957}. 
We note that different reconnection models can be adopted to calculate $\tau_{\rm rec}$ and our IKE model can be modified to study systems in different reconnection regimes. 

In the SP regime, the merging of two islands with different sizes ($A_1$,$A_2$) and fluxes ($\psi_1$,$\psi_2$) is a process of asymmetric reconnection (the symmetric case is a particular scenario in this description), where the outflow Alfv\'enic velocity $v_A$ and the pertinent Lundquist number $S$ are calculated using geometric averages~\citep{cassak2007scaling}, as follows:
\begin{equation}
    v_A = \left(B_1 B_2\right)^{1/2} = \left(\frac{\psi_1}{\sqrt{A_1/\pi}}\frac{\psi_2}{\sqrt{A_2/\pi}}\right)^{1/2} \simeq \frac{\left(\psi_1 \psi_2\right)^{1/2}}{\left(A_1A_2\right)^{1/4}}, 
\end{equation}
\begin{equation}
    S = \frac{\sqrt{\psi_1\psi_2}}{\eta}. \label{eq:define_S}
\end{equation}

The merging time between an island with ($\psi_1,A_1$) and island with ($\psi_2,A_2$) is calculated as:
\begin{equation}
\label{eq:tau_rec}
    \tau_{\rm rec}(\psi_1, A_1, \psi_2, A_2) \simeq \frac{\min \left( \sqrt{A_1},\sqrt{A_2}\right) } {v_{\rm rec}},
\end{equation}
where $v_{\rm rec}$ is the merging velocity of the two islands. 
The $v_{\rm rec}$ and outflow Alfv\'enic velocity $v_{A}$ are related by the dimensionless reconnection rate $\beta_{\rm rec}=v_{\rm rec}/v_A$. In the SP model, $\beta_{\rm rec}$ is related to the local Lundquist number of the merging islands $S$ by $\beta_{\rm rec} = S^{-1/2}$. 
Therefore:
\begin{equation}
\label{eq:v_rec}
    v_{\rm rec} = \beta_{\rm rec} v_A = S^{-1/2} v_A \simeq \left(\frac{\sqrt{\psi_1\psi_2}}{\eta}\right)^{-1/2} \frac{\left(\psi_1 \psi_2\right)^{1/2}}{\left(A_1A_2\right)^{1/4}} = \eta^{1/2}\left(\frac{\psi_1\psi_2}{A_1A_2}\right)^{1/4},
\end{equation}
and the reconnection time $\tau_{\rm rec}$ can be calculated.

We emphasize that the elements of the reconnection process that affect the IKE are the conservation of magnetic flux and area (both of which are quite generally valid), and the reconnection rate. Beyond this, the actual detailed reconnection physics --- such as, for example, the specific mechanisms that set the reconnection rate --- do not enter our model. Therefore, our IKE can be straightforwardly extended to other reconnection regimes where the reconnection rate is a constant. This includes collisionless plasmas or, indeed, any regime where Hall physics is important; in either case, the reconnection rate is $\beta_{\rm rec} \simeq 0.1$~[see, e.g., \cite{biskamp1995ion,shay2001alfvenic,birn2001,cassak2017review}].
For convenience, in the rest of the paper we refer to this regime with $\beta_{\rm rec} \simeq 0.1$ as the collisionless regime but, we stress, it is in fact more general than that.

\subsection{Accounting for islands undergoing merging processes}
\label{sec:missing_islands}
Due to the time delay between the sink term and the source term of the collision operator, the contributions of currently-merging islands to the distribution function $f(\psi,A,t)$ have already been subtracted by the sink term, while the components for the resulting new islands have not yet been added by the source term. This would lead to the problem that, when calculating global physical quantities using weighted averages with the distribution function, the islands in the merging process would not be taken into account. 

To fix this problem, we add the components of the distribution function that have been sunk but not yet re-added by the delayed source term.
We denote by $\tilde{f}(\psi,A,t)$ the distribution function of the islands that are undergoing the merging process; as stated above, under the assumption of binary merging, these islands do not participate in the interaction with additional islands.
 The total distribution function is then given by $f_{\rm tot}(\psi,A,t) = f(\psi,A,t) + \tilde{f}(\psi,A,t)$; all the physical quantities should be calculated using this total distribution function.
 
A kinetic equation for the time evolution of $\tilde{f}(\psi,A,t)$ may also be calculated from the merging of islands and can be decomposed into the source term $\mathcal{\tilde{C}}_{\rm Src}(\psi,A,t)$ and the sink term $\mathcal{\tilde{C}}_{\rm Snk}(\psi,A,t)$, as follows:
\begin{equation}
    \partial_t \tilde{f}(\psi,A,t) = \mathcal{\tilde{C}}_{\rm Src}(\psi,A,t) + \mathcal{\tilde{C}}_{\rm Snk}(\psi,A,t).
    \label{eq:ftilde_equation}
\end{equation}
We note that the advective terms $\partial_\psi(\dot{\psi}\tilde{f})$ and $\partial_A(\dot{A}\tilde{f})$ are neglected in Eq.~\eqref{eq:ftilde_equation}. That is, we do not consider the continuous evolution of the merging island parameters during the reconnection process. 
Instead, we assume that when islands start to merge, their parameters remain the same until the moment when islands finish merging. At the end of mergers, the islands' parameters jump from values of previous islands to values of the new islands (in particular, one of the islands disappears). This is implemented by the collisional integrals $\mathcal{\tilde{C}}_{\rm Src}$ and $\mathcal{\tilde{C}}_{\rm Snk}$. 
On a time scale much longer than the life time of one island generation, the replacement of the continuous change of islands parameters during mergers with a discontinuous change at the end of mergers will not significantly change the system evolution.

The merging islands distribution $\tilde{f}(\psi,A,t)$ increases when islands start to merge; therefore the source term $\mathcal{\tilde{C}}_{\rm Src}(\psi,A,t)$ should act immediately on ${\tilde{f}}(\psi,A,t)$.
Obviously $\mathcal{\tilde{C}}_{\rm Src}(\psi,A,t)=-\mathcal{C}_{\text{Snk}}(\psi,A,t)$, since the components of islands starting to merge should be subtracted from $f(\psi,A,t)$ and added to ${\tilde{f}}(\psi,A,t)$ instantaneously.
 
The sink term is more complex. 
We want to subtract from $\tilde{f}(\psi,A,t)$ once the islands finish merging. That is, at any given time $t$ when islands finish merging, we want to subtract the components that we added at $t-\tau_{\rm rec}$ when they started to merge. This can be implemented by introducing the time delay to the source term $\mathcal{\tilde{C}}_{\rm Src}(\psi,A,t)$:
\begin{equation}
\begin{aligned}
 &\mathcal{\tilde{C}}_{\rm Snk}(\psi,A,t) = -\int_0^t d\tau\, \mathcal{\tilde{C}}_{\rm Src}(\psi,A,t-\tau)\,\delta[\tau-\tau_{\rm rec}(\psi, A', \psi', A)]\\
 &=-\frac{1}{L^2}  \int_0^t d\tau f(\psi,A,t-\tau) \int_0^\infty dA' \int_0^\infty d\psi' f(\psi',A',t-\tau)\, \sigma(A,A^\prime)\, \delta v \ \delta[\tau-\tau_{\rm rec}(\psi, A', \psi', A)].
 \label{eq:sink_tilde}
\end{aligned}
\end{equation}

One of our merging rules says that the area of the resulting islands should be the sum of the areas of the two merging islands. Therefore, in this model, the total area of all the islands should be conserved. The total area can be calculated by:
\begin{equation}
    A_{\rm tot}=\int dA\ d\psi\ f_{\rm tot}(\psi,A,t)A.
\end{equation}
In the numerical implementation of our IKE that we discuss in Sections~\ref{sec:numerical} and~\ref{sec:NR}, the conservation of $A_{\rm tot}$ has been tested and confirmed numerically as a benchmark of the equation solver.

\subsection{Macroscopic quantities and magnetic spectrum}
\label{sec:macro}
We first introduce the parameter $V_{\rm fill} \equiv A_{\rm tot}/L^2 \in [0,1]$ to represent the volume-filling fraction of islands in the system. This quantity is determined by both the density and sizes of islands. As we discussed at the end of Section~\ref{sec:sinkterm}, larger $V_{\rm fill}$ implies smaller mean-free path $\lambda_{\rm mfp}$ for the islands. When the islands are volume filling, i.e., $V_{\rm fill} \approx 1$, the mean-free path is close to the typical length scale of islands $\lambda_{\rm mfp}\approx\sqrt{A_{\rm tot}/N}$. This is the case that we focus on in this paper, but we also show in Section~\ref{sec:delta} a scan in $V_{\rm fill}$ which indicates that this parameter does not control the evolution of the main quantities of interest in our study.

The other main macroscopic quantities of interest are the total number of islands $N$, the average area of islands $\braket{A}$, and the total magnetic energy $\mathcal{E}$. Their expressions are:
\begin{align}
    N(t) &= \int dA\ d\psi\ f_{\rm tot}(\psi,A,t);\\
    \braket{A}(t) &= \frac{A_{\rm tot}(t)}{N(t)}=\frac{1}{N(t)}\int dA\ d\psi\ f_{\rm tot}(\psi,A,t)A;\\
    \mathcal{E}(t) &=  \int dA\ d\psi\ f_{\rm tot}(\psi,A,t) \psi^2.
    \label{eq:macroscopic}
\end{align}
The scalings for the time evolution of these three quantities are the most important predictions from our hierarchical model~\citep{zhou2019magnetic}, which we repeat here:
\begin{align}
    N(t)\sim t^{-1}; \quad \braket{A}(t) \sim t; \quad \mathcal{E}(t) \sim t^{-1}.
\end{align}
One of the main goals of this study is to test if the above scaling laws remain valid for a system consisting of a non-trivial distribution of islands, which we describe in detail in Section~\ref{sec:NR}.

Apart from the evolution of macroscopic quantities, we are also interested in the magnetic energy distribution over different length scales. 
This can be first represented by the magnetic energy density associated with different areas $A$ of islands, denoted as $U(A,t)$, which can be calculated from the distribution function $f(\psi,A,t)$ as
\begin{equation}
\label{eq:AreaEnergyDist}
    U(A,t) = \int d\psi\ f_{\rm tot}(\psi,A,t) \psi^2,
\end{equation}
and is related to the magnetic energy as $\mathcal{E}(t)=\int U(A,t) dA$.
The conventional definition of magnetic energy spectrum in the wave number ($k$) space, denoted as $U(k,t)$, is related to $U(A,t)$.
The wave number $k$, corresponding to a length scale, is related to the area of islands $A$ as: 
\begin{equation}
\label{eq:kspace}
    k \equiv \frac{2\pi}{4\sqrt{A/\pi}} \simeq A^{-1/2},
\end{equation}
The $k$ space can therefore be constructed from the $A$ space.
The magnetic energy associated with each wave number and that associated with each value of area are related as $U(k)dk = U(A)dA$ where $A \sim k^{-2}$. 
Therefore, the magnetic energy spectrum $U(k,t)$ can be calculated from:
\begin{equation}
\label{eq:Uk_def}
   U(k,t) = U(A,t)\frac{dA}{dk} \sim k^{-3} U(A,t).
\end{equation}
Given an $A$-spectrum $U(A) \sim A^{-\alpha}$, we can obtain the corresponding $k$-spectrum $U(k) \sim k^{-\gamma}$,  where $\gamma=3-2\alpha$, according to Eq.~\eqref{eq:Uk_def}.
We note that in some circumstances, the magnetic energy spectrum calculated using the definition Eq.~\eqref{eq:Uk_def} can provide more accurate and meaningful information on the energy distribution over scales than the standard one based on Fourier transform of real-space configurations of magnetic fields. For example, in \citet{zhou2019magnetic}, a $k^{-2}$ magnetic energy spectrum was observed in 2D RMHD simulations of island mergers, but the origin of this spectrum was, in fact, traced simply to the sharp magnetic field reversals at the thin current sheets between merging islands~\citep{burgers1948mathematical}.

Finally, in order to visualise the island distribution function more easily, we calculate the 1D distribution functions $F(A,t)$ (the distribution function over $A$) and $F(\psi,t)$ (the distribution function over $\psi$):
\begin{equation}
\begin{aligned}
 F_A(A,t) = \int d\psi f_{\rm tot}(\psi,A,t),\\
 F_\psi(\psi,t) = \int dA f_{\rm tot}(\psi,A,t). 
 \end{aligned}
\end{equation}

\section{Numerical Implementation}
\label{sec:numerical}
We conduct a numerical study of this island-merging problem by numerically solving Eqs.~\eqref{eq:boltzmann} and~\eqref{eq:ftilde_equation}.
We use the following normalizations. 
We assume the islands are in a periodic square box with side $L = 2\pi$, and define a reference magnetic field $B_0 = 1$ and Alfv\'en velocity $v_{A0} = 1$. The reference magnetic flux is $\psi_0 = B_0 L$. The global Alfv\'en time is then defined as $L/v_{A0}$. 
This global Alfv\'en time is mainly for normalisation purposes and is different from the local Alfv\'en time defined with the size ($\sqrt{A}$) and magnetic field $(\psi/\sqrt{A})
$ of islands. 
For resistive-MHD cases, the resistivity is specified by setting the global Lundquist number $S_L=v_{A0}L/\eta$ defined with the system scale $L$ and reference magnetic field $B_0$.
With this fixed resistivity, the local Lundquist number of merging islands and the corresponding reconnection rate $\beta_{\rm rec}$ in the SP reconnection regime can be determined via Eq.~\eqref{eq:define_S}. 
For collisionless (or simply Hall-dominated) cases, the reconnection rate is not determined by the Lundquist number but is instead set as $\beta_{\rm rec} = 0.1$. 

The distribution $f(\psi,A,t)$ is discretized with $N_A, N_{\psi}$ points in the $A$ and $\psi$ domains respectively. The grid spacing for each quantity is uniform. In all simulations, $A$ ranges from $L^2/N_A$ to $L^2$ while $\psi$ ranges from $\psi_0/2$ to $3\psi_0/2$. The collision integrals in Eqs.~\eqref{eq:sink},~\eqref{eq:Csrc} and~\eqref{eq:sink_tilde} are calculated using a trapezoidal rule, while time evolution is carried out using a standard second-order predictor-corrector method.
The time step is constant and chosen to ensure numerical stability for the initial distribution, which imposes the strictest conditions because of the largest Alfv\'en speeds corresponding to the smallest islands.
The merging island distribution ${\tilde{f}}(\psi,A,t)$ is evolved in a similar manner.

Solving the IKE is numerically challenging because the collision operator is non-local in both time and phase space. As such, it requires storing the history of the evolution of $f$ and $\tilde{f}$, and performing calculations using quantities from up to the largest value of $\tau_{\rm rec}$ earlier. This increases both the memory requirements and the computational cost.
Additionally, the non-locality makes the parallelisation  of the computations inefficient. 

\section{Numerical Results}
\label{sec:NR}
We consider three different types of initial conditions (delta-distribution, Gaussian and power-law distribution), and study the following aspects of the system: (1) Time evolution of macroscopic quantities, comparing with the prediction of the hierarchical model $\mathcal{E}\sim t^{-1}, N\sim t^{-1}, \braket{A}\sim t$; (2) Evolution of 1D distribution functions $F_A(A,t), F_\psi(\psi,t)$; and (3) Evolution of magnetic energy spectrum $U(A,t)$ [which, as explained above, is directly related to $U(k,t)$]. 
The delta-distribution case is mainly used to benchmark the IKE model against the hierarchical model and (reduced-)MHD simulations that we have reported in~\citet{zhou2019magnetic}, while the Gaussian and power-law cases are more relevant to various astrophysical systems.

According to our assumption, the flux of the new islands will be the larger flux of the two merging islands. Therefore, the values of fluxes at any given time will be a subset of those at the beginning. 
However, new values of areas of islands will be continuously generated. 
Considering the high computational cost to solve the IKE, we employ this argument to justify the use of low resolution in the flux coordinate $\psi$ (small $N_\psi$) and relatively high resolution in the area coordinate $A$ (large $N_A$), focusing on studying the evolution of $F_A(A,t)$.

We introduce the parameter $R$ to quantify the maximum difference of areas and fluxes between islands that are allowed to interact:
\begin{equation}
    \label{eq:R}
    (\Delta A)_{\rm max} = R L^2, \quad (\Delta \psi)_{\rm max} = R B_0 L.
\end{equation}
We set $0 < R < 1$, where $R = 0$ gives only local interactions in phase space, as assumed by the hierarchical model~\citep{zhou2019magnetic}, and $R=1$ allows interactions between islands in all of phase space. 
We note that, in our model, increasing $R$ (i.e., allowing a wider range of interaction) does not necessarily lead to more merging events per unit time. As mentioned, our assumption of binary merging imposes an exclusion principle on merging events. Therefore, allowing the merger of islands within a wide range of distribution effectively reduces those between similar islands.
The motivations to introduce $R$ as a constraint on the interaction range are as follows. 
First, it allows us to study whether it is the interactions between similar-sized islands or those between large and small islands that dominate the evolution of the system. This will be important in the discussion of Section~\ref{sec:scaling_law}, when we draw connections between the present study and decaying 2D MHD turbulence (whose dynamics is controlled by island mergers) [e.g.,~\citep{matthaeus1986,zhou2019magnetic}]. In particular, the parameter $R$ maps to the question of whether the energy decay and associated inverse transfer is mostly caused by the local interaction (in Fourier space) of modes with similar wavenumbers, or if, instead, non-local effects play a significant role. Therefore, the investigation of the effect of this parameter can help to  justify, or refute, the use of single-scale models or scaling theories (that require assuming a characteristic scale and magnetic field) to study this problem.
Second, it creates a case that can be more closely compared to the hierarchical model, namely, allowing only local interactions, $R=0$, and starting with identical islands, i.e., delta-function initial distributions (see Section~\ref{sec:delta}).
Third, it reduces the computational cost of the calculations, making them more feasible to perform. 
The effect of varying $R$ on the results will be discussed below.

There are several factors that can cause discrepancies between the results of the IKE and the hierarchical model. For example, the hierarchical model restricts mergers to those between identical islands, whereas the IKE model relaxes this constraint and allows mergers between any pair of islands.
Also, the hierarchical model assumes mergers happen at discrete stages, while the IKE is a continuous model. Therefore, in the IKE, even if the system starts with identical islands (and only with local interaction), islands with different areas (but same flux) will be generated.
Yet another difference between the two approaches that is worth mentioning is that the hierarchical model assumes mergers to be successive immediately, while the IKE includes islands freely moving at the Alfv\'en speed between merging events. The inclusion of this effect thus causes the system to evolve slower in the IKE than in the hierarchical model. With high enough $S$ (or in the collisionless regime), $\tau_{\rm rec} \gg \tau_A$, the scaling with time in the IKE is expected to approach that of the hierarchical model.

\subsection{Delta-distribution initial condition} 
\label{sec:delta}
We first study the system with a delta-distribution initial condition, expressed as:
\begin{equation}
    f(\psi,A,t=0)=f_0\,\delta(\psi-\psi_{\rm ini})\,\delta(A-A_{\rm ini}).
\end{equation}
That is, the system starts with an ensemble of identical magnetic islands with flux $\psi_{\rm ini}$ and area $A_{\rm ini}$ [the same initial condition as used in~\citet{zhou2019magnetic}].
In the following calculations, we set $\psi_{\rm ini} = 5 B_0 L/6$ and area $A_{\rm ini} = L^2/N_A$ (the smallest area we can resolve). 
We use $N_A = 256$ and $N_\psi = 4$ to resolve the system. 
In this and subsequent studies, we vary the Lundquist number $S_L$ between $10^3$ and $10^4$ and study the effect of non-local interactions by varying $R$ from $0$ to $1$. 
For this initial condition, we also vary the value of $f_0$ to set the island volume-filling fraction $V_{\rm fill} \in \{0.125,0.25,0.5,1\}$.
We focus mostly on the case $V_{\rm fill}=1$ and study the effects of $S_L$ and $R$ with fixed $V_{\rm fill}=1$.

\begin{figure}
\centering
\begin{subfigure}{.5\textwidth}
  \centering
  \includegraphics[width=1.0\linewidth]{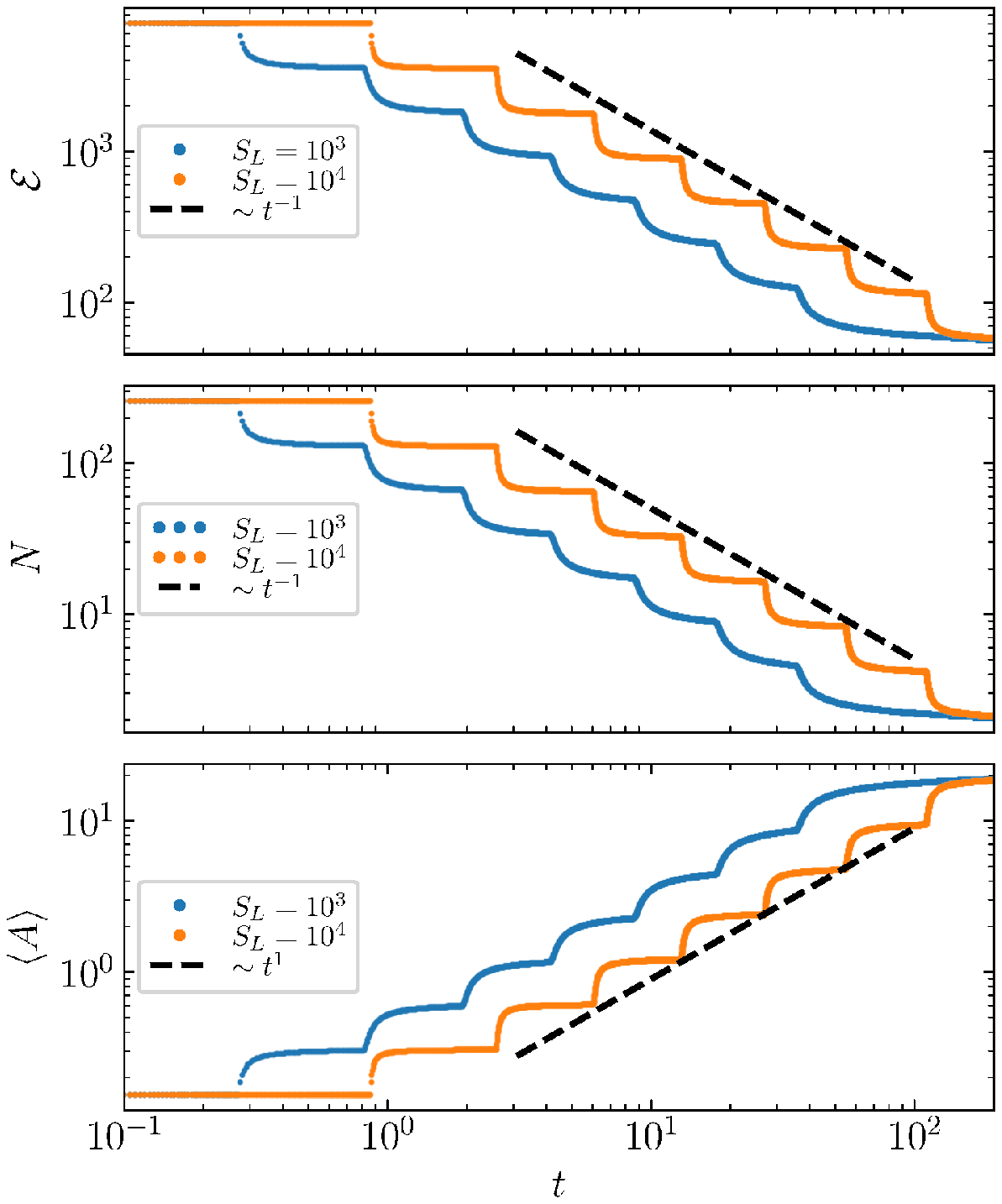}
\end{subfigure}%
\begin{subfigure}{.5\textwidth}
  \centering
  \includegraphics[width=1.0\linewidth]{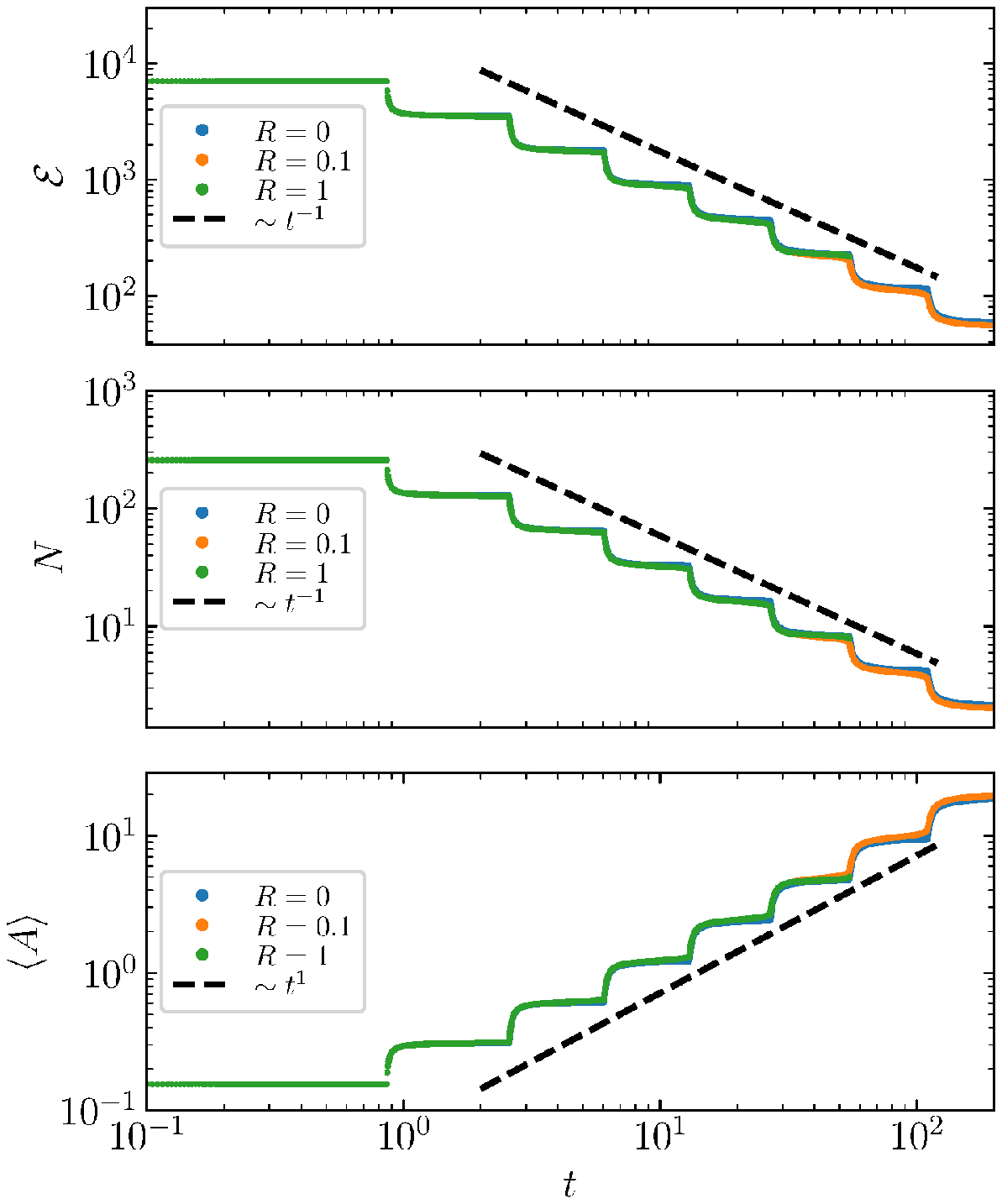}
\end{subfigure}
\begin{subfigure}{.5\textwidth}
  \centering 
  \includegraphics[width=1.0\linewidth]{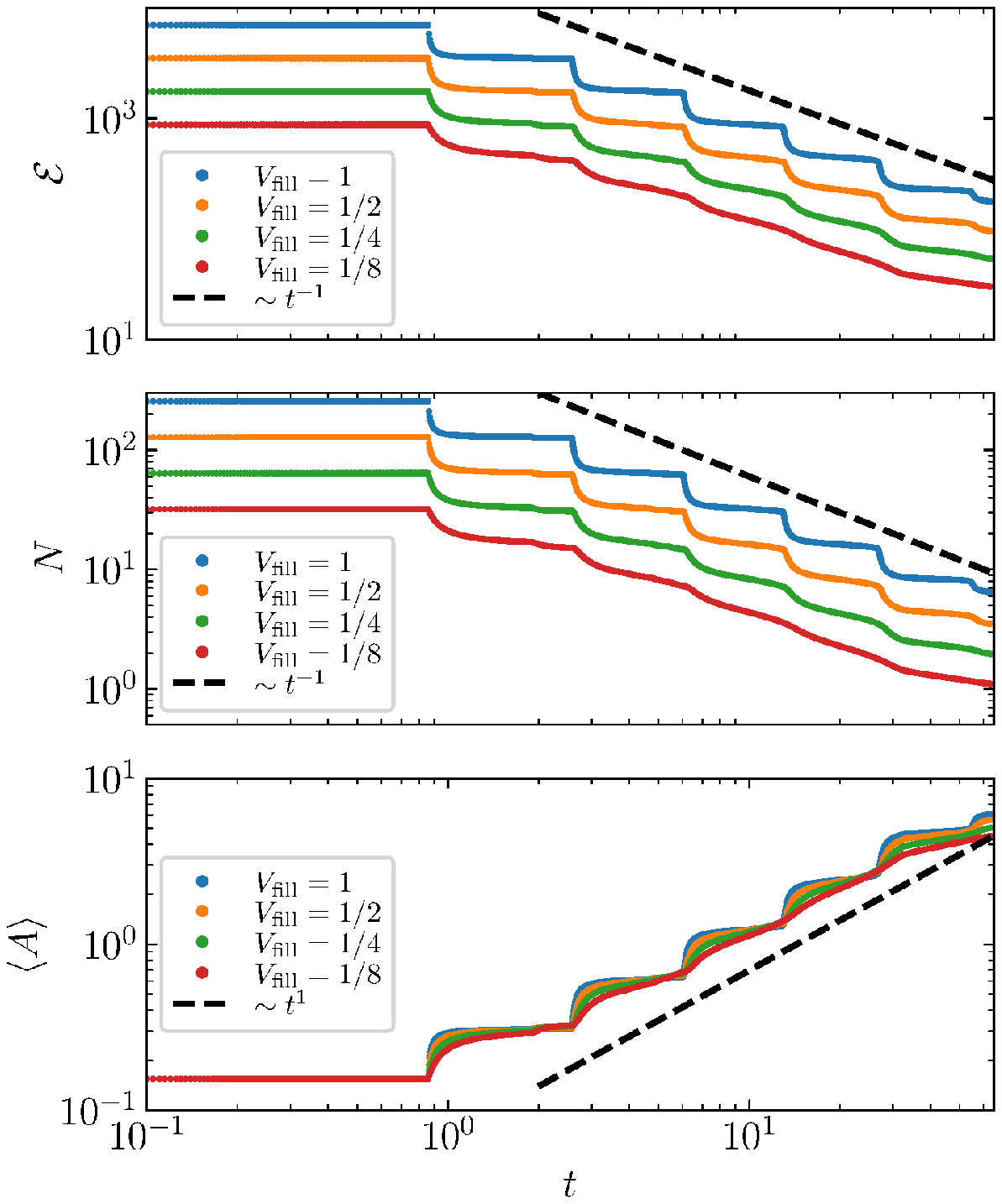}
\end{subfigure}
\caption{Delta-distribution initial conditions. Top left: The evolution of macroscopic quantities ($\mathcal{E}$, $N$ and $\braket{A}$) allowing only local interactions ($R=0$) between volume-filling islands ($V_{\rm fill}=1$) with varying $S_L$. The corresponding scaling laws from the hierarchical model are shown as reference (dashed lines). Top right: The evolution of macroscopic quantities with $S_L=10^4$, $V_{\rm fill}=1$ and varying $R$. Allowing interactions between non-identical islands does not significantly change the evolution of macroscopic quantities (all the curves overlap each other such that the black dots are not visible, with only the $R = 1$ case showing minor deviations for $t \gtrsim 10$). Bottom: The evolution of macroscopic quantities with fixed $S_L=10^4$, $R=1$, and varying the volume-filling fraction of islands, $V_{\rm fill}$. The evolution of the system does not depend strongly on  $V_{\rm fill}$.}
\label{fig:Deltatest}
\end{figure}

This is the ideal set up to test the scalings from the hierarchical model: total magnetic energy $\mathcal{E} \sim t^{-1}$, number of islands $N \sim t^{-1}$, and average area of islands $\langle A \rangle \sim t$~\citep{zhou2019magnetic}.
The evolution of these quantities is shown in Figure~\ref{fig:Deltatest}. The top left panels show the $R=0$ case (i.e.,~only identical islands interact) for $S_L = 10^3$, $10^4$. 
In this and subsequent figures, the unit of time is the global Alfv\'en time $L/v_{A0}$.
Both $\mathcal{E}$ and $N$ show a $t^{-1}$ scaling to a good approximation, while $\langle A \rangle$ scales as $t$. This shows agreement with the hierarchical model described in \citet{zhou2019magnetic}.
There are discrete ``steps'' that can be seen in the time evolution. The horizontal segments are due to the finite reconnection time, as the macroscopic quantities only change after merging is complete, while the steepness of the mostly vertical segments is controlled by the collision integral.

The effect of allowing non-identical islands to interact is shown in the top right panels of Figure~\ref{fig:Deltatest} that compare runs with $R \in \{0,0.1,1\}$ and fixed $S_L=10^4$. 
There is little effect, as can be seen by the overlapping traces as $R$ is varied. This is because merging in this system is dominated by identical islands. This result shows that the assumption of only considering merging between identical islands used in \citet{zhou2019magnetic} is valid for the system it analyses. 

The effect of the volume-filling fraction of islands is shown in the bottom plot of Figure~\ref{fig:Deltatest}, where $V_{\rm fill}$ is varied from $0.125$ to $1$. We see that the overall evolution does not strongly depend on how compactly the islands are distributed in the system: all curves exhibit approximately the same slope. This is consistent with the fact that reconnection is the dominant dynamical process governing the evolution of the system. The islands' mean-free path and volume-filling fraction only affect the convection time (i.e., the efficiency with which they pair up) while the merging process remains unchanged. 
Therefore, $V_{\rm fill}$ should not be a critical parameter for system evolution within the regime where the convection time, $\lambda_{\rm mfp}/v_A$, is much shorter than the merger time $\tau_{\rm rec} \sim \beta_{\rm rec}^{-1} \sqrt{\braket{A}}/v_A$, which sets the condition $\beta_{rec} (\lambda_{\rm mfp}/\sqrt{\braket{A}}) \ll 1$. 
Nevertheless, some small differences still occur between cases with different $V_{\rm fill}$. With decreasing $V_{\rm fill}$, the evolution of macroscopic quantities is slightly slower, which can be justified by the larger convection time (i.e., smaller pairing-up rate) of islands. The curve also becomes smoother with smaller $V_{\rm fill}$, because islands have to travel to pair with one another, as opposed to just merging with immediately adjacent ones in the volume-filling case. This leads to different merger starting times for different pairs of islands, thus smoothing out the ``step'' feature in the evolution curves. 

\begin{figure}
  \centering
  \includegraphics[width=0.65\linewidth]{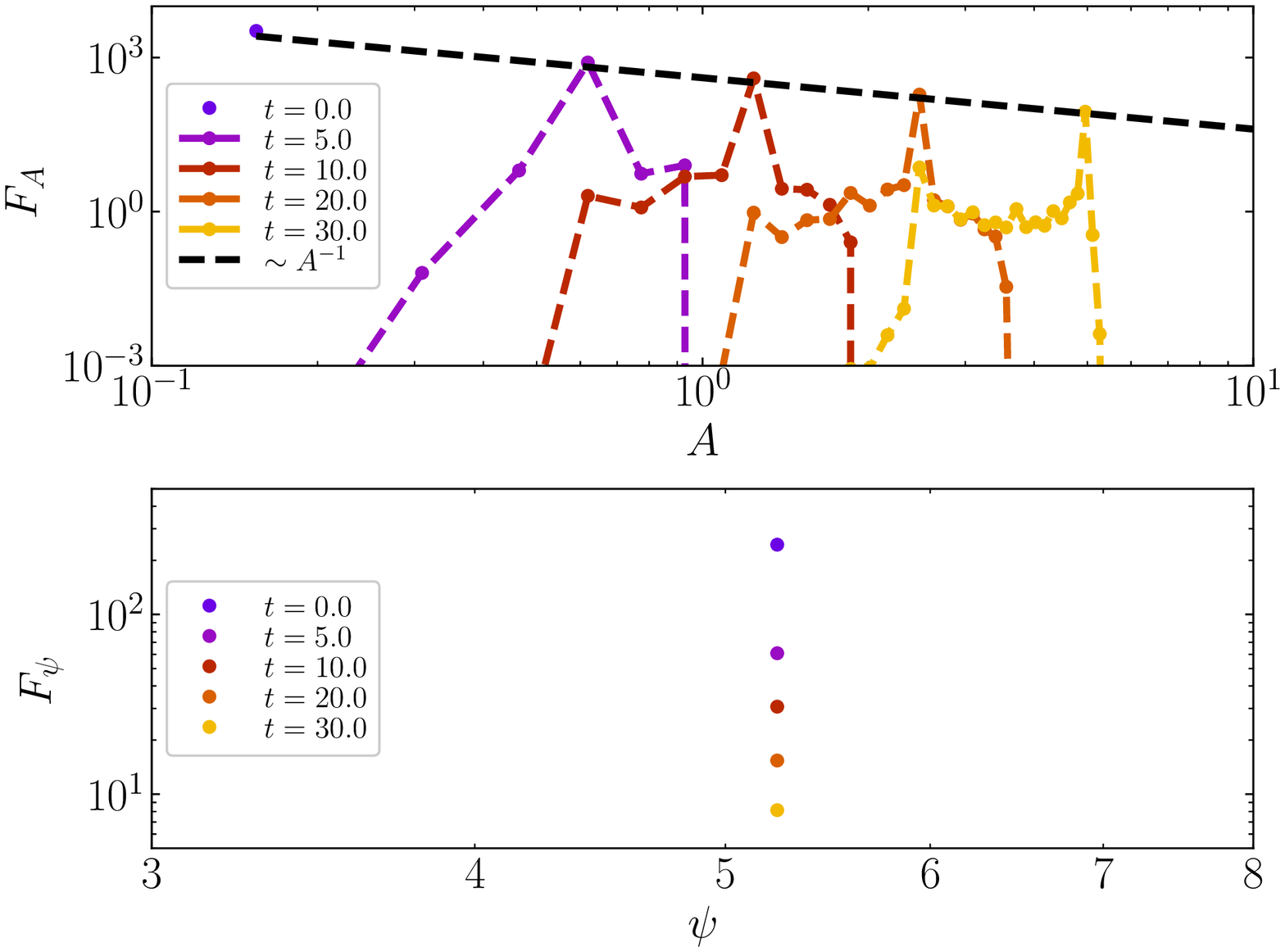}
\caption{The evolution of the 1D distribution function in area, $F_A$ (top), and in flux, $F_\psi$ (bottom), for the run with $\delta$-function initial distribution, with $S_L=10^4$ and unconstrained interactions ($R=1$).}
\label{fig:Deltatest1}
\end{figure}

The evolution of the distribution functions $F_A$ and $F_\psi$ for $S_L=10^4$ and $R=1$ is shown in Figure~\ref{fig:Deltatest1}. As the merging progresses, the peak of the area distribution moves from small to large scale (top panel). The peak value of $F_A$, which is approximately the number of islands $N$, is proportional to $\langle A\rangle^{-1}$ as the total area is conserved. Because merging of different-sized islands is allowed, the distribution becomes wider with time, though the value at the peak remains 1--2 orders of magnitude larger than in the rest of the distribution, confirming that the overall merging process is dominated by interactions between identical islands. 
The bottom panel shows the evolution of $F_\psi$. As there are only islands with one value of $\psi$ initially, the only change during evolution is the magnitude of the peak, which reflects the reduction in the number of islands.
The spectra (not shown) are qualitatively similar to the distribution functions, which have one dominant peak.

The results with the delta-distribution initial conditions provide a benchmark of the IKE model. We show that when the initial conditions and assumptions about merging are restricted to those of the model in \citet{zhou2019magnetic}, the evolution of macroscopic quantities such as $\mathcal{E}$, $N$, and $\langle A \rangle$ agrees with the analytic predictions and with the direct numerical solution of the RMHD equations reported in that reference. 

\subsection{Gaussian-distribution initial condition}
\label{sec:gaussian}
In a realistic system, magnetic islands will not be identical. For example, the population of large flux ropes (magnetic clouds) in the solar wind has a Gaussian distribution~\citep{janvier2014there}.
We thus study an initial Gaussian distribution with a spread around its mean area and flux, expressed as
\begin{equation}
\label{eq:gaussian}
    f(\psi,A,t=0)=f_0e^{-(A-\overline{A})^2/(2\sigma_A^2)}e^{-(\psi-\overline{\psi})^2/(2\sigma_{\psi}^2)}.
\end{equation}
Our fiducial run has $\bar{A}= L^2/40 \approx 0.99$ and $\bar{\psi}=B_0L/2=\pi$ , and we use $N_A = 256$ and $N_\psi = 4$ to resolve the system (a numerical convergence study is included in Appendix~\ref{sec:convergence}).

The standard deviations are set as $\sigma_{A}=3 L^2/N_{A} \approx 0.3\bar{A} \approx 0.47$ and $\sigma_\psi=3B_0L/N_\psi \approx 3/2\bar{\psi} \approx 4.71$. 
Other runs (not shown in this paper) with different values of $\sigma_A$ and $\sigma_\psi$ show qualitatively similar results; in particular, when the values of $\sigma_A$ and $\sigma_\psi$ are small enough, the evolution of the system is similar to that with the delta-distribution initial condition reported in Section~\ref{sec:delta}.
To study the difference between collisional reconnection, where $\beta_{\rm rec} \simeq S^{-1/2}$, and the collisionless case, where $\beta_{\rm rec} \simeq 0.1$, we perform an additional calculation in the collisionless regime with this same initial condition.

\begin{figure}
\centering
\begin{subfigure}{.5\textwidth}
  \centering
  \includegraphics[width=1.0\linewidth]{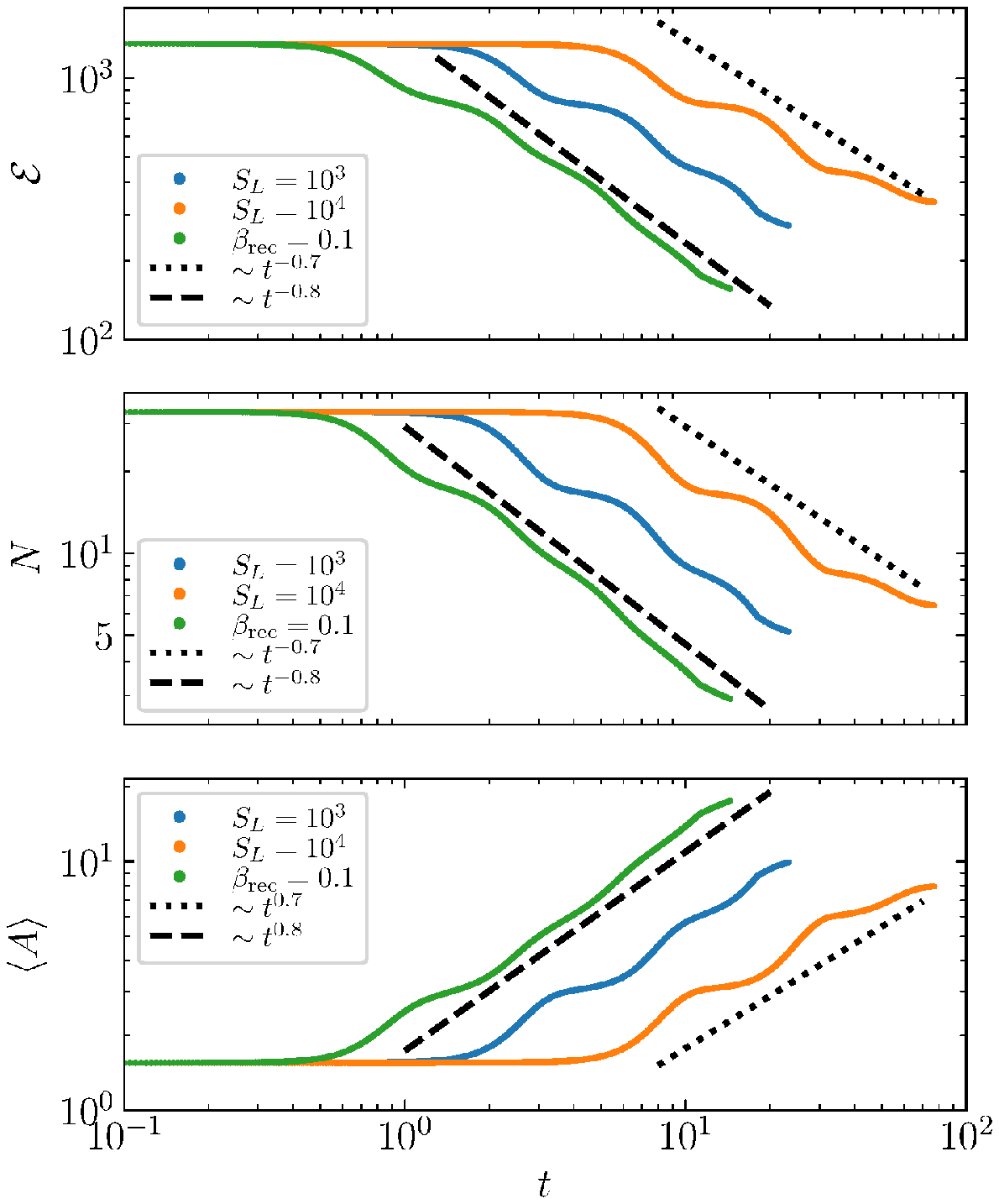}
\end{subfigure}%
\begin{subfigure}{.5\textwidth}
  \centering
  \includegraphics[width=1.0\linewidth]{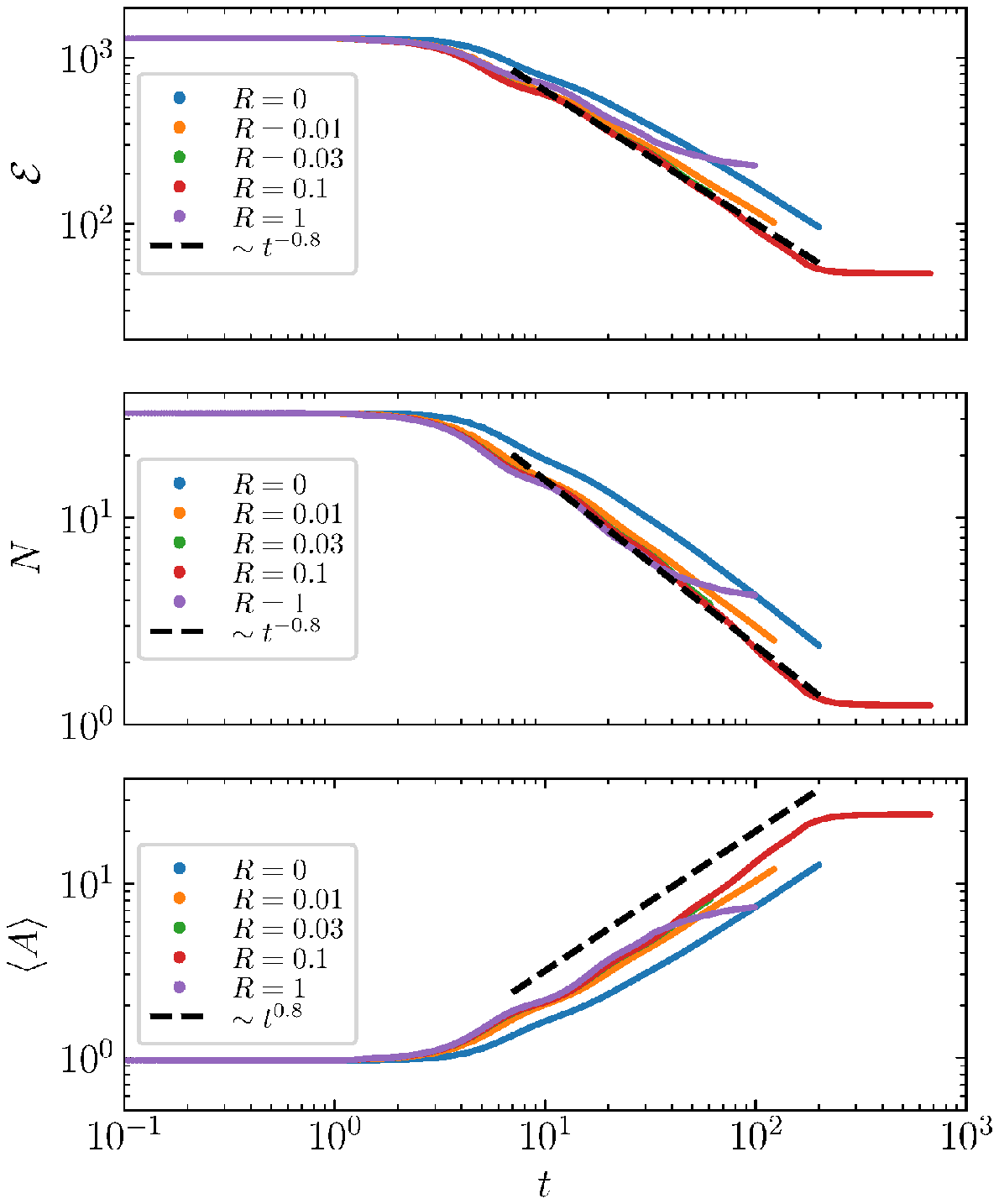}
\end{subfigure}
\caption{Gaussian-distribution initial condition. Left: Time evolution of macroscopic quantities ($\mathcal{E}$, $N$ and $\braket{A}$) with $R=1$ and $S_L=10^3$ (black curve), $S_L=10^4$ (orange curve) and the collisionless case (green curve). Right: Time evolution of macroscopic quantities at $S_L=10^4$ and varying~$R$.}
\label{fig:Gauss_ENA}
\end{figure}

\begin{figure}
\centering
\includegraphics[width=0.7\linewidth]{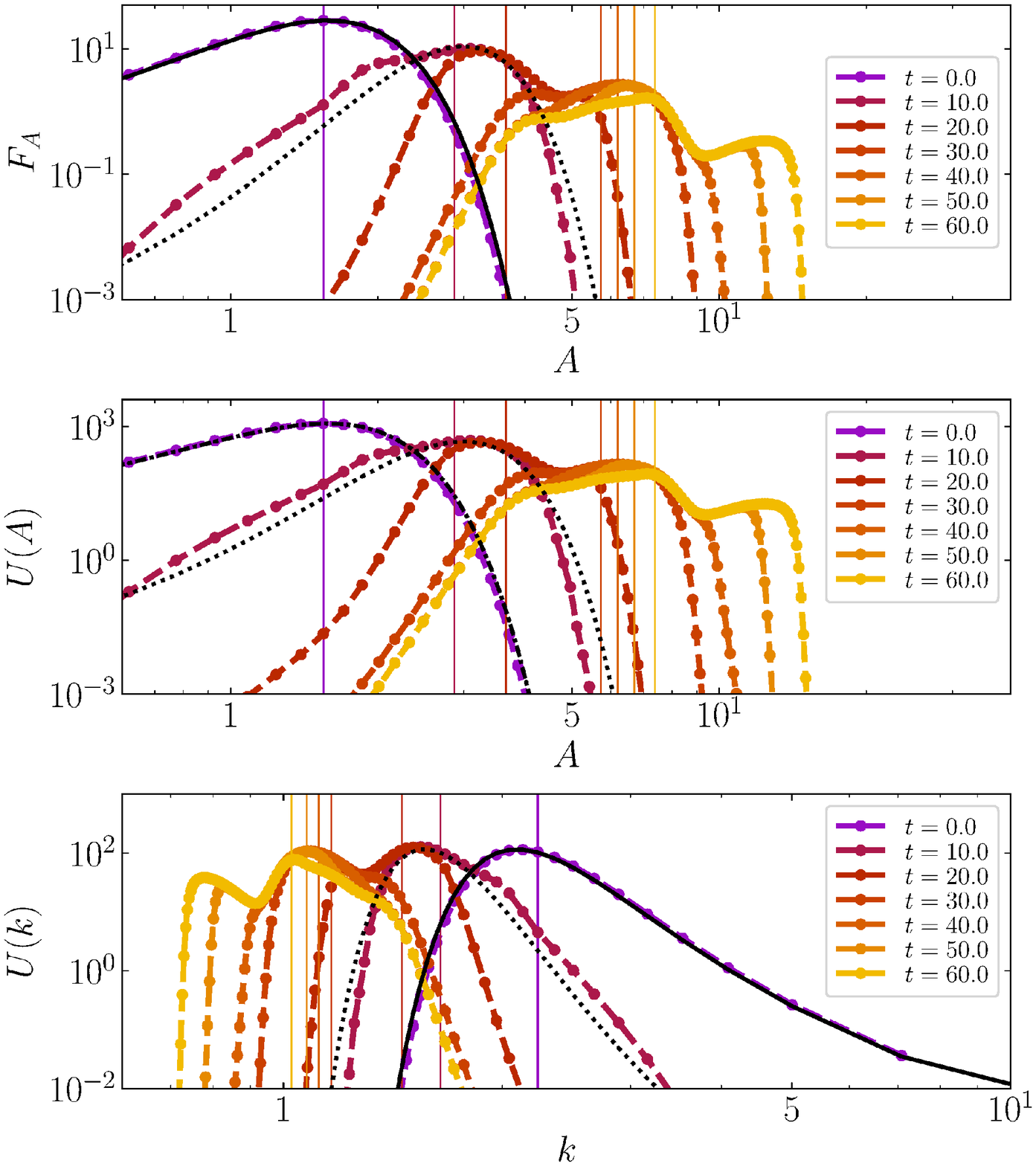}
\caption{Gaussian initial distribution for $R=1$ and $S_L=10^4$. Top: time evolution of~$F_A$; middle and bottom: time evolution of the energy distribution $U(A,t)$ and $U(k,t)$, respectively. Vertical lines indicate mean values of $A$ and $k$ at the corresponding moments of time. The solid (dotted) black curves in each panel indicate fittings of Gaussian distribution using $\sigma_A=0.47$ ($\sigma_A=0.6$) for $t=0$ ($t=10$).}
\label{fig:Gaus_sp_dis}
\end{figure}

Figure~\ref{fig:Gauss_ENA} shows the time evolution of the macroscopic quantities $\mathcal{E}, N$ and $\langle A\rangle$ for different values of $S_L$ (left column; the collisionless case is also included for comparison) and of $R$ (right column).
The time evolution of these quantities retains a power-law behaviour where the scaling of $\mathcal{E}$ and $N$ is $t^{-a}$ with $0.7 < a < 0.8$ and $\langle A\rangle$ scales as $t^a$. 
That is, an initial Gaussian distribution of islands leads to a somewhat slower evolution than the delta-distribution case. 
The ``steps'' are still present during the time evolution in this case, as the distribution is narrow enough that the difference in merging time between the smallest and largest islands is not too large, and a large fraction of the islands complete the same number of mergers.
The effect of changing $S_L$ or using the collisionless reconnection rate is shown on the left panels (at fixed $R=1$).
The reconnection rate, and hence the merging rate, increases going from $S_L = 10^4$ to $10^3$ to the collisionless regime. Merging starts earlier when the reconnection rate is higher. However, the slope of the power-law evolution is not strongly affected as the reconnection time scale is much longer than the Alfv\'enic time scale both in the SP and in the collisionless regimes.

The right panels of Figure~\ref{fig:Gauss_ENA} show the effect of allowing for interactions between non-identical islands (i.e., changing $R$ for fixed $S=10^4$).

The evolution of macroscopic quantities is found to be (marginally) faster for systems with finite $R$ than for the one with $R=0$, i.e., allowing for only identical-island interactions.
The merging time of islands is proportional to $\sqrt{A_s}(A_s A_l)^{1/4} = (A_s^3 A_l)^{1/4}$ [Eqs.~\eqref{eq:tau_rec} and \eqref{eq:v_rec}] where $s$ and $l$ refer to the smaller and larger islands in an interaction. 
Therefore, the merging time depends more strongly on the smaller island (with $A_s$).
In the case of the Gaussian distribution, islands with sizes close to the mean area have the largest population, and thus have a large effect on the overall dynamics. 
Compared to the merging time between two mean-sized islands, the merging time between a mean-sized island and a smaller island is shorter and dominated by the smaller island, while the merging time between a mean-sized island and a larger island is longer but only has a weak dependence on the area of the larger island. Therefore, allowing the interaction between non-identical islands essentially reduces the merging time.
Additionally, more islands are allowed to merge at any given time. The combination of these effects makes the overall dynamics faster when $R$ is finite. 

We also note that for the evolution of $N$ and $\langle A \rangle$, the curves for $R=0.01,0.03,0.1,1$ are essentially identical and are different from the $R=0$ curve. 
This suggests that the value of $R$ does not have a significant effect on the overall dynamics, as long as it is not zero. 
We conclude that highly non-local interactions (corresponding to mergers between islands differing a lot in size) are not dynamically important.
This observation has implications for the discussion of decaying turbulence found in Section.~\ref{sec:scaling_law}.
 
\begin{figure}
\centering
\includegraphics[width=0.7\linewidth]{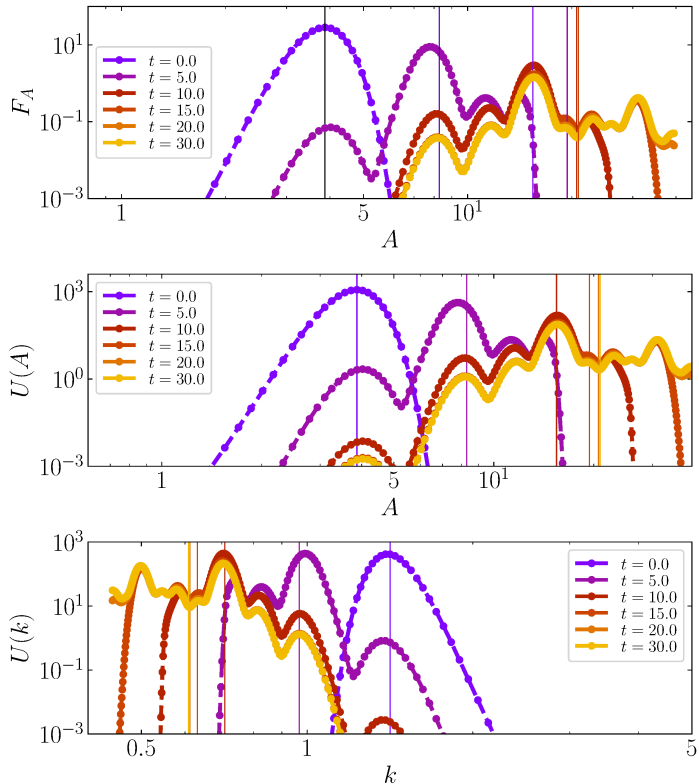}
\caption{Gaussian initial distribution with $R=1$ and the collisionless reconnection model. Top: time evolution of $F_A$; middle and bottom: time evolution of the energy distribution $U(A,t)$ and $U(k,t)$, respectively. Vertical lines indicate mean values of $A$ and $k$ at the corresponding moments of time. }
\label{fig:Gauss_collisionless_dis}
\end{figure}

The evolution of the distribution function $F_A$ and magnetic energy spectra are shown in Figure~\ref{fig:Gaus_sp_dis} for the case $S_L = 10^4, R=1$ (given the very limited resolution in the $\psi$ coordinate, as mentioned earlier, there is not much value in studying $F_\psi$ and therefore it is not shown).
At $t=10$, the initial distribution has decayed and a new peak grows at $A \approx 2 \bar{A}$ (top panel), as a result of the merging of islands with areas close to the first peak at $\bar{A}$ merging. The second peak of the distribution remains Gaussian, and has a somewhat larger width (comparing to the initial one), which we fit with $\sigma_A = 0.6$. After the merging of the initial islands, the peak of distribution continues to shift to larger area with a greater width. Due to the longer merging time, the old peaks have not always decayed when the new peaks form (e.g., at $t=30, t=60$).
We note that, at late times ($40<t<60$), the tail of $F_A$ at small $A$ remains (almost) unchanged and the mergers happen mainly between islands at the peak of $F_A$. This is because (i) the relatively fast mergers between (identical or similar) small islands are less frequent with larger values of $R$; and (ii) the probability for mergers between islands at the peak of $F_A$ is much higher than for mergers involving small islands because of their larger number density and cross sections.

The spectra evolution (middle and bottom panels) show similar behaviour as the distributions. \footnote{The energy spectrum is defined as $U(A) = \int f_{\rm tot}(\psi,A) \psi^2 d\psi$. Since the variation of $f_{\rm tot}$ with $\psi$ is very limited in this calculation due to the low $\psi$ resolution, the evolution of the spectra is very similar to the evolution of the area distribution $F_A\equiv \int f_{\rm tot}(\psi,A) d\psi$.}
We remark again that our definition of magnetic energy spectrum [Eqs.~\eqref{eq:AreaEnergyDist} and \eqref{eq:Uk_def}] is different from the conventional definition based on the Fourier transform of the magnetic field as a function of position in configuration space; the former straightforwardly presents magnetic energy distribution over islands with different scales, while the latter can be dominated by local features such as magnetic reversals at current sheets. The difference between the multi-peak spectra presented here and the $\sim k^{-2}$ spectrum in~\citet{zhou2019magnetic} is caused by the different definitions of spectrum, without incompatibility.
The collisionless case is shown in Figure~\ref{fig:Gauss_collisionless_dis}. The multiple peaks are still visible in the system both for $F_A$ and the spectra, but there is no clear transition from one Gaussian to another.
Instead, multiple peaks are present because of the smaller merging time. 
This means that the first islands in a given generation to complete merging can start their second merger while other islands of that original generation are still merging, giving rise to the multiple peaks.

\subsection{Power-law initial condition}
\label{sec:powerlaw}
Another situation of interest is that of an initial power-law distribution of islands, which may be relevant to various astrophysical systems such as small-scale ($< 0.1$ AU) flux ropes in the solar wind~\citep{janvier2014there} and magnetic structures in the downstream of a collisionless shock as a result of the Weibel instability~\citep{katz2007self}.
To reduce computational expense, we set up the initial islands with a power-law distribution over the size, but identical magnetic flux:
\begin{equation}
\label{eq:powerlaw}
    f(\psi,A,t=0)=f_0A^{-\alpha}\delta(\psi-\psi_{\rm ini}).
\end{equation}
The power-law exponent, $-\alpha$, is the same as that of its corresponding $A$-spectrum: $U(A) = \int d\psi f_0A^{-\alpha}\delta(\psi-\psi_{\rm ini}) \psi^2 =f_0 \psi_{\rm ini}^2 A^{-\alpha}$.
In our calculation, the power-law distribution is set up in the range  $A_{\rm min} < A < A_{\rm max}$, where $A_{\rm min} = L^2/N_A$ and $A_{\rm max} = 5 L^2/N_A$.
Initial conditions with $\alpha \in \{0.5,1,2 ,4\}$ are studied. We set $\psi_{\rm ini} = (5/6)LB_0$ and resolve the system with $N_A=256$.  

\begin{figure}
\centering
\begin{subfigure}{.5\textwidth}
  \centering
  \includegraphics[width=1.0\linewidth]{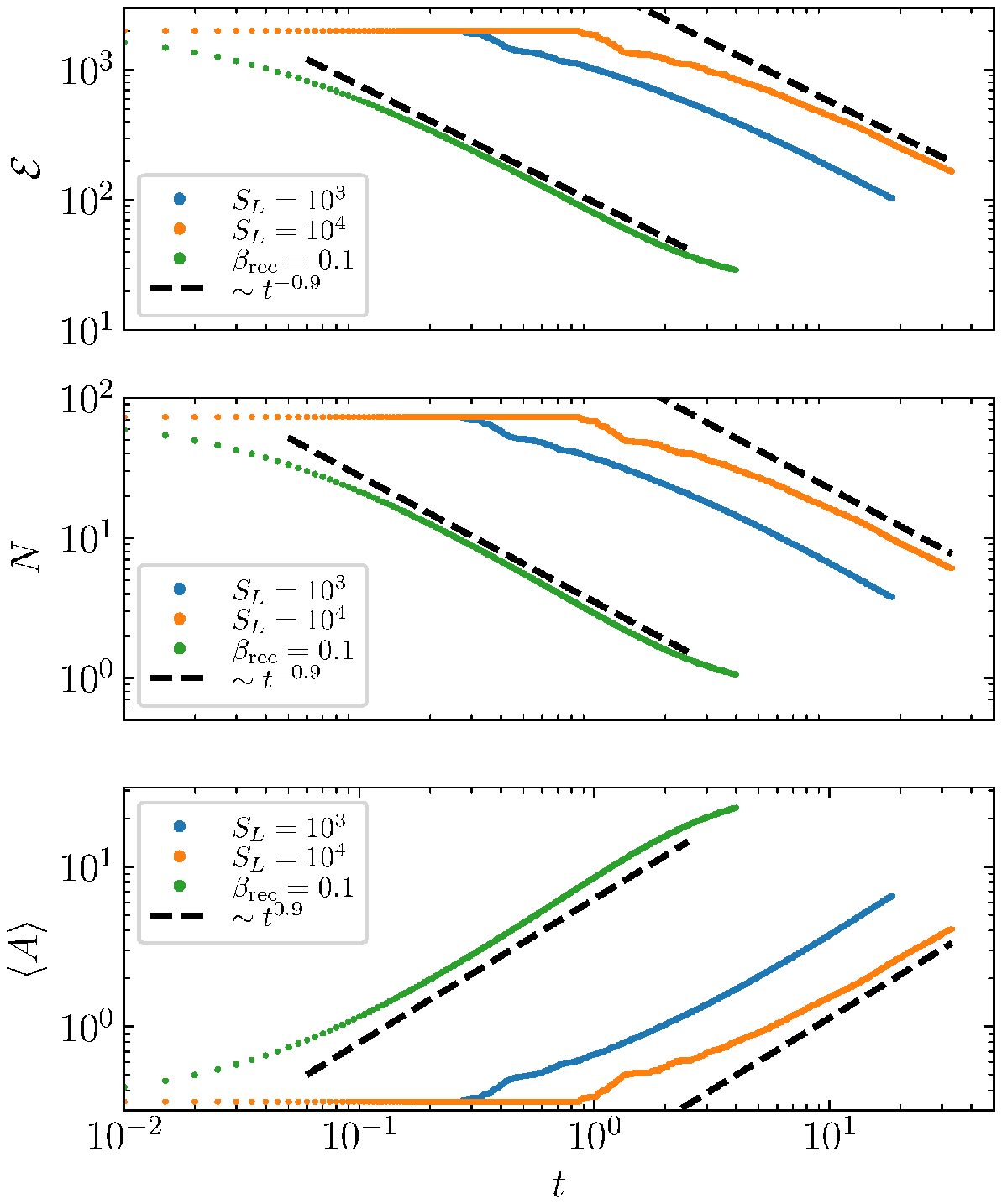}
\end{subfigure}%
\begin{subfigure}{.5\textwidth}
  \centering
  \includegraphics[width=1.0\linewidth]{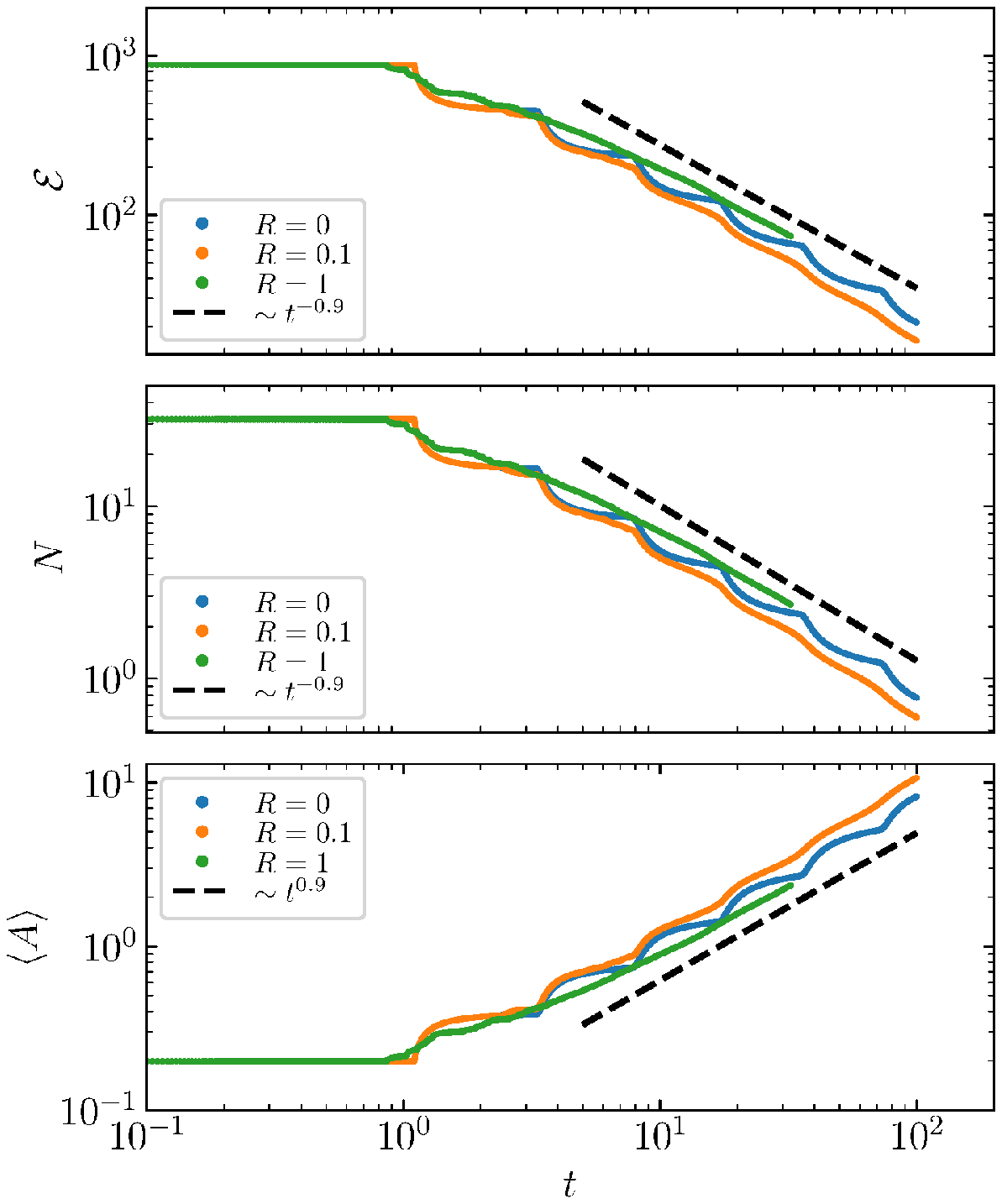}
\end{subfigure}
\caption{Power-law initial distribution. Left: The evolution of macroscopic quantities ($\mathcal{E}$, $N$ and $\braket{A}$) for unconstrained interaction ($R=1$) with $\alpha=1$ and $S_L=10^3,~10^4$ and the collisionless case. Right: The evolution of macroscopic quantities for $S_L=10^4$ and $\alpha=1$, with varying $R$.}
\label{fig:Powertest}
\end{figure}

\begin{figure}
  \centering
  \includegraphics[width=0.5\linewidth]{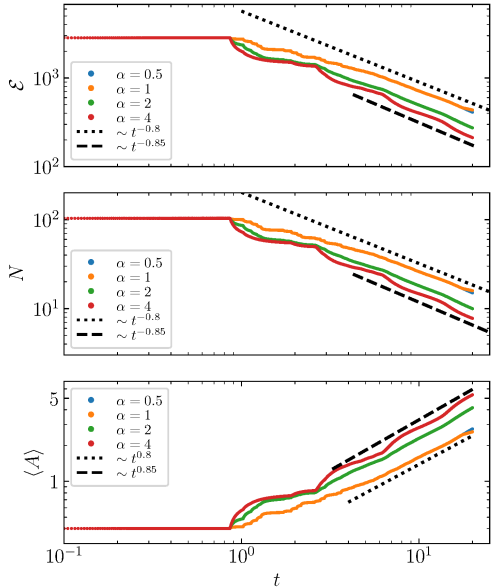}
\caption{Power-law initial distribution. Time evolution of macroscopic quantities ($\mathcal{E}$, $N$ and $\braket{A}$) for unconstrained interaction ($R=1$) and $S_L=10^4$, with varying $\alpha$. }
\label{fig:power_vary_alpha}
\end{figure}

The evolution of various macroscopic quantities with $\alpha = 1$ in the initial condition is shown in Figure~\ref{fig:Powertest}.
The left column shows cases with $S_L=10^3, 10^4$ and the collisionless case, for $R=1$.
The decay of $\mathcal{E}$ and $N$ shows a $t^{-a}$ power-law scaling, and $\langle A \rangle$ increases as $t^{a}$, where $a \approx 0.9$ and does not strongly depend on the reconnection rate.
This is only slightly different from our results for the delta and Gaussian initial distributions (reported in Sections~\ref{sec:delta} and~\ref{sec:gaussian}). 
We note that in this case, the curves of evolution of macroscopic quantities are smooth: the discrete ``steps'' that occur for the delta (Figure~\ref{fig:Deltatest}) and Gaussian (Figure \ref{fig:Gauss_ENA}) distributions are absent. 
This is because no dominant island size or merging time scale exists in the system due to the wide spread of island sizes pertaining to a power-law distribution.
Islands then merge over a wide range of time scales, causing the evolution to appear smoother. 
The effect of changing $R$ is shown in the right column for $S_L=10^4$. As $R$ increases, the decay of $\mathcal{E}$ and $N$ becomes slightly steeper, as does the corresponding increase of $\langle A\rangle$, which indicates a faster overall merging process. The explanation is the same as for the Gaussian case: allowing non-local interactions in the collision integral increases the number of islands that can merge and increases the average merging rate. 

The effect of varying the index of the initial power law from $\alpha = 0.5$ to $4$ on the evolution of the macroscopic quantities is shown in Figure~\ref{fig:power_vary_alpha}. 
The difference between the $\alpha=0.5$ and $\alpha=1.0$ runs is minor, indicating that for shallow initial power-law distributions, the evolution of macroscopic quantities depends only weakly on the initial power-law slopes.
As $\alpha$ increases, merging becomes faster, shown by somewhat steeper traces of $N$, $\mathcal{E}$ and $\langle A \rangle$.
This can be understood by noting that the distribution function becomes progressively more similar to a delta-function distribution with an increasing $\alpha$, and so this steepening is consistent with the results of Section~\ref{sec:delta} and should tend to~$\sim t^{-1}$.

\begin{figure}
\centering
\includegraphics[width=0.7\linewidth]{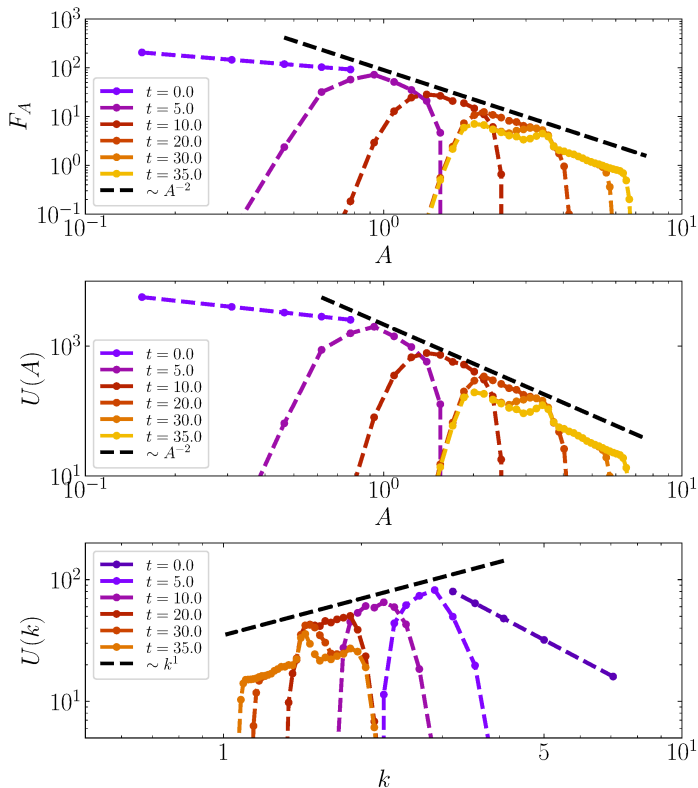}
\caption{Power-law initial distribution for $R=1,\alpha=1$, and $S_L=10^4$. Top: time evolution of the area distribution function $F(A)$; middle and bottom: time evolution of spectra $U(A,t)$ and $U(k,t)$, respectively. The dashed lines show reference $A^{-2}$, $k^{1}$ power-laws in the respective plots. }
\label{fig:power_spectrum}
\end{figure}

The evolution of the energy spectra and area distribution function is shown in Figure~\ref{fig:power_spectrum}, for $\alpha=1$ (corresponding to an initial $k^{-\gamma}$ spectrum where $\gamma = 3-2 \alpha=1$), $R=1$, and $S_L=10^4$. The distribution function $F_\psi$ is not shown because it starts as a delta-distribution and remains so, according to our merging rules. 
Interestingly, the area distribution $F_A$ (top panel) spontaneously forms a range of $A$ where a power-law distribution is maintained with $F_A \sim A^{-2}$, though the $-2$ index differs from the initial index. 
The formation of this $A^{-2}$ distribution is still present when the initial distribution has $\alpha = 0.5$ and $\alpha=2$\footnote{In the case of $\alpha = 4$ the distribution is so steep that it is closer to the delta-function case than the power-law case due to the limited dynamic range.}, implying that this $A^{-2}$ distribution is a ``attractor'' that the system tends to evolve to, independent of the initial conditions.  
Moreover, there is an extended envelope which covers the power-law regions of the curves at different times, showing the self-similar features of the system evolution.
Consistent with $F_A$, the energy spectra also transition to a $U(A) \sim A^{-2}$ [and correspondingly $U(k) \sim k$] power law as the system evolves (middle and bottom panels). 
Similarly, this behaviour is independent of the initial slope and the spectra tend to evolve to the~$\sim A^{-2}$~($\sim k$) fixed point.
For the collisionless case (not shown), the evolution of $F_A$ and spectra show similar behaviours and evolve to power laws with the same exponents as their resistive-MHD counterparts.
This self-similar evolution of $F_A$ and spectra can be demonstrated analytically by applying the hierarchical model of~\cite{zhou2019magnetic} to a power-law distribution of islands under the assumption that coalescence events most often occur between similar size islands --- see Appendix~\ref{sec:multi_scale_rules}. 
A derivation of the observed $k^1$ ($A^{-2}$) spectrum can be obtained based on the self-similar properties. 
However, we are not yet able to prove analytically that the $k^1$ ($A^{-2}$) scaling is an ``attractor'' of the spectra and $F_A$ that the system is expected to evolve to.
One final observation worth making is that, at later times, there is a peak at the center of the distributions and spectra, which is likely due to incomplete merging of islands within that range of $A$, while the rest of the distribution in regions of smaller and larger $A$ still shows a $-2$ power law.

\section{Connection to magnetically-dominated decaying turbulence and scaling theories}
\label{sec:scaling_law}
The system of a large number of coalescing islands is closely related to the problem of decaying turbulence in the magnetically-dominated regime.
On the one hand, the astrophysical systems of which magnetic fields can be conceptualised as interacting magnetic structures, or those where magnetic-island structures are produced, are usually in a turbulent state. 
On the other hand, turbulence is unavoidable as the random motion of a large ensemble of islands easily turns chaotic.
Without external energy sources, the turbulence will decay and its energy dissipation as well as the associated inverse cascade can be realised through island mergers.

The decay of MHD turbulence is believed to be of a self-similar nature.
By definition, a quantity $z$ is self-similar if it satisfies the relation $z(\ell x)=\ell^h z(x)$. The magnetic and velocity fields in MHD have such self-similarity, $\mathbf{B}(\ell \mathbf{x},\ell^{1-h}t)=\ell^h\mathbf{B}(\mathbf{x},t)$, originated from the rescaling symmetry of the MHD equations~\citep{olesen1997inverse}.
We note that our hierarchical model~\citep{zhou2019magnetic,zhou2020multi}, or the $R=0$ limit of the IKE, can reproduce such self-similarity. Considering the evolution of a 2D single-scale (identical islands) system, the change of quantities after mergers can be represented by simultaneously rescaling the following quantities: $B'=\ell^h B$, $A'=\ell^2 A$, and $t'=\ell^{1-h} t$, where $\ell=(t'/t)^{1/2}$ is the scaling factor that maps between two arbitrarily chosen moments of time $t$ and $t'$, and $h=-1$ is chosen based on the conservation of magnetic flux: $B' \sqrt{A'} \sim B\sqrt{A}$.
The decay of a 2D turbulent system can be considered as a consecutive sequence of the above rescaling operations~\citep{olesen1997inverse}; in the magnetically-dominated regime, this sequence is materialised through successive island mergers.

It is reassuring that our heuristic island-based model follows the fundamental rescaling symmetry of MHD equations in the limit of $R=0$, i.e., local (in Fourier space) interactions only. 
It is not clear, however, if the self-similar properties remain valid when non-local interactions (mergers between non-identical islands) are allowed.
This can be partially answered by the scan in the parameter $R$ (allowed interaction range) in our IKE model.
It is shown in Figs.~\ref{fig:Deltatest}, ~\ref{fig:Gauss_ENA}, and~\ref{fig:Powertest} that, for different island distributions, the evolution of the system does not change significantly with values of $R$ ranging from the local ($R=0$) to the unconstrained ($R=1$) interaction case. For the case of a Gaussian initial distribution (Fig.~\ref{fig:Gauss_ENA}), the effect of $R$ is more noticeable than that with the other two initial distributions. However, even in this case, the evolution of the system with unconstrained interactions ($R=1$) is almost identical to when only interactions between nearly similar islands are allowed ($R=0.01$, $0.03$, and $0.1$), with a non-significant but noticeable deviation from the strict local-interaction case ($R=0$). 
This implies that highly non-local interactions (big-small island mergers) do not strongly affect the evolution of the system.
\footnote{One caveat of this conclusion is that for most of our simulations, due to the limited dynamical range, islands with significant energy are separated from the peak of the distribution function only by a factor of order unity. 
Therefore, it is possible that the island distributions are still to some extent represented by a characteristic scale, and the non-local interaction is weak because of the lack of scale separation for islands containing significant energy. Simulations with a larger dynamical range and wider island distribution are required to test whether non-local interactions are dynamically-important to a system of coalescing islands.}

The absence of dynamically-important highly non-local interactions allows us to adopt scaling arguments for this problem and to assume a statistical self-similarity, because scaling arguments mainly consider the local (in Fourier space) dynamics around the assumed characteristic quantities.
For a system with a characteristic length scale $l$ and a characteristic magnetic field $B$ at this scale, islands can be distributed and interact in a non-trivial way around this $l$ and $B$.
The characteristic time scale for these mergers will be $\tau_{\rm rec} \sim \beta_{\rm rec}^{-1} l/B$ [Eqs.~\eqref{eq:tau_rec} and~\eqref{eq:v_rec}], and the decay of magnetic energy can be expressed as $d B^2/dt \sim -B^2/\tau_{\rm rec} \sim \beta_{\rm rec} B^3/l$. 
During the reconnection process, the magnetic flux $\psi \sim Bl$ is approximated conserved. 
This causes the dimensionless reconnection rate of the merging islands in the SP regime to remain the same during the evolution $\beta_{\rm rec} = S^{-1/2} \sim (Bl/\eta)^{-1/2} \sim \text{const}$. ($\beta_{\rm rec}$ is also a constant in other reconnection regimes as we discussed in Section~\ref{sec:tau_rec}.)
Combined with the constancy of both the flux $Bl$ and $\beta_{\rm rec}$, we obtain the scaling laws of energy decay and growth of characteristic length scale $B^2 \sim \beta_{\rm rec} t^{-1}$ and $l \sim \beta_{\rm rec}^{-1/2} t^{1/2}$ [\citep{zhou2019magnetic,zhou2020multi}; see also~\citet{schekochihin2020mhd}]. In the case of volume-filling islands, the number of islands scales as $N \sim \beta_{\rm rec} t^{-1}$ following $N l^{2} \sim \text{const}$.
These scaling laws are identical to those predicted by our hierarchical model.
Both the scaling arguments and the hierarchical model adopt the concept of reconnection-controlled ``structure merger'', but the former does not rely on the restrictive assumptions of pairwise and identical-island mergers, which are assumed in the latter. 

We note that the scaling laws $B^2 \sim t^{-1}$ and $l \sim t^{1/2}$ in 2D decaying turbulence have been established for decades without invoking the idea of reconnection~\citep{hatori1984kolmogorov,biskamp1989dynamics}. In those studies, the decay time scale of magnetic fields is assumed to be the eddy-turn-over time $l/B$ if equipartition between kinetic and magnetic energies holds.
In reconnection-based models discussed above, the conservation of magnetic flux is invoked to relate the evolution of $B$ and $l$.
In a 2D MHD system, an equivalent role is played by the conservation of anastrophy $\int d^2 \mathbf{r} A_z^2$, where $A_z$ is the component of the vector potential perpendicular to the 2D plane and therefore, $A_z \sim Bl \sim \text{const}$.
It is remarkable that the same scaling laws hold for both a decay controlled by reconnection and one on the ideal Afv\'enic time scale. 
This ``coincidence'' occurs because the reconnection time scale tracks the Afv\'enic time scale when $\beta_{\rm rec}$ is a constant. [This does not hold for systems with hyper-resistivity where different scaling laws are found~\citep{hosking2021}.]
Physically, reconnection is the main mechanism responsible for magnetic energy dissipation and that should be what sets the time scale, as confirmed by direct numerical simulations in~\citet{zhou2019magnetic} where energy decay curves from systems with different values of resistivity overlap once their time-axes are normalised to their characteristic reconnection time scale.

The above scaling theory based on the conservation of magnetic flux can be straightforwardly applied to a 3D system with a strong guide field by invoking an additional element of critical balance~\citep{zhou2020multi}. 
However, the 2D magnetic flux, or the anastrophy, is not an invariant in a generic 3D system. 
It is well-known that the conservation of magnetic helicity (in replacement of anastrophy) is responsible for the decay and inverse transfer of helical MHD turbulence~\citep{taylor1974relaxation,pouquet_1978,christensson2001}.
Recently, the puzzling inverse-transfer phenomenon that occurs in magnetically-dominated turbulence with zero net magnetic helicity [e.g., ~\citep{zrake2014inverse,brandenburg2015nonhelical,brandenburg2017classes,bhat2021inverse}] has also been solved with a powerful tool of the Saffman helicity invariant~\citep{hosking2021}. In~\citet{hosking2021}, the physical picture of merging magnetic structures is shown to be compatible with their formal theory based on the self-similar argument and the Saffman helicity invariant. In~\citet{bhat2021inverse}, the energy decay of 3D nonhelical turbulence is shown to occur on the reconnection time scale. The above evidence indicates the wide applicability of the structure-merger type of dynamics to magnetically-dominated turbulent systems.

\section{Conclusion}
\label{sec:conclusion}
This paper investigates the evolution of a 2D system of merging magnetic islands using a statistical description. The islands are characterised by their areas and magnetic fluxes, and their evolution is governed by a Boltzmann-type kinetic equation which we derive (dubbed the island kinetic equation, IKE). In this equation, island mergers are accounted for via a collision integral whose key feature is to allow for finite-time (rather than instantaneous) mergers --- thereby informing the system about the reconnection time scale, which governs the mergers, and which is large compared to the Alfv\'enic time scale.

We use this IKE to study the inverse transfer process enabled by island merging, focusing on the scaling with time of the growth of the magnetic field length scale and the associated decay of the magnetic energy, as well as the evolution of the distribution of islands and magnetic spectra. 
By solving the IKE numerically for different initial island distributions, we find that the time evolution of global quantities is insensitive to the initial distribution, and is close to the predictions of~\citet{zhou2019magnetic,zhou2020multi}: magnetic energy $\mathcal{E} \sim \tilde{t}^{-1}$, the number of islands $N \sim \tilde{t}^{-1}$, and the averaged area of islands $\braket{A} \sim \tilde{t}$, where $\tilde{t}$ is the time normalized to the reconnection time scale.
This weak dependence of the evolution of global quantities on initial island distribution is consistent with our conclusion in Appendix~\ref{sec:multi_scale_rules} where we generalise the analytical model for identical islands in~\citet{zhou2019magnetic} to describe islands with distributions of sizes and fluxes.
Although the predictions from our hierarchical model are only confirmed numerically using a relatively limited range of Lundquist number $S_L \in \{10^3, 10^4\}$ (as well as the $\beta_{\rm rec}=0.1$ case), we believe that they should also hold in the plasmoid-dominated regimes with higher Lundquist number, as we discuss in footnote \ref{footnote_plasmoids} of page~\pageref{footnote_plasmoids}.

We study the system evolution in detail with three different types of initial island distributions: identical islands, Gaussian distributions and power-law distributions. We also introduce a dimensionless parameter, $R$, to quantify the maximum difference of areas and fluxes between islands that are allowed to interact.
In the limiting case of an initial distribution with only one island size and merging between identical islands, corresponding to the hierarchical model and the set-up of the reduced-MHD simulations in \citet{zhou2019magnetic}, the time evolution of the macroscopic quantities predicted by the model is reproduced. In this case, increasing $R$ does not change the evolution, showing that the assumption used in \cite{zhou2019magnetic} that only identical islands merge is valid for adequately describing the overall system dynamics.
In more general cases with non-trivial initial (Gaussian and power-law) distributions of islands, we find that the time evolution of macroscopic quantities still remains in the form of a power-law, and can be described by the hierarchical model of~\citet{zhou2019magnetic} to a reasonably good approximation;
the evolution observed in those cases is only slightly slower (the power laws are slightly shallower) than the results of the delta-distribution study and the predictions of the hierarchical model.
This holds true for both the Gaussian and power-law initial conditions, and indicates that a distribution of islands may merge somewhat more slowly than a population of identical islands. 
Aside from the global quantities, with Gaussian initial conditions, the distribution functions (spectra) are spread over a wider range of areas (wave numbers) and form multiple peaks at later times. With the power-law initial conditions, the distributions (spectra) evolve to a fixed $\sim A^{-2}$ ($\sim k$) power law at late times, regardless of the initial slope, as a result of self-similar evolution.  

These results are directly relevant to space and astrophysical systems for which the overall dynamics can be conceptualized as a turbulent sea of interacting magnetic islands (or flux tubes in 3D).
While direct numerical simulations of such systems might be hard to interpret due to complicated multi-scale interactions among various physical processes, the present study isolates the self-dynamics of magnetic islands and its contribution to the overall evolution of the system, thus providing useful insights into how such systems might organize themselves.
Furthermore, our IKE model can also be used as a building block in the study of other problems such as particle acceleration and plasma heating. 
The main physics process that we study here --- magnetic island merging enabled by magnetic reconnection --- is essentially an energy transfer process. 
Understanding the statistics of island mergers is key to deriving the statistics of dissipation processes and particle energisation.
Therefore, our IKE model can be combined with models of other dissipative physical processes and shed light on long-standing problems such as the heating of the solar corona and accretion disk coronae, and the production of high-energy particles in solar wind and heliosheath.

Finally, while the extrapolation of these results to 3D geometries is not straightforward, we think that, at least in situations where a strong guide field is present, the evolution of the system will not be changed significantly by the dynamics in the third dimension (parallel to the guide field). 
Indeed, in \citet{zhou2020multi}, we showed that the parallel (to the guide field) dynamics are essentially passive, dictated by the perpendicular dynamics through a critical-balance-like relation. 
In the weak guide-field limit, however, the system dynamics can be qualitatively different.
Kink-type modes may play a significant role and disrupt the flux-rope structures.
In addition, if the system has zero net magnetic helicity (i.e., roughly equal number of structures with opposite polarities in helicity), roughly half of the mergers would happen between structures with opposite helicities, resulting in non-helical structures that will relax to zero magnetic energy on the ideal time scale. 
This feature has been discussed in detail in~\citet{hosking2021}, based on which they derive decay laws for nonhelical (as well as helical) MHD turbulence.
\\

\paragraph{\textbf{Acknowledgements.}}
MZ thanks David Hosking for his insightful comments and the suggestion to add the discussion in Section 5 to the manuscript. She also thanks Marc Swisdak for his insightful comments on this work.
MZ and DHW thank Jonathan Ng and Manasvi Lingam for useful discussions during APS-DPP 2019.\\

\paragraph{\textbf{Funding.}} This work was supported by NSF CAREER award No.~1654168 (MZ and NFL), NASA award NNH19ZDA001N-FINESST (MZ), NSF grants AST-1411879 and AST-1806084 and NASA ATP grants NNX16AB28G and NNX17AK57G (DAU), and the Caltech SURF fellowship (DHW). 
This research used resources of the MIT-PSFC partition of the Engaging cluster at the MGHPCC facility, funded by DOE award No.~DE-FG02-91-ER54109. \\

\paragraph{\textbf{Declaration of Interests. }} The authors report no conflict of interest.\\

\appendix

\section{Comparison with previous studies}
\label{sec:comparison}
A statistical model describing the distribution of plasmoids in a large current sheet with hierarchical structures has been developed by~\citet{fermo2010statistical}. 
In Fermo's model, the islands are described by a distribution function $f(\psi, A,t)$ with the same phase space variables $\psi$ and $A$ that we employ in this paper, and the evolution of the distribution function is governed by a Boltzmann-type kinetic equation:
\begin{equation}
\begin{split}
    \frac{\partial f}{\partial t} + \frac{\partial}{\partial \psi}\left(\dot{\psi}f\right)& + \frac{\partial}{\partial A}\left(\dot{A}f\right) =\ S\left(\psi, A\right)-\frac{v_A}{L}f\\
    &+ \frac{1}{L}\int_0^A dA^\prime f\left(\psi, A^\prime\right)\int_0^\psi d\psi^\prime v\left(\psi, A^\prime, \psi^\prime, A-A^\prime\right)f\left(\psi^\prime, A-A^\prime\right)\\
    &-\frac{1}{L}f\left(\psi,A\right)\int_0^\infty dA^\prime\int_0^\infty d\psi^\prime v\left(\psi, A, \psi^\prime, A^\prime\right)f\left(\psi^\prime, A^\prime\right).
\end{split}
\end{equation}
 Fermo's model is essentially 1D in real space. The plasmoid distribution evolves in time due to the following effects. The growth of sizes and fluxes of plasmoids is caused by the reconnection in secondary current sheets, represented by the two terms with $\dot{A}$ and $\dot{\psi}$ in the \textit{l.h.s} of the equation. These two terms increase the size of islands but do not change their numbers. On the \textit{r.h.s}, the generation of new plasmoids is represented by the source term $S(\psi,A)$ and the ejection of plasmoids out of the system is represented by the sink term $v_Af/L$. The coalescence of plasmoids is represented by the two integral terms and implicitly assumed to be instantaneous. Certain selection rules for plasmoid coalescence are implemented in the integrals.
 The detailed description of the model can be found in~\citet{fermo2010statistical}, where the steady state solution has been studied.
 A similar kinetic model for plasmoids in current sheets is discussed in \cite{huang2012distribution} in which the growth, merging and ejection of plasmoids are modeled in the phase space of magnetic flux. While these models are designed to capture different physical effects, they share a common assumption that the coalescence of plasmoids is an instantaneous process. 
  
 The main differences between those previous models and ours stem from the fact that we are interested in astrophysical systems that can be conceptualised as a sea of interacting islands in 2D, while those models consider the dynamics of plasmoids in a 1D reconnecting current sheet. This leads to the following key differences: (1) The time scale for island coalescence is long compared to the Alfv\'en time in our system, and we account for this by using a non-instantaneous collision operator to represent coalescence.~\footnote{Accounting for the finite coalescence time could also improve the accuracy when studying the plasmoid dynamics in a 1D reconnecting current sheet, and should be further explored. In~\citet{uzdensky2010fast}, an analytical model predicting the distribution function of plasmoids is provided, where the coalescence of plasmoids is assumed to be instantaneous. However, direct numerical simulations by~\cite{loureiro2012magnetic} show some discrepancies relative to what is predicted in~\citet{uzdensky2010fast}, which are identified as being precisely due to the fact that coalescence is not instantaneous.}
  (2) We only consider the change of the distribution function $f(\psi,A)$ caused by coalescence of islands. That is, we keep only the time derivative of the distribution function and the two collision integrals in the kinetic equation. The term for new island generation from secondary tearing-unstable current sheets, the term for islands ejecting out of the system, and the two terms for the growth of sizes and fluxes of islands due to large-scale reconnection are neglected. The justification for neglecting those terms can be found in Section~\ref{sec:merger_IKE}.
  (3) The collision probability and cross-section for islands are calculated in two spatial dimensions.

\section{Convergence study}
\label{sec:convergence}
\begin{figure}
\centering
  \includegraphics[width=1.0\linewidth]{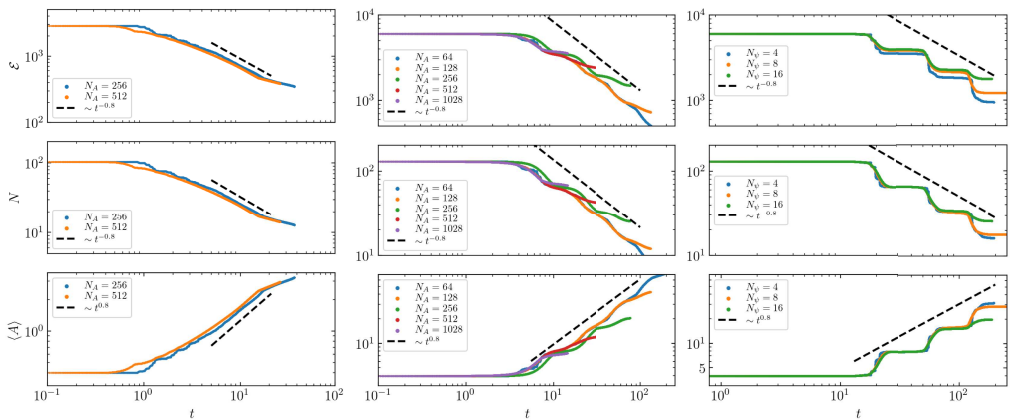}
\caption{Time evolution of $\mathcal{E}$ (top panels), $N$ (middle panels), and $\braket{A}$ (bottom panels) from a numerical convergence study. All simulations are initialized with unconstrained interaction ($R=1$) and $S_L=10^4$. Explicit numerical parameters are given in Table \ref{table:ConvStudy}. Left: power law initial distribution ($\alpha=1$): varying $N_A$ with fixed $N_{\psi}=4$. Middle: Gaussian initial distribution: varying $N_A$ with fixed $N_{\psi}=4$. Right: Gaussian initial distribution: varying $N_{\psi}$ with fixed $N_A=32$.}
\label{fig:Contest1}
\end{figure}

We report here a convergence study on the numerical solutions of the IKE.
We employ the power-law and Gaussian initial distributions to study the convergence of phase space resolution, $N_A$ and $N_{\psi}$, and focus on the evolution of macroscopic quantities $\mathcal{E}$, $N$ and $\langle A \rangle$.
The parameters used in our convergence studies are presented in Table \ref{table:ConvStudy} and results are shown in \figref{fig:Contest1}.

\setlength{\tabcolsep}{12pt}
\begin{table}
\centering
\def\arraystretch{1.25}
\begin{tabular}{cccc}
 \hhline{====}
 Parameters & Power law & Gaussian & Gaussian\\
 \noalign{\hrule height 0.5pt}
 $N_A$   & $256,512$ &$64,128,256,512,1024$   &$32$\\
 $N_{\psi}$&   $4$  &$4$& $4,8,16$   \\
 \hhline{====}
\end{tabular}
\caption{Summary of parameters of runs in Figure~\ref{fig:Contest1} for the IKE convergence studies. All runs are performed with fixed $R=1$ and $S_L=10^4$.
}
\label{table:ConvStudy}
\end{table}

For the power-law distribution case, we initialized the distributions using Eq.~\eqref{eq:powerlaw} with fixed
$\alpha=1$ and $\psi_{\rm ini}=(5/6)B_0L$. 
We studied $N_A \in \{256,512 \}$, and the initial power law distribution is set up in the range $L^2/N_A<A<5L^2/N_A$.
From the top panel of \figref{fig:Contest1}, we observe that 
the islands in the higher-resolution runs for the power-law case start merging slightly earlier, but the evolution of the macroscopic quantities with time converges to the same power-law. 
The evolution of the quantities is also smoother because the improved area-resolution allows better resolution of merging times. 

For the Gaussian case, we initialized the distributions with fixed $\bar{A}=L^2/25$ and $\bar{\psi}=B_0L/2$.
The standard deviations are set as $\sigma_{A}= 0.3\bar{A}=0.47$ and $\sigma_\psi= 3/2\bar{\psi}=4.71$. 
Two groups of runs are performed: one with fixed $N_\psi=4$ and varying $N_A \in \{64,128,256,512,1024 \}$, and one with fixed $N_A=32$ and varying $N_\psi \in \{4,8,16 \}$.
In the bottom left panel of \figref{fig:Contest1} we observe that the results converge well for macroscopic quantities of the system for various $N_A$ values.
The higher resolution introduces smoother evolution, as already commented on for the power-law case.
In the bottom right panel of \figref{fig:Contest1}, we see that the evolution of $\mathcal{E}$, $N$, and $\langle A \rangle$ are almost identical for various values of $N_\psi$.
The convergence of results with small values of $N_\psi$ is not surprising given our assumption of flux conservation during mergers.

These results show the convergence of the numerical solution with increasing resolution, and that the resolutions used in Section~\ref{sec:delta}-\ref{sec:powerlaw} are adequate to represent the evolution of the macroscopic quantities of interest in this study.

\section{Multi-scale rules of merger and spectrum evolution}
\label{sec:multi_scale_rules}
In \citet{zhou2019magnetic}, we consider the situation where the system contains an ensemble of identical islands, and derive the following scaling laws for the evolution of the system:
\begin{align}
 k &= k_0 \tilde{t}^{-1/2},  \quad B = B_0 \tilde{t}^{-1/2},\label{eq:expkB}\\
\mathcal{E} &= \mathcal{E}_0 \tilde{t}^{-1}, \quad N = N_0 \tilde{t}^{-1}, \quad \psi = \psi_0,\label{eq:expENpsi}
\end{align}
where $k \equiv \pi/2R$, with $R$ the island radius, and $\tilde{t}$ is time normalized to the reconnection time scale, $\tilde{t}=t/\tau_0$. 
In the above expressions, the subscript $0$ denotes quantities evaluated at $\tilde{t}=t_{\rm ini}$, an ``initial'' time that is chosen arbitrarily as long as the evolution of quantities follows the power laws. Note that Eqs.~\eqref{eq:expkB} and~\eqref{eq:expENpsi} are valid to describe the dynamics of the system only when the time scale we study is much longer than one merger time of islands, i.e., in the ``continuous'' limit. Therefore, the early-time dynamics at $\tilde{t} \lesssim 1$ cannot be described by a power-law evolution. 
The self-similar features we are about to derive based on Eqs.~\eqref{eq:expkB} and~\eqref{eq:expENpsi} only apply to the asymptotic long-time limit of the system ($\tilde{t} \geq t_{\rm ini} \gg 1$).

The above time evolution of single-scale quantities [Eqs.~\eqref{eq:expkB} and \eqref{eq:expENpsi}] describes a system with islands that are identical at any given time. However, it can be generalized to describe a multi-scale system, i.e., a system consisting of islands with distributions of sizes and fluxes, as we will explain in the following.

We first note that the value of the time normalization $\tau_0$, which is the merger time of the initial islands, depends on the size and magnetic field of the islands. In the resistive MHD Sweet-Parker regime, it can be written as
\begin{equation}
   \tau_0 = \beta_{rec}^{-1}\frac{R_0}{v_{A,0}}=S^{1/2} \frac{R_0}{v_{A,0}} \propto (R_0 B_0)^{1/2}\frac{R_0}{B_0}\propto k_0^{-3/2}B_0^{-1/2}.
   \label{eq:tau0}
\end{equation}
Therefore, for islands with different $k_0$, though the form of the time evolution expression with normalized time $\tilde{t}$ appears identical, the values of their normalization factor, $\tau_0(k_0)$, are different. 
At any given physical time $t$, islands with different $k_0$ reach different generations; and the $k_0$ dependence in $\tau_0$ will change the scaling of the magnetic energy spectrum $U(k,t)$ over $k$ and $t$.

Assume, therefore, that we have initial islands with different sizes, denoted as $k_0$. By adding the subscripts $k_0$, we have the variables: $\psi_{k_0}(t)$, $R_{k_0}(t)$, $B_{k_0}(t)$, $N_{k_0}(t)$, $\mathcal{E}_{k_0}(t)$, $S_{k_0}$ and $\tau_{0,k_0}$, corresponding to the physical quantities associated to the islands with initial wave number $k_0$.
Any global quantity, $Q(t)$, can thus be written as $Q(t)=\int dk_0 \ Q_{k_0}(t)$.
The above notation involves an implicit assumption that the initial wave number $k_0$ is the only parameter needed to characterise an island. That is, for islands with same $k_0$, their other quantities, such as magnetic field and flux, are all identical.
The multi-scale description will be given in $k$ space, where the general idea is to apply the single-scale description separately to each initial scale $k_0$, with an assumption that interactions are preferentially local in $k$-space, i.e., that coalescence events occur mainly between identical-size islands.
This assumption is indeed corroborated by our numerical results reported in Section~\ref{sec:delta}-\ref{sec:powerlaw}, which show that relaxing the constraint of identical-island-merger only causes slight differences in the evolution of global quantities.

We first present this forward in time: from $k_0$ to $k$ at some later time $t$. It is similar to a Lagrangian approach in this $k_0$-space, where we view various quantities as functions of $k_0$ and $t$. 
Subsequently, we invert these relationships and go to an Euler representation where we describe quantities as functions of present wave number $k$ and $t$. 
We note that the evolution of islands with different initial sizes $k_0$ proceeds at different rates, determined in part by their different initial reconnection time scales $\tau_{0,k_0}$:
\begin{equation}
\begin{aligned}
\label{eq:tau0k0}
  \tau_{0,k_0} = &\beta_{rec}^{-1}\frac{R_{k_0}}{v_{A0,k_0}}=S_{k_0}^{1/2} \frac{R_{k_0}}{v_{A0,k_0}} \propto k_0^{-3/2}B_{0,k_0}^{-1/2}.
\end{aligned}
\end{equation}
 Thus, since $\tau_{0,k_0}$ depends on $k_0$, by the time we observe these islands at some given later time $t$, they have undergone different numbers of coalescence stages.
 That is, they have reached different generation numbers, $n_{k_0}(t)$, determined as  $n_{k_0}(t) = \log_2 (t/\tau_{0,k_0})$. 
Combining Eqs.~\eqref{eq:expkB} and \eqref{eq:tau0k0}, the wave number that islands with an initial $k_0$ possess at time $t$ can be written as:
\begin{equation}
\begin{aligned}
k_{k_0}(t) = k_0 (t/\tau_{0,k_0})^{-1/2}  \propto
k_0^{1/4}  t^{-1/2}  B_{0,k_0}^{-1/4}. 
\end{aligned}
\end{equation}

We first look at the time evolution of global quantities in this multi-scale hierarchical model. 
Here we use $N$ as an example.
The initial condition is $N(t_{\rm ini}) = \int dk_0 N_{k_0}(t_{\rm ini})$. 
We have assumed that the single-scale description can be applied separately to each initial scale $k_0$.
Therefore, each $N_{k_0}$ is expected to follow Eq.~\eqref{eq:expENpsi} independently, and the evolution of $N$ can thus be written as:
\begin{equation}
\begin{aligned}
    N(t) &= \int dk_0\ N_{k_0}(t) = \int dk_0\ N_{k_0}(t_{\rm ini})\left( \frac{t}{\tau_{0,k_0}}\right)^{-1} \\
    &= t^{-1} \int dk_0 \ N_{k_0}(t_{\rm ini}) k_0^{3/2}B_{k_0}^{1/2} \propto t^{-1}.
\end{aligned}
\end{equation}
Similarly, we obtain $\mathcal{E} \propto t^{-1}$.
That is, in a multi-scale system, the indices of the power-law time dependence of global quantities are the same as in a single-scale system, while the normalization factors of the time traces of global quantities become nontrivial functions of $k_0$ and are determined by the initial island distribution.
This conclusion can be applied to systems with an arbitrary initial island distribution function and it agrees with our numerical results in Section~\ref{sec:NR} with three different types of initial distribution.

We proceed to discuss the magnetic energy spectrum, $U(k,t)$, in this multi-scale system.
We assume that the initial magnetic field of an island with initial wave number $k_0$ satisfies the scaling  $B_{0,k_0}\propto k_0^\theta$. 
The magnetic flux thus satisfies the scaling $\psi_{0,k_0} \propto B_{0,k_0}/k_0 \propto k_0^{\theta-1}$, and
the initial merger time is $\tau_{0,k_0} \propto k_0^{-3/2}B_{0,k_0}^{-1/2} \propto k_0^{-(3+\theta)/2}$.
We also consider an ensemble of islands with a initial power-law distribution over size $f(\psi,A,t=t_{\rm ini})\propto A^{-\alpha}\sim k_0^{2\alpha-3}$. 
With our assumption that $k_0$ (and thus the initial area of islands $A \sim k_0^{-2}$) is the only parameter to characterize an island, the distribution over flux $\psi$ is determined by that over $A$.
Therefore, the initial distribution can be written as 
\begin{equation}
\label{eq:powerlaw_both}
    f(\psi, A, t=t_{\rm ini})=f_0 A^{-\alpha}  \delta(\psi- C A^{-(\theta-1)/2}),
\end{equation}
where $C$ is a geometric constant.
The corresponding initial $A$-spectrum is $U(A,t_{\rm ini}) = \int d\psi f(\psi,A,t_{\rm ini}) \psi^2 \sim A^{-\alpha-\theta + 1}$ and the initial $k$-spectrum is $U(k,t_{\rm ini}) \sim k^{2\alpha+2\theta-5}$ [using Eq.~\eqref{eq:Uk_def}]. The initial number density spectrum $N(A,t_{\rm ini}) = \int d\psi f(\psi,A,t_{\rm ini})\sim A^{-\alpha}$ has the same exponent as the initial area distribution function.

The initial magnetic energy spectrum, by its definition, can be related to $B_{k_0}$ and $N_{k_0}$ as following:
\begin{equation}
    U(k,t_{\rm ini}) = \frac{1}{8 \pi}\int dk_0\ \left[B_{k_0}^2(t_{\rm ini})/k_0^2 \right] N_{k_0}(t_{\rm ini})  \delta(k-k_0).
\end{equation}

Islands with different initial sizes $k_0$ evolve independently, following the single-scale description [Eq.~\eqref{eq:expkB} and Eq.~\eqref{eq:expENpsi}]. Hence the time evolution of the spectrum can be expressed as:
\begin{equation}
\begin{aligned}
\label{eq:U_integral}
        U(k,t) &= {1\over {8 \pi}} \int d k_0 \bigl[ B_{k_0}^2(t) / k_{k_0}^2(t) \bigr] N_{k_0}(t) \delta[k - k_{k_0}(t)] \\  
    	&= \frac{1}{8 \pi} \int dk_0\ B_{k_0}^2(t_{\rm ini}) \left(\frac{t}{\tau_{0,k_0}}\right)^{-1} k_0^{-2} \left(\frac{t}{\tau_{0,k_0}}\right) N_{k_0}(t_{\rm ini})\left(\frac{t}{\tau_{0,k_0}}\right)^{-1}\delta \left[k-k_0 \left(\frac{t}{\tau_{0,k_0}}\right)^{-\frac{1}{2}} \right].
\end{aligned}
\end{equation}
The initial magnetic spectrum has a power-law inertial range [consistent with the $U(k,t_{\rm ini})$ that we obtained earlier by integrating the distribution function Eq.\eqref{eq:powerlaw_both}]:
\begin{equation}
    \left[B_{k_0}^2(t_{\rm ini})/k_0^2 \right] N_{k_0}(t_{\rm ini}) \propto k_0^{2\theta+2\alpha-5},
\end{equation}
and $\tau_{0,k_0} \propto k_0^{-(3+\theta)/2}$.
The delta-function $\delta [k-k_0 (t/\tau_{0,k_0})^{-1/2}]$ is strictly valid as a function of $k_0$ only when $\theta \neq 1$ [for $\theta=1$, $\tau_{0,k_0}\propto k_0^{-2}$ and $k_0(t/\tau_{0,k_0})^{-1/2} \propto t^{-1/2}$ no longer depends on $k_0$], and then the integral in Eq.~\eqref{eq:U_integral} can be evaluated at later times as
\begin{equation}
\label{eq:U_at_t}
    	U(k,t)= t^{-\frac{1}{2}} \tau_{0,k_0}^{\frac{1}{2}} \left[B_{k_0}^2(t_{\rm ini})/k_0^2 \right] N_{k_0}(t_{\rm ini}) \propto t^{-\kappa} k^{-\gamma},
\end{equation}
where $k_0=k(t/\tau_{0,k_0})^{1/2}$, and the exponents $\kappa$ and $\gamma$ are functions of $\alpha$ and $\theta$:
\begin{equation}
\label{eq:define_alpha_gamma}
   \kappa =\frac{4\alpha+4\theta-12}{\theta-1}, \qquad 
    \gamma =\frac{8\alpha+7\theta-23}{\theta-1}.
\end{equation}
These two exponents are related as
\begin{equation}
  2\kappa=\gamma+1. 
  \label{eq:theta_gamma_relation}
\end{equation}
The $t^{-\kappa} k^{-\gamma}$ expression explicitly shows that the evolution of magnetic energy spectrum $U(k,t)$ in a multi-scale system is also self-similar. 
The initial spectrum $U(k,t_{\rm ini})\propto k^{2\alpha+2\theta-5}$ is already undergoing a self-similar evolution and should also be described by Eq.~\eqref{eq:U_at_t}.
By equating its exponent and that of the self-similar spectrum $U(k,t)$, i.e., $2\theta+2\alpha-5=-\gamma$, we obtain $\alpha=3-\theta$.
Using this relation and Eqs.~\eqref{eq:define_alpha_gamma}, we find the solution $\gamma=-(\theta-1)/(\theta-1)=-1$.
This corresponds to $\kappa=0$ [Eq.~\eqref{eq:theta_gamma_relation}].
However, it does not mean the islands are not evolving. The spectrum as a whole is moving to smaller $k$; but at every $k$ within the self-similar range, the evolution is such that the value of $U(k)$ remains the same. 
This calculation suggests that the long-time behaviour of the spectrum evolution is self-similar and the spectrum is expected to exhibit a $k^{1}$ (corresponding to $A^{-2}$ ) inertial range. 

The above result is independent of the value of $\theta$ and thus remains valid when $\theta$ is arbitrarily close to 1.
In the case of $\theta$ approaching 1, $\psi_{0,k_0}\propto B_{0,k_0}/k_0 \propto k_0^\theta/k_0$ is independent of $k_0$, and so all the initial islands have identical flux, independent of their sizes. This is indeed the case considered in our numerical study in Section~\ref{sec:powerlaw}.
In our numerical results, the magnetic energy spectra starting with different initial indices eventually evolve to power-law spectra with index $\gamma=-1$ (shown in Figure~\ref{fig:power_spectrum}, bottom panel).
It is consistent with the idea that, at later times, the evolution of the spectrum is self-similar and follows the $\gamma=-1$ solution. 
We note, however, that this calculation does not prove that this $\gamma=-1$ solution is an ``attractor''; that is, how systems with different initial spectra enter this self-similar phase after a relatively short early stage is still unclear.

 In the collisionless reconnection regime, the normalized reconnection rate $\beta_{\rm rec} \simeq 0.1$ is a constant. 
Therefore, the dependence of $\tau_0$ on the size and magnetic field of islands [Eq.~\eqref{eq:tau0}] becomes:
\begin{align}
    \tau_{0,k_0}=\beta^{-1}_{\rm rec} \frac{R_{k_0}}{v_{A0,k_0}} \propto k_0^{-1} B_{0,k_0}^{-1} \propto k_0^{-(\theta+1)}.
\end{align}
Following the same procedure as laid out above for the resistive-MHD Sweet-Parker case, we obtain the spectrum evolution $U(k,t) \propto t^{-\kappa}k^{-\gamma}$ where in this case
\begin{equation}
\label{eq:define_alpha_gamma_collisioneless}
   \kappa =\frac{2\alpha+2\theta-6}{\theta-1}, \qquad 
    \gamma =\frac{4\alpha+3\theta-11}{\theta-1}.
\end{equation}
Similarly to the resistive MHD case, we obtain the solution $\gamma=-1$ for this self-similar evolution. Our numerical simulations of the collisionless case agree with this result (not shown here).

\bibliographystyle{jpp}
\bibliography{ref}

\begin{thebibliography}{118}
\expandafter\ifx\csname natexlab\endcsname\relax\def\natexlab#1{#1}\fi
\def\au#1{#1} \def\ed#1{#1} \def\yr#1{#1}\def\at#1{#1}\def\jt#1{\textit{#1}}
  \def\bt#1{#1}\def\bvol#1{\textbf{#1}} \def\vol#1{#1} \def\pg#1{#1}
  \def\publ#1{#1}\def\arxiv#1{#1}\def\org#1{#1}\def\st#1{\textit{#1}}

\bibitem[Ball {\em et~al.\/}(2018)Ball, Sironi \& {\"O}zel]{ball2018electron}
{\sc \au{Ball, D.}, \au{Sironi, L.} \& \au{{\"O}zel, F.}} \yr{2018}
  \at{Electron and proton acceleration in trans-relativistic magnetic
  reconnection: dependence on plasma beta and magnetization}.  \jt{The
  Astrophysical Journal}  \bvol{862}~(1),  \pg{80}.

\bibitem[Bhat {\em et~al.\/}(2021)Bhat, Zhou \& Loureiro]{bhat2021inverse}
{\sc \au{Bhat, P.}, \au{Zhou, M.} \& \au{Loureiro, N.~F.}} \yr{2021}
  \at{Inverse energy transfer in decaying, three-dimensional, non-helical
  magnetic turbulence due to magnetic reconnection}.  \jt{Monthly Notices of
  the Royal Astronomical Society}  \bvol{501}~(2),  \pg{3074--3087}.

\bibitem[{Bhattacharjee} {\em et~al.\/}(2009){Bhattacharjee}, {Huang}, {Yang}
  \& {Rogers}]{bhattacharjee2009fast}
{\sc \au{{Bhattacharjee}, A.}, \au{{Huang}, Y.}, \au{{Yang}, H.} \&
  \au{{Rogers}, B.}} \yr{2009}  \at{{Fast reconnection in
  high-{Lundquist}-number plasmas due to the plasmoid Instability}}.
  \jt{Physics of Plasmas}  \bvol{16}~(11),  \pg{112102}.

\bibitem[{Birn} {\em et~al.\/}(2001){Birn}, {Drake}, {Shay}, {Rogers},
  {Denton}, {Hesse}, {Kuznetsova}, {Ma}, {Bhattacharjee}, {Otto} \&
  {Pritchett}]{birn2001}
{\sc \au{{Birn}, J.}, \au{{Drake}, J.~F.}, \au{{Shay}, M.~A.}, \au{{Rogers},
  B.~N.}, \au{{Denton}, R.~E.}, \au{{Hesse}, M.}, \au{{Kuznetsova}, M.},
  \au{{Ma}, Z.~W.}, \au{{Bhattacharjee}, A.}, \au{{Otto}, A.} \&
  \au{{Pritchett}, P.~L.}} \yr{2001}  \at{{Geospace Environmental Modeling
  (GEM) magnetic reconnection challenge}}.  \jt{Journal of Geophysical
  Research}  \bvol{106}~(A3),  \pg{3715--3720}.

\bibitem[Biskamp {\em et~al.\/}(1995)Biskamp, Schwarz \& Drake]{biskamp1995ion}
{\sc \au{Biskamp, D.}, \au{Schwarz, E.} \& \au{Drake, J.~F.}} \yr{1995}
  \at{Ion-controlled collisionless magnetic reconnection}.  \jt{Physical Review
  Letters}  \bvol{75}~(21),  \pg{3850}.

\bibitem[Biskamp \& Welter(1989)]{biskamp1989dynamics}
{\sc \au{Biskamp, Dieter} \& \au{Welter, Helmut}} \yr{1989}  \at{Dynamics of
  decaying two-dimensional magnetohydrodynamic turbulence}.  \jt{Physics of
  Fluids B: Plasma Physics}  \bvol{1}~(10),  \pg{1964--1979}.

\bibitem[Boldyrev \& Loureiro(2017)]{boldyrev2017magnetohydrodynamic}
{\sc \au{Boldyrev, S.} \& \au{Loureiro, N.~F.}} \yr{2017}
  \at{Magnetohydrodynamic turbulence mediated by reconnection}.  \jt{The
  Astrophysical Journal}  \bvol{844}~(2),  \pg{125}.

\bibitem[Borg {\em et~al.\/}(2012)Borg, Taylor \& Eastwood]{borg2012}
{\sc \au{Borg, A.~L.}, \au{Taylor, M.} \& \au{Eastwood, J.~P.}} \yr{2012}
  \at{Observations of magnetic flux ropes during magnetic reconnection in the
  earth's magnetotail.}  \jt{Annales Geophysicae (09927689)}  \bvol{30}~(5).

\bibitem[Brandenburg \& Kahniashvili(2017)]{brandenburg2017classes}
{\sc \au{Brandenburg, A.} \& \au{Kahniashvili, T.}} \yr{2017}  \at{Classes of
  hydrodynamic and magnetohydrodynamic turbulent decay}.  \jt{Physical Review
  Letters}  \bvol{118}~(5),  \pg{055102}.

\bibitem[Brandenburg {\em et~al.\/}(2015)Brandenburg, Kahniashvili \&
  Tevzadze]{brandenburg2015nonhelical}
{\sc \au{Brandenburg, A.}, \au{Kahniashvili, T.} \& \au{Tevzadze, A.~G.}}
  \yr{2015}  \at{Nonhelical inverse transfer of a decaying turbulent magnetic
  field}.  \jt{Physical Review Letters}  \bvol{114}~(7),  \pg{075001}.

\bibitem[Burgers(1948)]{burgers1948mathematical}
{\sc \au{Burgers, J.M.}} \yr{1948}  \bt{ \at{A mathematical model illustrating
  the theory of turbulence}}.  \st{Advances in Applied Mechanics},
  \vol{vol.~1},  \pg{pp. 171--199}.  \publ{Elsevier}.

\bibitem[Cartwright \& Moldwin(2010)]{cartwright2010heliospheric}
{\sc \au{Cartwright, M.~L.} \& \au{Moldwin, M.~B.}} \yr{2010}  \at{Heliospheric
  evolution of solar wind small-scale magnetic flux ropes}.  \jt{Journal of
  Geophysical Research: Space Physics}  \bvol{115}~(A8).

\bibitem[{Cassak} {\em et~al.\/}(2017){Cassak}, {Liu} \&
  {Shay}]{cassak2017review}
{\sc \au{{Cassak}, P.~A.}, \au{{Liu}, Y.~H.} \& \au{{Shay}, M.~A.}} \yr{2017}
  \at{{A review of the 0.1 reconnection rate problem}}.  \jt{Journal of Plasma
  Physics}  \bvol{83}~(5),  \pg{715830501}.

\bibitem[Cassak \& Shay(2007)]{cassak2007scaling}
{\sc \au{Cassak, P.~A.} \& \au{Shay, M.~A.}} \yr{2007}  \at{Scaling of
  asymmetric magnetic reconnection: General theory and collisional
  simulations}.  \jt{Physics of Plasmas}  \bvol{14}~(10),  \pg{102114}.

\bibitem[Cazzola {\em et~al.\/}(2016)Cazzola, Curreli, Markidis \&
  Lapenta]{cazzola2016}
{\sc \au{Cazzola, E.}, \au{Curreli, D.}, \au{Markidis, S.} \& \au{Lapenta, G.}}
  \yr{2016}  \at{On the ions acceleration via collisionless magnetic
  reconnection in laboratory plasmas}.  \jt{Physics of Plasmas}
  \bvol{23}~(11),  \pg{112108}.

\bibitem[{Cerutti} {\em et~al.\/}(2013){Cerutti}, {Werner}, {Uzdensky} \&
  {Begelman}]{cerutti2013}
{\sc \au{{Cerutti}, B.}, \au{{Werner}, G.~R.}, \au{{Uzdensky}, D.~A.} \&
  \au{{Begelman}, M.~C.}} \yr{2013}  \at{{Simulations of Particle Acceleration
  beyond the Classical Synchrotron Burnoff Limit in Magnetic Reconnection: An
  Explanation of the Crab Flares}}.  \jt{The Astrophysical Journal}
  \bvol{770},  \pg{147}.

\bibitem[{Christensson} {\em et~al.\/}(2001){Christensson}, {Hindmarsh} \&
  {Brandenburg}]{christensson2001}
{\sc \au{{Christensson}, Mattias}, \au{{Hindmarsh}, Mark} \& \au{{Brandenburg},
  Axel}} \yr{2001}  \at{{Inverse cascade in decaying three-dimensional
  magnetohydrodynamic turbulence}}.  \jt{Physical Review E}  \bvol{64}~(5),
  \pg{056405}.

\bibitem[Comisso {\em et~al.\/}(2018)Comisso, Huang, Lingam, Hirvijoki \&
  Bhattacharjee]{comisso2018magnetohydrodynamic}
{\sc \au{Comisso, L.}, \au{Huang, Y-M.}, \au{Lingam, M.}, \au{Hirvijoki, E.} \&
  \au{Bhattacharjee, A.}} \yr{2018}  \at{Magnetohydrodynamic turbulence in the
  plasmoid-mediated regime}.  \jt{The Astrophysical Journal}  \bvol{854}~(2),
  \pg{103}.

\bibitem[{Dahlin} {\em et~al.\/}(2014){Dahlin}, {Drake} \&
  {Swisdak}]{dahlin2014}
{\sc \au{{Dahlin}, J.~T.}, \au{{Drake}, J.~F.} \& \au{{Swisdak}, M.}} \yr{2014}
   \at{{The mechanisms of electron heating and acceleration during magnetic
  reconnection}}.  \jt{Physics of Plasmas}  \bvol{21}~(9),  \pg{092304}.

\bibitem[{Dahlin} {\em et~al.\/}(2017){Dahlin}, {Drake} \&
  {Swisdak}]{dahlin2017}
{\sc \au{{Dahlin}, J.~T.}, \au{{Drake}, J.~F.} \& \au{{Swisdak}, M.}} \yr{2017}
   \at{{The role of three-dimensional transport in driving enhanced electron
  acceleration during magnetic reconnection}}.  \jt{Physics of Plasmas}
  \bvol{24}~(9),  \pg{092110}.

\bibitem[Daughton \& Karimabadi(2007)]{daughton2007collisionless}
{\sc \au{Daughton, W.} \& \au{Karimabadi, H.}} \yr{2007}  \at{Collisionless
  magnetic reconnection in large-scale electron-positron plasmas}.  \jt{Physics
  of Plasmas}  \bvol{14}~(7),  \pg{072303}.

\bibitem[{Daughton} {\em et~al.\/}(2011){Daughton}, {Roytershteyn},
  {Karimabadi}, {Yin}, {Albright}, {Bergen} \& {Bowers}]{daughton2011}
{\sc \au{{Daughton}, W.}, \au{{Roytershteyn}, V.}, \au{{Karimabadi}, H.},
  \au{{Yin}, L.}, \au{{Albright}, B.~J.}, \au{{Bergen}, B.} \& \au{{Bowers},
  K.~J.}} \yr{2011}  \at{{Role of electron physics in the development of
  turbulent magnetic reconnection in collisionless plasmas}}.  \jt{Nature
  Physics}  \bvol{7},  \pg{539--542}.

\bibitem[{Dmitruk} \& {G{\'o}mez}(1999)]{dmitruk1999}
{\sc \au{{Dmitruk}, P.} \& \au{{G{\'o}mez}, D.~O.}} \yr{1999}  \at{{Scaling Law
  for the Heating of Solar Coronal Loops}}.  \jt{The Astrophysical Journal
  Letters}  \bvol{527},  \pg{L63--L66}.

\bibitem[Dong {\em et~al.\/}(2018)Dong, Wang, Huang, Comisso \&
  Bhattacharjee]{dong2018role}
{\sc \au{Dong, C.}, \au{Wang, L.}, \au{Huang, Y-M.}, \au{Comisso, L.} \&
  \au{Bhattacharjee, A.}} \yr{2018}  \at{Role of the plasmoid instability in
  magnetohydrodynamic turbulence}.  \jt{Physical Review Letters}
  \bvol{121}~(16),  \pg{165101}.

\bibitem[{Drake} {\em et~al.\/}(2010){Drake}, {Opher}, {Swisdak} \&
  {Chamoun}]{drake2010magnetic}
{\sc \au{{Drake}, J.~F.}, \au{{Opher}, M.}, \au{{Swisdak}, M.} \&
  \au{{Chamoun}, J.~N.}} \yr{2010}  \at{{A Magnetic Reconnection Mechanism for
  the Generation of Anomalous Cosmic Rays}}.  \jt{The Astrophysical Journal}
  \bvol{709},  \pg{963--974}.

\bibitem[{Drake} {\em et~al.\/}(2012){Drake}, {Swisdak} \&
  {Fermo}]{drake2012power}
{\sc \au{{Drake}, J.~F.}, \au{{Swisdak}, M.} \& \au{{Fermo}, R.}} \yr{2012}
  \at{The power-law spectra of energetic particles during multi-island magnetic
  reconnection}.  \jt{The Astrophysical Journal Letters}  \bvol{763}~(1),
  \pg{L5}.

\bibitem[Drake {\em et~al.\/}(2006)Drake, Swisdak, Schoeffler, Rogers \&
  Kobayashi]{drake2006island}
{\sc \au{Drake, J.~F.}, \au{Swisdak, M.}, \au{Schoeffler, K.~M.}, \au{Rogers,
  B.~N.} \& \au{Kobayashi, S.}} \yr{2006}  \at{Formation of secondary islands
  during magnetic reconnection}.  \jt{Geophysical Research Letters}
  \bvol{33}~(13).

\bibitem[{Einaudi} \& {Velli}(1999)]{velli1999}
{\sc \au{{Einaudi}, G.} \& \au{{Velli}, M.}} \yr{1999}  \at{{The distribution
  of flares, statistics of magnetohydrodynamic turbulence and coronal
  heating}}.  \jt{Physics of Plasmas}  \bvol{6},  \pg{4146--4153}.

\bibitem[Fermo {\em et~al.\/}(2010)Fermo, Drake \&
  Swisdak]{fermo2010statistical}
{\sc \au{Fermo, R.~L.}, \au{Drake, J.~F.} \& \au{Swisdak, M.}} \yr{2010}  \at{A
  statistical model of magnetic islands in a current layer}.  \jt{Physics of
  Plasmas}  \bvol{17}~(1),  \pg{010702}.

\bibitem[Fried(1959)]{burton1959}
{\sc \au{Fried, B.~D.}} \yr{1959}  \at{Mechanism for instability of transverse
  plasma waves}.  \jt{The Physics of Fluids}  \bvol{2}~(3),  \pg{337--337}.

\bibitem[Furno {\em et~al.\/}(2005)Furno, Intrator, Hemsing, Hsu, Abbate, Ricci
  \& Lapenta]{furno2005coalescence}
{\sc \au{Furno, I.}, \au{Intrator, T.~P.}, \au{Hemsing, E.~W.}, \au{Hsu,
  S.~C.}, \au{Abbate, S.}, \au{Ricci, P.} \& \au{Lapenta, G.}} \yr{2005}
  \at{Coalescence of two magnetic flux ropes via collisional magnetic
  reconnection}.  \jt{Physics of Plasmas}  \bvol{12}~(5),  \pg{055702}.

\bibitem[{Galeev} {\em et~al.\/}(1979){Galeev}, {Rosner} \&
  {Vaiana}]{Galeev1979structured}
{\sc \au{{Galeev}, A.~A.}, \au{{Rosner}, R.} \& \au{{Vaiana}, G.~S.}} \yr{1979}
   \at{{Structured coronae of accretion disks.}}  \jt{The Astrophysical
  Journal}  \bvol{229},  \pg{318--326}.

\bibitem[{Galsgaard} \& {Nordlund}(1996)]{galsgaard1996a}
{\sc \au{{Galsgaard}, K.} \& \au{{Nordlund}, {\AA}.}} \yr{1996}  \at{{Heating
  and activity of the solar corona 1. Boundary shearing of an initially
  homogeneous magnetic field}}.  \jt{Journal of Geophysical Research: Space
  Physics}  \bvol{101},  \pg{13445--13460}.

\bibitem[Gekelman {\em et~al.\/}(2016)Gekelman, De~Haas, Daughton,
  Van~Compernolle, Intrator \& Vincena]{gekelman2016pulsating}
{\sc \au{Gekelman, W.}, \au{De~Haas, T.}, \au{Daughton, W.},
  \au{Van~Compernolle, B.}, \au{Intrator, T.} \& \au{Vincena, S.}} \yr{2016}
  \at{Pulsating magnetic reconnection driven by three-dimensional flux-rope
  interactions}.  \jt{Physical Review Letters}  \bvol{116}~(23),  \pg{235101}.

\bibitem[Gingell {\em et~al.\/}(2019)Gingell, Schwartz, Eastwood, Burch, Ergun,
  Fuselier, Gershman, Giles, Khotyaintsev, Lavraud {\em
  et~al.\/}]{gingell2019observations}
{\sc \au{Gingell, I.}, \au{Schwartz, S.~J.}, \au{Eastwood, J.~P.}, \au{Burch,
  J.~L.}, \au{Ergun, R.~E.}, \au{Fuselier, S.}, \au{Gershman, D.~J.},
  \au{Giles, B.~L.}, \au{Khotyaintsev, Y.~V.}, \au{Lavraud, B.} \& \au{others}}
  \yr{2019}  \at{Observations of magnetic reconnection in the transition region
  of quasi-parallel shocks}.  \jt{Geophysical Research Letters}  \bvol{46}~(3),
   \pg{1177--1184}.

\bibitem[Gruzinov(2001)]{gruzinov2001gamma}
{\sc \au{Gruzinov, A.}} \yr{2001}  \at{Gamma-ray burst phenomenology, shock
  dynamo, and the first magnetic fields}.  \jt{The Astrophysical Journal
  Letters}  \bvol{563}~(1),  \pg{L15}.

\bibitem[{Guo} {\em et~al.\/}(2019){Guo}, {Li}, {Daughton}, {Kilian}, {Li},
  {Liu}, {Yan} \& {Ma}]{Guo2019}
{\sc \au{{Guo}, F.}, \au{{Li}, X.}, \au{{Daughton}, W.}, \au{{Kilian}, P.},
  \au{{Li}, H.}, \au{{Liu}, Y.}, \au{{Yan}, W.} \& \au{{Ma}, D.}} \yr{2019}
  \at{{Determining the Dominant Acceleration Mechanism during Relativistic
  Magnetic Reconnection in Large-scale Systems}}.  \jt{The Astrophysical
  Journal Letters}  \bvol{879}~(2),  \pg{L23}.

\bibitem[Guo {\em et~al.\/}(2020)Guo, Li, Daughton, Li, Kilian, Liu, Zhang \&
  Zhang]{guo2020magnetic}
{\sc \au{Guo, F.}, \au{Li, X.}, \au{Daughton, W.}, \au{Li, H.}, \au{Kilian,
  P.}, \au{Liu, Y.}, \au{Zhang, Q.} \& \au{Zhang, H.}} \yr{2020}  \at{Magnetic
  energy release, plasma dynamics and particle acceleration during relativistic
  turbulent magnetic reconnection}.  \jt{arXiv:2008.02743} .

\bibitem[{Guo} {\em et~al.\/}(2015){Guo}, {Liu}, {Daughton} \& {Li}]{Guo2015}
{\sc \au{{Guo}, F.}, \au{{Liu}, Y.}, \au{{Daughton}, W.} \& \au{{Li}, H.}}
  \yr{2015}  \at{{Particle Acceleration and Plasma Dynamics during Magnetic
  Reconnection in the Magnetically Dominated Regime}}.  \jt{The Astrophysical
  Journal}  \bvol{806}~(2),  \pg{167}.

\bibitem[{Hakobyan} {\em et~al.\/}(2020){Hakobyan}, {Petropoulou}, {Spitkovsky}
  \& {Sironi}]{Hakobyan2020a}
{\sc \au{{Hakobyan}, H.}, \au{{Petropoulou}, M.}, \au{{Spitkovsky}, A.} \&
  \au{{Sironi}, L.}} \yr{2020}  \at{{Secondary Energization in Compressing
  Plasmoids during Magnetic Reconnection}}.  \jt{arXiv:2006.12530} .

\bibitem[Hatori(1984)]{hatori1984kolmogorov}
{\sc \au{Hatori, T.}} \yr{1984}  \at{Kolmogorov-style argument for the decaying
  homogeneous mhd turbulence}.  \jt{Journal of the Physical Society of Japan}
  \bvol{53}~(8),  \pg{2539--2545}.

\bibitem[{Holman} {\em et~al.\/}(2003){Holman}, {Sui}, {Schwartz} \&
  {Emslie}]{holman2003electron}
{\sc \au{{Holman}, G.~D.}, \au{{Sui}, L.}, \au{{Schwartz}, R.~A.} \&
  \au{{Emslie}, A.~G.}} \yr{2003}  \at{{Electron Bremsstrahlung Hard X-Ray
  Spectra, Electron Distributions, and Energetics in the 2002 July 23 Solar
  Flare}}.  \jt{The Astrophysical Journal Letters}  \bvol{595},
  \pg{L97--L101}.

\bibitem[{Hosking} \& {Schekochihin}(2021)]{hosking2021}
{\sc \au{{Hosking}, D.~N.} \& \au{{Schekochihin}, A.~A.}} \yr{2021}
  \at{{Reconnection-Controlled Decay of Magnetohydrodynamic Turbulence and the
  Role of Invariants}}.  \jt{Physical Review X}  \bvol{11}~(4),  \pg{041005}.

\bibitem[Hu {\em et~al.\/}(2019)Hu, Chen \& le~Roux]{hu2019}
{\sc \au{Hu, Q.}, \au{Chen, Y.} \& \au{le~Roux, J.}} \yr{2019}  \at{Radial
  evolution of the properties of small-scale magnetic flux ropes in the solar
  wind}.  \jt{Journal of Physics: Conference Series}  \bvol{1332},
  \pg{012005}.

\bibitem[Hu {\em et~al.\/}(2018)Hu, Zheng, Chen, le~Roux \&
  Zhao]{hu2018automated}
{\sc \au{Hu, Q.}, \au{Zheng, J.}, \au{Chen, Y.}, \au{le~Roux, J.} \& \au{Zhao,
  L.}} \yr{2018}  \at{Automated detection of small-scale magnetic flux ropes in
  the solar wind: First results from the wind spacecraft measurements}.
  \jt{The Astrophysical Journal Supplement Series}  \bvol{239}~(1),  \pg{12}.

\bibitem[Huang \& Bhattacharjee(2010)]{huang2010scaling}
{\sc \au{Huang, Y.} \& \au{Bhattacharjee, A}} \yr{2010}  \at{Scaling laws of
  resistive magnetohydrodynamic reconnection in the high-{Lundquist}-number,
  plasmoid-unstable regime}.  \jt{Physics of Plasmas}  \bvol{17}~(6),
  \pg{062104}.

\bibitem[{Huang} \& {Bhattacharjee}(2012)]{huang2012distribution}
{\sc \au{{Huang}, Y.} \& \au{{Bhattacharjee}, A.}} \yr{2012}  \at{{Distribution
  of Plasmoids in High-{Lundquist}-Number Magnetic Reconnection}}.
  \jt{Physical Review Letters}  \bvol{109}~(26),  \pg{265002}.

\bibitem[Intrator {\em et~al.\/}(2013)Intrator, Sun, Dorf, Sears, Feng, Weber
  \& Swan]{intrator2013flux}
{\sc \au{Intrator, T.~P.}, \au{Sun, X.}, \au{Dorf, L.}, \au{Sears, J.~A.},
  \au{Feng, Y.}, \au{Weber, T.~E.} \& \au{Swan, H.~O.}} \yr{2013}  \at{Flux
  ropes and {3D} dynamics in the relaxation scaling experiment}.  \jt{Plasma
  Physics and Controlled Fusion}  \bvol{55}~(12),  \pg{124005}.

\bibitem[Janvier {\em et~al.\/}(2014)Janvier, D{\'e}moulin \&
  Dasso]{janvier2014there}
{\sc \au{Janvier, M.}, \au{D{\'e}moulin, P.} \& \au{Dasso, S.}} \yr{2014}
  \at{Are there different populations of flux ropes in the solar wind?}
  \jt{Solar Physics}  \bvol{289}~(7),  \pg{2633--2652}.

\bibitem[Jara-Almonte {\em et~al.\/}(2016)Jara-Almonte, Ji, Yamada, Yoo \&
  Fox]{jara2016laboratory}
{\sc \au{Jara-Almonte, J.}, \au{Ji, H.}, \au{Yamada, M.}, \au{Yoo, J.} \&
  \au{Fox, W.}} \yr{2016}  \at{Laboratory observation of resistive electron
  tearing in a two-fluid reconnecting current sheet}.  \jt{Physical Review
  Letters}  \bvol{117}~(9),  \pg{095001}.

\bibitem[Ji \& Daughton(2011)]{ji2011phase}
{\sc \au{Ji, Hantao} \& \au{Daughton, William}} \yr{2011}  \at{Phase diagram
  for magnetic reconnection in heliophysical, astrophysical, and laboratory
  plasmas}.  \jt{Physics of Plasmas}  \bvol{18}~(11),  \pg{111207}.

\bibitem[Kato(2005)]{kato2005saturation}
{\sc \au{Kato, T.~N.}} \yr{2005}  \at{Saturation mechanism of the {Weibel}
  instability in weakly magnetized plasmas}.  \jt{Physics of Plasmas}
  \bvol{12}~(8),  \pg{080705}.

\bibitem[{Kato} \& {Takabe}(2008)]{kato2008}
{\sc \au{{Kato}, T.~N.} \& \au{{Takabe}, H.}} \yr{2008}  \at{{Nonrelativistic
  Collisionless Shocks in Unmagnetized Electron-Ion Plasmas}}.  \jt{The
  Astrophysical Journal Letters}  \bvol{681}~(2),  \pg{L93}.

\bibitem[Katz {\em et~al.\/}(2007)Katz, Keshet \& Waxman]{katz2007self}
{\sc \au{Katz, B.}, \au{Keshet, U.} \& \au{Waxman, E.}} \yr{2007}
  \at{Self-similar collisionless shocks}.  \jt{The Astrophysical Journal}
  \bvol{655}~(1),  \pg{375}.

\bibitem[{Klimchuk}(2006)]{Klimchuck2006solving}
{\sc \au{{Klimchuk}, J.~A.}} \yr{2006}  \at{{On Solving the Coronal Heating
  Problem}}.  \jt{Solar Physics}  \bvol{234}~(1),  \pg{41--77}.

\bibitem[{Klimchuk} {\em et~al.\/}(2008){Klimchuk}, {Patsourakos} \&
  {Cargill}]{Klimchuk2008}
{\sc \au{{Klimchuk}, J.~A.}, \au{{Patsourakos}, S.} \& \au{{Cargill}, P.~J.}}
  \yr{2008}  \at{{Highly Efficient Modeling of Dynamic Coronal Loops}}.
  \jt{The Astrophysical Journal}  \bvol{682}~(2),  \pg{1351--1362}.

\bibitem[Lapenta(2008)]{lapenta2008self}
{\sc \au{Lapenta, Giovanni}} \yr{2008}  \at{Self-feeding turbulent magnetic
  reconnection on macroscopic scales}.  \jt{Physical Review Letters}
  \bvol{100}~(23),  \pg{235001}.

\bibitem[{Lazar} {\em et~al.\/}(2009){Lazar}, {Schlickeiser}, {Wielebinski} \&
  {Poedts}]{Lazar2009}
{\sc \au{{Lazar}, M.}, \au{{Schlickeiser}, R.}, \au{{Wielebinski}, R.} \&
  \au{{Poedts}, S.}} \yr{2009}  \at{{Cosmological Effects of Weibel-Type
  Instabilities}}.  \jt{The Astrophysical Journal}  \bvol{693}~(2).

\bibitem[{Lazarian} \& {Opher}(2009)]{lazarian2009model}
{\sc \au{{Lazarian}, A.} \& \au{{Opher}, M.}} \yr{2009}  \at{{A Model of
  Acceleration of Anomalous Cosmic Rays by Reconnection in the Heliosheath}}.
  \jt{The Astrophysical Journal}  \bvol{703},  \pg{8--21}.

\bibitem[Le {\em et~al.\/}(2013)Le, Egedal, Ohia, Daughton, Karimabadi \&
  Lukin]{le2013}
{\sc \au{Le, A.}, \au{Egedal, J.}, \au{Ohia, O.}, \au{Daughton, W.},
  \au{Karimabadi, H.} \& \au{Lukin, V.~S.}} \yr{2013}  \at{Regimes of the
  electron diffusion region in magnetic reconnection}.  \jt{Physical Review
  Letters}  \bvol{110}~(13),  \pg{135004}.

\bibitem[Li {\em et~al.\/}(2019)Li, Guo, Li, Stanier \&
  Kilian]{li2019formation}
{\sc \au{Li, X.}, \au{Guo, F.}, \au{Li, H.}, \au{Stanier, A.} \& \au{Kilian,
  P.}} \yr{2019}  \at{Formation of power-law electron energy spectra in
  three-dimensional low-$\beta$ magnetic reconnection}.  \jt{The Astrophysical
  Journal}  \bvol{884}~(2),  \pg{118}.

\bibitem[Lingam \& Comisso(2018)]{lingam2018maximum}
{\sc \au{Lingam, M.} \& \au{Comisso, L.}} \yr{2018}  \at{A maximum entropy
  principle for inferring the distribution of 3d plasmoids}.  \jt{Physics of
  Plasmas}  \bvol{25}~(1),  \pg{012114}.

\bibitem[Linton(2006)]{linton2006reconnection}
{\sc \au{Linton, M.~G.}} \yr{2006}  \at{Reconnection of nonidentical flux
  tubes}.  \jt{Journal of Geophysical Research: Space Physics}
  \bvol{111}~(A12).

\bibitem[Linton {\em et~al.\/}(2001)Linton, Dahlburg \&
  Antiochos]{linton2001reconnection}
{\sc \au{Linton, M.~G.}, \au{Dahlburg, R.~B.} \& \au{Antiochos, S.~K.}}
  \yr{2001}  \at{Reconnection of twisted flux tubes as a function of contact
  angle}.  \jt{The Astrophysical Journal}  \bvol{553}~(2),  \pg{905}.

\bibitem[Liu {\em et~al.\/}(2014)Liu, Birn, Daughton, Hesse \&
  Schindler]{liu2014}
{\sc \au{Liu, Y.}, \au{Birn, J.}, \au{Daughton, W.}, \au{Hesse, M.} \&
  \au{Schindler, K.}} \yr{2014}  \at{Onset of reconnection in the near
  magnetotail: \text{PIC} simulations}.  \jt{Journal of Geophysical Research:
  Space Physics}  \bvol{119}~(12),  \pg{9773--9789}.

\bibitem[Liu {\em et~al.\/}(2013)Liu, Daughton, Karimabadi, Li \&
  Roytershteyn]{liu2013bifurcated}
{\sc \au{Liu, Y.}, \au{Daughton, W.}, \au{Karimabadi, H.}, \au{Li, H.} \&
  \au{Roytershteyn, V.}} \yr{2013}  \at{Bifurcated structure of the electron
  diffusion region in three-dimensional magnetic reconnection}.  \jt{Physical
  Review Letters}  \bvol{110}~(26),  \pg{265004}.

\bibitem[Loureiro \& Boldyrev(2017{\natexlab{{\em
  a\/}}})]{loureiro2017collisionless}
{\sc \au{Loureiro, N.~F.} \& \au{Boldyrev, S.}} \yr{2017{\natexlab{{\em a\/}}}}
   \at{Collisionless reconnection in magnetohydrodynamic and kinetic
  turbulence}.  \jt{The Astrophysical Journal}  \bvol{850}~(2),  \pg{182}.

\bibitem[Loureiro \& Boldyrev(2017{\natexlab{{\em b\/}}})]{loureiro2017role}
{\sc \au{Loureiro, N.~F.} \& \au{Boldyrev, S.}} \yr{2017{\natexlab{{\em b\/}}}}
   \at{Role of magnetic reconnection in magnetohydrodynamic turbulence}.
  \jt{Physical Review Letters}  \bvol{118}~(24),  \pg{245101}.

\bibitem[{Loureiro} {\em et~al.\/}(2012){Loureiro}, {Samtaney}, {Schekochihin}
  \& {Uzdensky}]{loureiro2012magnetic}
{\sc \au{{Loureiro}, N.~F.}, \au{{Samtaney}, R.}, \au{{Schekochihin}, A.~A.} \&
  \au{{Uzdensky}, D.~A.}} \yr{2012}  \at{Magnetic reconnection and stochastic
  plasmoid chains in high-lundquist-number plasmas}.  \jt{Physics of Plasmas}
  \bvol{19}~(4),  \pg{042303}.

\bibitem[{Loureiro} {\em et~al.\/}(2007){Loureiro}, {Schekochihin} \&
  {Cowley}]{loureiro2007instability}
{\sc \au{{Loureiro}, N.~F.}, \au{{Schekochihin}, A.~A.} \& \au{{Cowley},
  S.~C.}} \yr{2007}  \at{{Instability of current sheets and formation of
  plasmoid chains}}.  \jt{Physics of Plasmas}  \bvol{14}~(10),
  \pg{100703--100703}.

\bibitem[{Lyutikov} {\em et~al.\/}(2017){Lyutikov}, {Sironi}, {Komissarov} \&
  {Porth}]{lyutikov2017a}
{\sc \au{{Lyutikov}, M.}, \au{{Sironi}, L.}, \au{{Komissarov}, S.~S.} \&
  \au{{Porth}, Oliver}} \yr{2017}  \at{{Explosive X-point collapse in
  relativistic magnetically dominated plasma}}.  \jt{Journal of Plasma Physics}
   \bvol{83}~(6),  \pg{635830601}.

\bibitem[Lyutikov {\em et~al.\/}(2017)Lyutikov, Sironi, Komissarov \&
  Porth]{lyutikov2017b}
{\sc \au{Lyutikov, M.}, \au{Sironi, L.}, \au{Komissarov, S.~S.} \& \au{Porth,
  O.}} \yr{2017}  \at{Particle acceleration in relativistic magnetic
  flux-merging events}.  \jt{Journal of Plasma Physics}  \bvol{83}~(6),
  \pg{635830602}.

\bibitem[Mallet {\em et~al.\/}(2017)Mallet, Schekochihin \&
  Chandran]{mallet2017disruption}
{\sc \au{Mallet, A.}, \au{Schekochihin, A.~A.} \& \au{Chandran, B. D.~G.}}
  \yr{2017}  \at{Disruption of sheet-like structures in alfv{\'e}nic turbulence
  by magnetic reconnection}.  \jt{Monthly Notices of the Royal Astronomical
  Society}  \bvol{468}~(4),  \pg{4862--4871}.

\bibitem[{Matthaeus} \& {Lamkin}(1986)]{matthaeus1986}
{\sc \au{{Matthaeus}, W.~H.} \& \au{{Lamkin}, S.~L.}} \yr{1986}  \at{{Turbulent
  magnetic reconnection}}.  \jt{Physics of Fluids}  \bvol{29},
  \pg{2513--2534}.

\bibitem[Medvedev {\em et~al.\/}(2004)Medvedev, Fiore, Fonseca, Silva \&
  Mori]{medvedev2004long}
{\sc \au{Medvedev, M.~V.}, \au{Fiore, M.}, \au{Fonseca, R.~A.}, \au{Silva,
  L.~O.} \& \au{Mori, W.~B.}} \yr{2004}  \at{Long-time evolution of magnetic
  fields in relativistic gamma-ray burst shocks}.  \jt{The Astrophysical
  Journal Letters}  \bvol{618}~(2),  \pg{L75}.

\bibitem[Medvedev \& Loeb(1999)]{medvedev1999generation}
{\sc \au{Medvedev, Mikhail~V} \& \au{Loeb, Abraham}} \yr{1999}  \at{Generation
  of magnetic fields in the relativistic shock of gamma-ray burst sources}.
  \jt{The Astrophysical Journal}  \bvol{526}~(2),  \pg{697}.

\bibitem[Moldwin {\em et~al.\/}(2000)Moldwin, Ford, Lepping, Slavin \&
  Szabo]{moldwin2000small}
{\sc \au{Moldwin, M.~B.}, \au{Ford, S.}, \au{Lepping, R.}, \au{Slavin, J.} \&
  \au{Szabo, A.}} \yr{2000}  \at{Small-scale magnetic flux ropes in the solar
  wind}.  \jt{Geophysical Research Letters}  \bvol{27}~(1),  \pg{57--60}.

\bibitem[Moldwin {\em et~al.\/}(1995)Moldwin, Phillips, Gosling, Scime,
  McComas, Bame, Balogh \& Forsyth]{moldwin1995ulysses}
{\sc \au{Moldwin, M.~B.}, \au{Phillips, J.~L.}, \au{Gosling, J.~T.}, \au{Scime,
  E.~E.}, \au{McComas, D.~J.}, \au{Bame, S.~J.}, \au{Balogh, A.} \&
  \au{Forsyth, R.~J.}} \yr{1995}  \at{Ulysses observation of a noncoronal mass
  ejection flux rope: Evidence of interplanetary magnetic reconnection}.
  \jt{Journal of Geophysical Research: Space Physics}  \bvol{100}~(A10),
  \pg{19903--19910}.

\bibitem[{\O}ieroset {\em et~al.\/}(2016){\O}ieroset, Phan, Haggerty, Shay,
  Eastwood, Gershman, Drake, Fujimoto, Ergun, Mozer {\em
  et~al.\/}]{oieroset2016}
{\sc \au{{\O}ieroset, M.}, \au{Phan, T.~D.}, \au{Haggerty, C.}, \au{Shay,
  M.~A.}, \au{Eastwood, J.~P.}, \au{Gershman, D.~J.}, \au{Drake, J.~F.},
  \au{Fujimoto, M.}, \au{Ergun, R.~E.}, \au{Mozer, F.~S.} \& \au{others}}
  \yr{2016}  \at{Mms observations of large guide field symmetric reconnection
  between colliding reconnection jets at the center of a magnetic flux rope at
  the magnetopause}.  \jt{Geophysical Research Letters}  \bvol{43}~(11),
  \pg{5536--5544}.

\bibitem[Olesen(1997)]{olesen1997inverse}
{\sc \au{Olesen, P.}} \yr{1997}  \at{Inverse cascades and primordial magnetic
  fields}.  \jt{Physics Letters B}  \bvol{398}~(3-4),  \pg{321--325}.

\bibitem[{Opher} {\em et~al.\/}(2011){Opher}, {Drake}, {Swisdak}, {Schoeffler},
  {Richardson}, {Decker} \& {Toth}]{opher2011magnetic}
{\sc \au{{Opher}, M.}, \au{{Drake}, J.~F.}, \au{{Swisdak}, M.},
  \au{{Schoeffler}, K.~M.}, \au{{Richardson}, J.~D.}, \au{{Decker}, R.~B.} \&
  \au{{Toth}, G.}} \yr{2011}  \at{{Is the Magnetic Field in the Heliosheath
  Laminar or a Turbulent Sea of Bubbles?}}  \jt{The Astrophysical Journal}
  \bvol{734},  \pg{71}.

\bibitem[{Parker}(1957)]{parker_sweet_1957}
{\sc \au{{Parker}, E.~N.}} \yr{1957}  \at{{Sweet's Mechanism for Merging
  Magnetic Fields in Conducting Fluids}}.  \jt{Journal of Geophysical Research}
   \bvol{62},  \pg{509--520}.

\bibitem[{Parker}(1972)]{parker1972}
{\sc \au{{Parker}, E.~N.}} \yr{1972}  \at{{Topological Dissipation and the
  Small-Scale Fields in Turbulent Gases}}.  \jt{The Astrophysical Journal}
  \bvol{174},  \pg{499}.

\bibitem[{Parker}(1983{\natexlab{{\em a\/}}})]{parker1983}
{\sc \au{{Parker}, E.~N.}} \yr{1983{\natexlab{{\em a\/}}}}  \at{{Magnetic
  neutral sheets in evolving fields. I - General theory.}}  \jt{The
  Astrophysical Journal}  \bvol{264},  \pg{635--647}.

\bibitem[{Parker}(1983{\natexlab{{\em b\/}}})]{parker1983part2}
{\sc \au{{Parker}, E.~N.}} \yr{1983{\natexlab{{\em b\/}}}}  \at{{Magnetic
  Neutral Sheets in Evolving Fields. II - Formation of the Solar Corona}}.
  \jt{The Astrophysical Journal}  \bvol{264},  \pg{642}.

\bibitem[Parker(1988)]{parker1988nanoflares}
{\sc \au{Parker, E.~N.}} \yr{1988}  \at{Nanoflares and the solar x-ray corona}.
   \jt{The Astrophysical Journal}  \bvol{330},  \pg{474--479}.

\bibitem[Petropoulou {\em et~al.\/}(2016)Petropoulou, Giannios \&
  Sironi]{petropoulou2016blazar}
{\sc \au{Petropoulou, M.}, \au{Giannios, D.} \& \au{Sironi, L.}} \yr{2016}
  \at{Blazar flares powered by plasmoids in relativistic reconnection}.
  \jt{Monthly Notices of the Royal Astronomical Society}  \bvol{462}~(3),
  \pg{3325--3343}.

\bibitem[{Phan} {\em et~al.\/}(2018){Phan}, {Eastwood}, {Shay}, {Drake},
  {Sonnerup}, {Fujimoto}, {Cassak}, {{\O}ieroset}, {Burch}, {Torbert}, {Rager},
  {Dorelli}, {Gershman}, {Pollock}, {Pyakurel}, {Haggerty}, {Khotyaintsev},
  {Lavraud}, {Saito}, {Oka}, {Ergun}, {Retino}, {Le Contel}, {Argall}, {Giles},
  {Moore}, {Wilder}, {Strangeway}, {Russell}, {Lindqvist} \&
  {Magnes}]{phan2018electron}
{\sc \au{{Phan}, T.~D.}, \au{{Eastwood}, J.~P.}, \au{{Shay}, M.~A.},
  \au{{Drake}, J.~F.}, \au{{Sonnerup}, B.~U.~{\"O}.}, \au{{Fujimoto}, M.},
  \au{{Cassak}, P.~A.}, \au{{{\O}ieroset}, M.}, \au{{Burch}, J.~L.},
  \au{{Torbert}, R.~B.}, \au{{Rager}, A.~C.}, \au{{Dorelli}, J.~C.},
  \au{{Gershman}, D.~J.}, \au{{Pollock}, C.}, \au{{Pyakurel}, P.~S.},
  \au{{Haggerty}, C.~C.}, \au{{Khotyaintsev}, Y.}, \au{{Lavraud}, B.},
  \au{{Saito}, Y.}, \au{{Oka}, M.}, \au{{Ergun}, R.~E.}, \au{{Retino}, A.},
  \au{{Le Contel}, O.}, \au{{Argall}, M.~R.}, \au{{Giles}, B.~L.}, \au{{Moore},
  T.~E.}, \au{{Wilder}, F.~D.}, \au{{Strangeway}, R.~J.}, \au{{Russell},
  C.~T.}, \au{{Lindqvist}, P.~A.} \& \au{{Magnes}, W.}} \yr{2018}
  \at{{Electron magnetic reconnection without ion coupling in Earth's turbulent
  magnetosheath}}.  \jt{Nature}  \bvol{557},  \pg{202--206}.

\bibitem[Pouquet(1978)]{pouquet_1978}
{\sc \au{Pouquet, A.}} \yr{1978}  \at{On two-dimensional magnetohydrodynamic
  turbulence}.  \jt{Journal of Fluid Mechanics}  \bvol{88}~(1),  \pg{1–16}.

\bibitem[{Retin{\`o}} {\em et~al.\/}(2007){Retin{\`o}}, {Sundkvist}, {Vaivads},
  {Mozer}, {Andr{\'e}} \& {Owen}]{retino2007situ}
{\sc \au{{Retin{\`o}}, A.}, \au{{Sundkvist}, D.}, \au{{Vaivads}, A.},
  \au{{Mozer}, F.}, \au{{Andr{\'e}}, M.} \& \au{{Owen}, C.~J.}} \yr{2007}
  \at{{In situ evidence of magnetic reconnection in turbulent plasma}}.
  \jt{Nature Physics}  \bvol{3},  \pg{236--238}.

\bibitem[Ruyer \& Fiuza(2018)]{ruyer2018}
{\sc \au{Ruyer, C.} \& \au{Fiuza, F.}} \yr{2018}  \at{Disruption of current
  filaments and isotropization of the magnetic field in counterstreaming
  plasmas}.  \jt{Physical Review Letters}  \bvol{120},  \pg{245002}.

\bibitem[{Samtaney} {\em et~al.\/}(2009){Samtaney}, {Loureiro}, {Uzdensky},
  {Schekochihin} \& {Cowley}]{samtaney2009formation}
{\sc \au{{Samtaney}, R.}, \au{{Loureiro}, N.~F.}, \au{{Uzdensky}, D.~A.},
  \au{{Schekochihin}, A.~A.} \& \au{{Cowley}, S.~C.}} \yr{2009}  \at{{Formation
  of Plasmoid Chains in Magnetic Reconnection}}.  \jt{Physical Review Letters}
  \bvol{103}~(10),  \pg{105004}.

\bibitem[Schekochihin(2020)]{schekochihin2020mhd}
{\sc \au{Schekochihin, A.~A.}} \yr{2020}  \at{Mhd turbulence: A biased review}.
   \jt{arXiv preprint arXiv:2010.00699} .

\bibitem[{Schlickeiser} \& {Shukla}(2003)]{schlicheiser2003}
{\sc \au{{Schlickeiser}, R.} \& \au{{Shukla}, P.~K.}} \yr{2003}
  \at{{Cosmological Magnetic Field Generation by the Weibel Instability}}.
  \jt{The Astrophysical Journal Letters}  \bvol{599}~(2),  \pg{L57--L60}.

\bibitem[Schoeffler {\em et~al.\/}(2011)Schoeffler, Drake \&
  Swisdak]{scheoffler2011}
{\sc \au{Schoeffler, Kevin~M}, \au{Drake, JF} \& \au{Swisdak, M}} \yr{2011}
  \at{The effects of plasma beta and anisotropy instabilities on the dynamics
  of reconnecting magnetic fields in the heliosheath}.  \jt{The Astrophysical
  Journal}  \bvol{743}~(1),  \pg{70}.

\bibitem[Shay {\em et~al.\/}(2001)Shay, Drake, Rogers \&
  Denton]{shay2001alfvenic}
{\sc \au{Shay, M.~A.}, \au{Drake, J.~F.}, \au{Rogers, B.~N.} \& \au{Denton,
  R.~E.}} \yr{2001}  \at{Alfv{\'e}nic collisionless magnetic reconnection and
  the hall term}.  \jt{Journal of Geophysical Research: Space Physics}
  \bvol{106}~(A3),  \pg{3759--3772}.

\bibitem[Silva {\em et~al.\/}(2003)Silva, Fonseca, Tonge, Dawson, Mori \&
  Medvedev]{silva2003interpenetrating}
{\sc \au{Silva, L.~O.}, \au{Fonseca, R.~A.}, \au{Tonge, J.~W.}, \au{Dawson,
  J.~M.}, \au{Mori, W.~B.} \& \au{Medvedev, M.~V.}} \yr{2003}
  \at{Interpenetrating plasma shells: near-equipartition magnetic field
  generation and nonthermal particle acceleration}.  \jt{The Astrophysical
  Journal Letters}  \bvol{596}~(1),  \pg{L121}.

\bibitem[{Sironi} {\em et~al.\/}(2016){Sironi}, {Giannios} \&
  {Petropoulou}]{Sironi2016plasmoids}
{\sc \au{{Sironi}, L.}, \au{{Giannios}, D.} \& \au{{Petropoulou}, M.}}
  \yr{2016}  \at{{Plasmoids in relativistic reconnection, from birth to
  adulthood: first they grow, then they go}}.  \jt{Monthly Notices of the Royal
  Astronomical Society}  \bvol{462}~(1),  \pg{48--74}.

\bibitem[{Sironi} \& {Spitkovsky}(2014)]{sironi2014}
{\sc \au{{Sironi}, L.} \& \au{{Spitkovsky}, A.}} \yr{2014}  \at{{Relativistic
  Reconnection: An Efficient Source of Non-thermal Particles}}.  \jt{The
  Astrophysical Journal Letters}  \bvol{783}~(1),  \pg{L21}.

\bibitem[{Spitkovsky}(2008)]{spitkovsky2008}
{\sc \au{{Spitkovsky}, A.}} \yr{2008}  \at{{On the Structure of Relativistic
  Collisionless Shocks in Electron-Ion Plasmas}}.  \jt{The Astrophysical
  Journal Letters}  \bvol{673}~(1),  \pg{L39}.

\bibitem[{Stone} {\em et~al.\/}(2005){Stone}, {Cummings}, {McDonald},
  {Heikkila}, {Lal} \& {Webber}]{stone2005voyager}
{\sc \au{{Stone}, E.~C.}, \au{{Cummings}, A.~C.}, \au{{McDonald}, F.~B.},
  \au{{Heikkila}, B.~C.}, \au{{Lal}, N.} \& \au{{Webber}, W.~R.}} \yr{2005}
  \at{{Voyager 1 Explores the Termination Shock Region and the Heliosheath
  Beyond}}.  \jt{Science}  \bvol{309},  \pg{2017--2020}.

\bibitem[{Stone} {\em et~al.\/}(2008){Stone}, {Cummings}, {McDonald},
  {Heikkila}, {Lal} \& {Webber}]{stone2008asymmetric}
{\sc \au{{Stone}, E.~C.}, \au{{Cummings}, A.~C.}, \au{{McDonald}, F.~B.},
  \au{{Heikkila}, B.~C.}, \au{{Lal}, N.} \& \au{{Webber}, W.~R.}} \yr{2008}
  \at{{An asymmetric solar wind termination shock}}.  \jt{Nature}  \bvol{454},
  \pg{71--74}.

\bibitem[{Sweet}(1958)]{sweet_neutral_1958}
{\sc \au{{Sweet}, P.~A.}} \yr{1958} {The Neutral Point Theory of Solar Flares}.
   \bt{In {\em Electromagnetic Phenomena in Cosmical Physics\/} (ed.
  \ed{B.~{Lehnert}})},  \st{IAU Symposium},  \vol{vol.~6},  \pg{p. 123}.

\bibitem[Taylor(1974)]{taylor1974relaxation}
{\sc \au{Taylor, J~Brian}} \yr{1974}  \at{Relaxation of toroidal plasma and
  generation of reverse magnetic fields}.  \jt{Physical Review Letters}
  \bvol{33}~(19),  \pg{1139}.

\bibitem[Uzdensky(2020)]{uzdensky2020relativistic}
{\sc \au{Uzdensky, D.~A.}} \yr{2020}  \at{Relativistic nonthermal particle
  acceleration in two-dimensional collisionless magnetic reconnection}.
  \jt{arXiv preprint arXiv:2007.09533} .

\bibitem[{Uzdensky} \& {Goodman}(2008)]{uzdensky2008statistical}
{\sc \au{{Uzdensky}, D.~A.} \& \au{{Goodman}, J.}} \yr{2008}  \at{{Statistical
  Description of a Magnetized Corona above a Turbulent Accretion Disk}}.
  \jt{The Astrophysical Journal}  \bvol{682},  \pg{608--629}.

\bibitem[Uzdensky {\em et~al.\/}(2010)Uzdensky, Loureiro \&
  Schekochihin]{uzdensky2010fast}
{\sc \au{Uzdensky, D.~A.}, \au{Loureiro, N.~F.} \& \au{Schekochihin, A.~A.}}
  \yr{2010}  \at{Fast magnetic reconnection in the plasmoid-dominated regime}.
  \jt{Physical Review Letters}  \bvol{105}~(23),  \pg{235002}.

\bibitem[Wang \& Sheeley~Jr(2006)]{wang2006}
{\sc \au{Wang, Y.} \& \au{Sheeley~Jr, N.}} \yr{2006}  \at{Observations of flux
  rope formation in the outer corona}.  \jt{The Astrophysical Journal}
  \bvol{650}~(2),  \pg{1172}.

\bibitem[Weibel(1959)]{weibel1959}
{\sc \au{Weibel, Erich~S}} \yr{1959}  \at{Spontaneously growing transverse
  waves in a plasma due to an anisotropic velocity distribution}.  \jt{Physical
  Review Letters}  \bvol{2}~(3),  \pg{83}.

\bibitem[{Werner} {\em et~al.\/}(2018){Werner}, {Uzdensky}, {Begelman},
  {Cerutti} \& {Nalewajko}]{Werner2018}
{\sc \au{{Werner}, G.~R.}, \au{{Uzdensky}, D.~A.}, \au{{Begelman}, M.~C.},
  \au{{Cerutti}, B.} \& \au{{Nalewajko}, K.}} \yr{2018}  \at{{Non-thermal
  particle acceleration in collisionless relativistic electron-proton
  reconnection}}.  \jt{Monthly Notices of the Royal Astronomical Society}
  \bvol{473}~(4),  \pg{4840--4861}.

\bibitem[{Werner} {\em et~al.\/}(2016){Werner}, {Uzdensky}, {Cerutti},
  {Nalewajko} \& {Begelman}]{werner2016}
{\sc \au{{Werner}, G.~R.}, \au{{Uzdensky}, D.~A.}, \au{{Cerutti}, B.},
  \au{{Nalewajko}, K.} \& \au{{Begelman}, M.~C.}} \yr{2016}  \at{{The Extent of
  Power-law Energy Spectra in Collisionless Relativistic Magnetic Reconnection
  in Pair Plasmas}}.  \jt{The Astrophysical Journal Letters}  \bvol{816},
  \pg{L8}.

\bibitem[Yamada {\em et~al.\/}(1997)Yamada, Ji, Hsu, Carter, Kulsrud, Bretz,
  Jobes, Ono \& Perkins]{yamada1997study}
{\sc \au{Yamada, M.}, \au{Ji, H.}, \au{Hsu, S.}, \au{Carter, T.}, \au{Kulsrud,
  R.}, \au{Bretz, N.}, \au{Jobes, F.}, \au{Ono, Y.} \& \au{Perkins, F.}}
  \yr{1997}  \at{Study of driven magnetic reconnection in a laboratory plasma}.
   \jt{Physics of Plasmas}  \bvol{4}~(5),  \pg{1936--1944}.

\bibitem[Zhang {\em et~al.\/}(2012)Zhang, Cheng {\em et~al.\/}]{zhang2012}
{\sc \au{Zhang, J.}, \au{Cheng, X.} \& \au{others}} \yr{2012}  \at{Observation
  of an evolving magnetic flux rope before and during a solar eruption}.
  \jt{Nature Communications}  \bvol{3}~(1),  \pg{1--6}.

\bibitem[Zhou {\em et~al.\/}(2019)Zhou, Bhat, Loureiro \&
  Uzdensky]{zhou2019magnetic}
{\sc \au{Zhou, M.}, \au{Bhat, P.}, \au{Loureiro, N.~F.} \& \au{Uzdensky,
  D.~A.}} \yr{2019}  \at{Magnetic island merger as a mechanism for inverse
  magnetic energy transfer}.  \jt{Physical Review Research}  \bvol{1},
  \pg{012004}.

\bibitem[Zhou {\em et~al.\/}(2020)Zhou, Loureiro \& Uzdensky]{zhou2020multi}
{\sc \au{Zhou, M.}, \au{Loureiro, N.~F.} \& \au{Uzdensky, D.~A.}} \yr{2020}
  \at{Multi-scale dynamics of magnetic flux tubes and inverse magnetic energy
  transfer}.  \jt{Journal of Plasma Physics}  \bvol{86}~(4),  \pg{535860401}.

\bibitem[Zhou {\em et~al.\/}(2021)Zhou, Zhdankin, Kunz, Loureiro \&
  Uzdensky]{zhou2021spontaneous}
{\sc \au{Zhou, M.}, \au{Zhdankin, V.}, \au{Kunz, M.~W.}, \au{Loureiro, N.~F.}
  \& \au{Uzdensky, D.~A.}} \yr{2021}  \at{Spontaneous magnetization of
  collisionless plasma through the action of a shear flow}.  \jt{arXiv preprint
  arXiv:2110.01134} .

\bibitem[Zrake(2014)]{zrake2014inverse}
{\sc \au{Zrake, J.}} \yr{2014}  \at{Inverse cascade of nonhelical magnetic
  turbulence in a relativistic fluid}.  \jt{The Astrophysical Journal Letters}
  \bvol{794}~(2),  \pg{L26}.

\bibitem[Zrake \& Arons(2017)]{zrake2017turbulent}
{\sc \au{Zrake, J.} \& \au{Arons, J.}} \yr{2017}  \at{Turbulent magnetic
  relaxation in pulsar wind nebulae}.  \jt{The Astrophysical Journal}
  \bvol{847}~(1),  \pg{57}.

\end{thebibliography}

\makeatletter
\let\jnl@style=\rm
\def\ref@jnl#1{{\jnl@style#1}}
\def\aj{\ref@jnl{AJ}}                   
\def\actaa{\ref@jnl{Acta Astron.}}      
\def\araa{\ref@jnl{ARA\&A}}             
\def\apj{\ref@jnl{ApJ}}                 
\def\apjl{\ref@jnl{ApJ}}                
\def\apjs{\ref@jnl{ApJS}}               
\def\ao{\ref@jnl{Appl.~Opt.}}           
\def\apss{\ref@jnl{Ap\&SS}}             
\def\aap{\ref@jnl{A\&A}}                
\def\aapr{\ref@jnl{A\&A~Rev.}}          
\def\aaps{\ref@jnl{A\&AS}}              
\def\azh{\ref@jnl{AZh}}                 
\def\baas{\ref@jnl{BAAS}}               
\def\bac{\ref@jnl{Bull. astr. Inst. Czechosl.}} 
\def\caa{\ref@jnl{Chinese Astron. Astrophys.}}
\def\cjaa{\ref@jnl{Chinese J. Astron. Astrophys.}}
\def\icarus{\ref@jnl{Icarus}}           
\def\jcap{\ref@jnl{J. Cosmology Astropart. Phys.}}
\def\jrasc{\ref@jnl{JRASC}}             
\def\memras{\ref@jnl{MmRAS}}            
\def\mnras{\ref@jnl{MNRAS}}             
\def\na{\ref@jnl{New A}}                
\def\nar{\ref@jnl{New A Rev.}}          
\def\pra{\ref@jnl{Phys.~Rev.~A}}        
\def\prb{\ref@jnl{Phys.~Rev.~B}}        
\def\prc{\ref@jnl{Phys.~Rev.~C}}        
\def\prd{\ref@jnl{Phys.~Rev.~D}}        
\def\pre{\ref@jnl{Phys.~Rev.~E}}        
\def\prl{\ref@jnl{Phys.~Rev.~Lett.}}    
\def\pasa{\ref@jnl{PASA}}               
\def\pasp{\ref@jnl{PASP}}               
\def\pasj{\ref@jnl{PASJ}}               
\def\rmxaa{\ref@jnl{Rev. Mexicana Astron. Astrofis.}}
\def\qjras{\ref@jnl{QJRAS}}             
\def\skytel{\ref@jnl{S\&T}}             
\def\solphys{\ref@jnl{Sol.~Phys.}}      
\def\sovast{\ref@jnl{Soviet~Ast.}}      
\def\ssr{\ref@jnl{Space~Sci.~Rev.}}     
\def\zap{\ref@jnl{ZAp}}                 
\def\nat{\ref@jnl{Nature}}              
\def\iaucirc{\ref@jnl{IAU~Circ.}}       
\def\aplett{\ref@jnl{Astrophys.~Lett.}} 
\def\apspr{\ref@jnl{Astrophys.~Space~Phys.~Res.}}     
\def\bain{\ref@jnl{Bull.~Astron.~Inst.~Netherlands}}   
\def\fcp{\ref@jnl{Fund.~Cosmic~Phys.}}  
\def\gca{\ref@jnl{Geochim.~Cosmochim.~Acta}}   
\def\grl{\ref@jnl{Geophys.~Res.~Lett.}} 
\def\jcp{\ref@jnl{J.~Chem.~Phys.}}      
\def\jgr{\ref@jnl{J.~Geophys.~Res.}}    
\def\jqsrt{\ref@jnl{J.~Quant.~Spec.~Radiat.~Transf.}}   
\def\memsai{\ref@jnl{Mem.~Soc.~Astron.~Italiana}}    
\def\nphysa{\ref@jnl{Nucl.~Phys.~A}}   
\def\physrep{\ref@jnl{Phys.~Rep.}}   
\def\physscr{\ref@jnl{Phys.~Scr}}   
\def\planss{\ref@jnl{Planet.~Space~Sci.}}   
\def\procspie{\ref@jnl{Proc.~SPIE}}   

\let\astap=\aap
\let\apjlett=\apjl
\let\apjsupp=\apjs
\let\applopt=\ao
\makeatother

\end{document}